\documentstyle[12pt,worldsci]{article}

\def\thefootnote{\fnsymbol{footnote}}
\def\bea {\begin{eqnarray}}
\def\eea {\end{eqnarray}}
\def\be {\begin{equation}}
\def\ee {\end{equation}}
\def\ben{\begin{enumerate}}
\def\een{\end{enumerate}}
\def\bi{\begin{itemize}}
\def\ei{\end{itemize}}
\def\ie{{\it i.e.}}
\def\viz{{\it viz.}\ }
\def\eg{{\it e.g.}}
\def\etal{{\it et al.}}
\def\del{\partial}
\def\L{{\cal L}}
\def\O{{\cal O}}
\def\F{{\cal F}}

\def\prl {Phys. Rev. Lett.\ }
\def\pl {Phys. Lett.\ }
\def\pr {Phys. Rev.\ }
\def\np {Nucl. Phys.\ }
\def\gA{g_{\mbox{\tiny A}}}
\def\rA{r_{\mbox{\tiny A}}}
\def\gAeff{g_{\mbox{\tiny A,eff}}}
\def\gV{g_{\mbox{\tiny V}}}
\def\gM{g_{\mbox{\tiny M}}}
\def\gS{g_{\mbox{\tiny S}}}
\def\gT{g_{\mbox{\tiny T}}}
\def\gP{g_{\mbox{\tiny P}}}
\def\gPeff{g_{\mbox{\tiny P,eff}}}
\def\GA{G_{\mbox{\tiny A}}}
\def\GV{G_{\mbox{\tiny V}}}
\def\GF{G_{\mbox{\tiny F}}}
\def\DRV{\Delta_{\mbox{\tiny R}}^{\mbox{\tiny V}}}
\def\DRA{\Delta_{\mbox{\tiny R}}^{\mbox{\tiny A}}}
\def\mW{m_{\mbox{\tiny W}}}
\def\mA{m_{\mbox{\tiny A}}}
\def\mD{m_{\mbox{\tiny D}}}
\def\mZ{m_{\mbox{\tiny Z}}}
\def\MA{M_{\mbox{\tiny A}}}
\def\MV{M_{\mbox{\tiny V}}}
\def\thW{\theta_{\mbox{\tiny W}}}

\def\gpNN{g_{\pi{\mbox{\tiny NN}}}}

\def\FpNN{F_{\pi{\mbox{\tiny NN}}}}
\def\FANN{F_{\mbox{\tiny ANN}}}
\def\cc{\mbox{\tiny C}}
\def\tbar{\mid \! \mid \! \mid }
\def\mids{\! \mid \! }
\def\hyphen{{\mbox{-}}}
\def\nl{$\,\!$}
\newcommand{\sfrac}[2]{\mbox{\small{$\frac{#1}{#2}$}}}

\begin{document}
\title{CURRENTS AND THEIR COUPLINGS IN THE WEAK SECTOR OF THE
STANDARD MODEL}
\author{I. S. Towner and J. C. Hardy \\
{\em AECL, Chalk River Laboratories, Chalk River \\
Ontario K0J 1J0, Canada}}
\maketitle

\begin{abstract}
{\footnotesize
Beta-decay and muon-capture experiments in nuclei are reviewed.
The conserved vector current hypothesis is confirmed through the
observed constancy of the vector coupling constant determined
from the superallowed Fermi transitions and from the measurement
of the weak-magnetism term in mirror Gamow-Teller transitions.
The axial-vector and pseudoscalar coupling constants in the
nucleon are determined from neutron decay and muon capture
on the proton respectively.  In finite nuclei, evidence
for these coupling constants being reduced relative to their
free-nucleon values is discussed.  Meson-exchange currents are
shown to be an important correction to the time-like part
of the axial current as evident in first-forbidden beta decays.
Tests of the Standard Model are discussed, as well as extensions
beyond it involving right-hand currents
and scalar interactions.
}

\end{abstract}

\renewcommand{\thefootnote}{\#\arabic{footnote}}
\setcounter{footnote}{0}

\section{Introduction} \label{intro}
\subsection{Standard Model for Quarks and Leptons} \label{smql}

The standard electroweak model is based on the gauge group \cite {WSG}
$SU(2) \times U(1)$, with four vector gauge bosons $W^a_{\mu}$, $a = 1,2,3$,
and $B_{\mu}$ for the $SU(2)$ and $U(1)$ sectors respectively, and the
corresponding gauge coupling constants $g$ and $g^\prime$.  The left-handed
fermion fields,
$\psi_i = (\nu_i, \ell^-_i)$ and $(u_i, d_i^\prime\,)$,
comprise neutrinos, $\nu_i$, leptons, $\ell^{-}_i$,
up and down quarks, $u_i$ and $d^{\prime}_i$ for the
$i^{th}$ fermion family, and
transform as doublets under $SU(2)$, where
$d^\prime_i \equiv \sum_j V_{ij} d_j$ and $V$ is the
Cabibbo-Kobayashi-Maskawa (CKM) matrix, which we discuss shortly.  The
right-handed fields are $SU(2)$ singlets.  The interaction part of the
Lagrangian for quarks and leptons is \footnote { We use the Pauli metric as
discussed by deWit and Smith \cite {deWS86} }

\bea
\L_{\rm int} & = & \frac{ig}{2 \sqrt 2} \sum_i \overline{\psi}_i
\gamma_{\mu}(1 + \gamma_5) (T^+ W^+_{\mu} + T^- W^-_{\mu}) \psi_i
\nonumber \\
& + &
ie \sum_i \overline{\psi}_i q_i \gamma_{\mu} \psi_i A_{\mu}
\nonumber \\
& + &
\frac{ie}{2 \cos \thW} \sum_i \overline{\psi}_i \gamma_{\mu}
(V^i + A^i \gamma_5) \psi_i Z_{\mu},
\label{Lint}
\eea

\noindent where $\thW \equiv \tan^{-1}(g^\prime/g)$ is the weak angle, $e
= g \sin \thW$ is the positron electric charge (and hence is positive) and
$A_{\mu} \equiv B_{\mu} \cos \thW + W^3_{\mu} \sin \thW$ is the
massless photon field.  Further, $W^{\pm} \equiv (W^1 \mp iW^2)/2$ and
$Z = -B \sin \thW + W^3 \cos \thW$ are the massive charged and neutral
weak boson fields, respectively, while $T^+$ and $T^-$ are the weak isospin
raising and lowering operators.  The vector and axial couplings are

\bea
V^i &\equiv & t_{3L}(i) - 2q_i \sin^2 \thW
\nonumber \\
A^i &\equiv & t_{3L}(i),
\label{ZVA}
\eea

\noindent where $t_{3L}(i)$ is the weak isospin of fermion $i$ (
$+\mbox{\small{$\frac{1}{2}$}}$ for
$u_i$ and $\nu_i$; $-\mbox{\small{$\frac{1}{2}$}}$
for $d_i$ and $\ell_i$) and $q_i$ is the charge of
$\psi_i$ in units of $e$.
\par
The three terms in the interaction Lagrangian of Eq.\,\ref {Lint} represent
the charged-current weak interaction, the electromagnetic interaction and the
neutral-current weak interaction respectively.  Note that the Lorentz
structure
involves only vectors and axial vectors; there is no compelling experimental
evidence for scalars, pseudoscalars or tensor constructions.  The minimal
Standard Model as described by Eq.\,\ref{Lint} successfully \cite{LLM92}
explains $W$ and $Z$ decays, neutrino-hadron scattering, neutrino-electron
scattering and parity-violating electron-hadron neutral-current
experiments,
providing that  radiative corrections to order $\alpha$,
the fine-structure
constant, are applied.  In this chapter we will focus our discussion on
the charged-current weak interaction in semi-leptonic decays involving quarks
and leptons from the first family.  As an example, consider the decay $d
\rightarrow u e^- \overline{\nu}_e$, for which
the $T$-matrix is
\cite{deWS86}

\be
T_{fi} = {g^2 \over 8} V_{ud}\; \overline{\psi}_u
\gamma_{\mu}(1 + \gamma_5) \psi_d \;
{\delta_{\mu \nu} + k_{\mu}k_{\nu}/\mW^2 \over k^2 + \mW^2} \;
\overline{\psi}_{e^-} \gamma_{\nu} (1 + \gamma_5)
\psi_{\overline{\nu}_e},
\label{Tfik}
\ee

\noindent where $k$ is the momentum transferred between the hadrons and
leptons.  For momenta small compared to the $W$-boson mass, $\mW$, which is
typically the case for leptons and quarks of the first family, $T_{fi}$
reduces to

\bea
T_{fi}& = &{\GF \over \sqrt 2} V_{ud}\; \overline{\psi}_u
\gamma_{\mu}(1 + \gamma_5) \psi_d \;
\overline{\psi}_{e^-} \gamma_{\mu}(1 + \gamma_5) \psi_{\overline{\nu}_e}
\nonumber \\
& \equiv & - {\GF \over \sqrt 2} V_{ud} J^h_{\mu} J^{\ell}_{\mu},
\label{Tfi0}
\eea

\noindent which is the familiar effective four-fermion interaction with the
Fermi constant given by $\GF/\sqrt 2 = g^2/(8\mW^2)$.  Here $J^h_{\mu}$ and
$J_{\mu}^{\ell}$ are known as the hadron and lepton currents, where

\be
J^h_{\mu} = i \overline{\psi}_u \gamma_{\mu} (1 + \gamma_5) \psi_d
\equiv V^h_{\mu} + A^h_{\mu}
\label{Jmuh}
\ee

\noindent  and analogously for
$J_{\mu}^{\ell}$.
\bigskip

\subsection{Cabibbo-Kobayashi-Maskawa Matrix} \label{secCKM}

The up-quark, $u$, obtained in Eq.\,\ref{Tfik} by the explicit
operation of the weak interaction, $\gamma_{\mu}(1+\gamma_5)$, upon the
down-quark, $d$, may not be the physically observed up-quark.  That is, the
quark mass eigenstates are not the same as the weak interaction
eigenstates
\footnote {In the limit that quark masses are zero, the two bases will be
identical and no CKM matrix is required.  In the lepton sector we will not
consider any ramifications of the possibility that neutrino masses are
non-zero and hence not introduce any CKM matrix in the lepton sector.}.
The matrix relating the two bases was defined for
three families and given an explicit parameterization by
Kobayashi and Maskawa
\cite {KM73}\nl .  It generalizes the two-family case where the matrix is
parameterized by a single angle: the Cabibbo angle \cite {Cab63}\nl .
\par
By convention, the three quarks with charge 2/3 ($u$, $c$ and $t$) are unmixed
and
all the mixing is expressed in terms of a $3 \times 3$ unitary matrix $V$
operating on the quarks with charge $-1/3$ $(d,s,b)$:

\be
\left( \begin{array}{c}
d^{\prime} \\ s^{\prime} \\b^{\prime}
\end{array} \right)
= \left( \begin{array}{ccc}
V_{ud} & V_{us} & V_{ub} \\
V_{cd} & V_{cs} & V_{cb} \\
V_{td} & V_{ts} & V_{tb}
\end{array} \right)
\left( \begin{array}{c}
d \\ s \\ b
\end{array} \right) .
\label{CKM}
\ee

\noindent The values of individual matrix elements can in principle be
determined from weak decays of the relevant quarks.  The Particle Data
Group's recommended values \cite{PDG94} are listed in
Table \ref{KMvalu} except for $V_{ud}$, which we
will have more to say about later in this chapter.  By
construction, the CKM matrix is unitary so its nine elements are not
independent of each other.  Typical parameterizations of the matrix involve
three angles and one phase.
In the Standard Model, the explanation of the observed
CP violation requires such a phase
and hence requires at least three families of quarks.
\par
The unitarity of the CKM matrix is most stringently tested
experimentally in the first
row of the matrix:

\be
\mids V_{ud} \mids ^2 + \mids V_{us} \mids ^2 + \mids V_{ub}\mids ^2 = 1.
\label{Unit}
\ee

\begin{table}[t]
\begin{center}
\caption{Parameters of the CKM matrix
\label{KMvalu} }
\vskip 1mm
\begin{tabular}{ll}
\hline
Value & Source \\[1mm]
\hline
$\mids V_{ud} \mids \ = 0.9741 \pm 0.0017$ & Nuclear beta
decay relative to muon decay \\
$\mids V_{us} \mids \ = 0.2205 \pm 0.0018$ & $K_{e3}$ and hyperon
decays \\
$\mids V_{cd} \mids \ = 0.204 \pm 0.017$ & Neutrino production of charm
from valence $d$ quarks \\
$\mids V_{cs} \mids \ = 1.01 \pm 0.18$ & $D \rightarrow
\overline{K}e^+\nu_e$ and $D$ lifetimes \\
$\mids V_{ub}/V_{cb} \mids \ = 0.08 \pm 0.02$ & Energy spectrum
for semileptonic decay of $B$ mesons \\
$\mids V_{cb} \mids \ = 0.040 \pm 0.005$ & $\overline{B} \rightarrow D^*
\ell \overline{\nu}_{\ell}$ and $B$ lifetimes \\
\hline
\end{tabular}
\end{center}
\end{table}

\noindent This test is dominated by the accuracy with which $V_{ud}$ can be
determined.  A failure of the test could signal an extension to the
Standard Model, such as extra $U(1)$ groups or right-handed,
$\gamma_{\mu}(1-\gamma_5)$, weak-interaction couplings.  We return to these
points later.
\bigskip

\subsection{Lagrangian for Nucleons} \label{lforn}

In nuclear physics, a similar Lagrangian for nucleons is required with the
nucleons treated as elementary spin-1/2 fermions.  The Standard Model
interaction Lagrangian of Eq.\,\ref{Lint} may not be immediately applicable
for this case, however, since this elementary Lagrangian is explicitly
constructed for, and tested in, interactions of point-like fermions.
Nucleons, by contrast,
are composite particles of three quarks.  To treat them as
elementary spin-1/2 fermions requires some modification to the elementary
Lagrangian as can be seen from the following argument.

Consider the time component of the vector part of the hadron current
(see Eq.\,\ref{Jmuh}), $V_0^h = \overline{\psi}_u \gamma_4 \psi_d =
\psi^{\dagger}_u \psi_d$,
and evaluate its expectation value between proton and neutron states.  In the
naive quark model, the nucleon  comprises three first-generation
quarks arranged in
a fully symmetric way with orbital angular momentum zero and spin 1/2.
The expectation
value is easily evaluated \cite{Ho89}:

\be
\langle p \mid V^h_0 \mid n \rangle = 1.
\label{qvc}
\ee

\noindent The result of unity implies the vector current normalization in the
nucleon is unchanged from that in the quark Lagrangian.  This, as we shall
see, is as it should be and is a consequence of the conservation of
the vector current.
In contrast to this, consider the space component of the
axial-vector part of the hadron current:
{\bf A}$^h = i \overline{\psi}_u \mbox{\boldmath $\gamma$} \gamma_5
\psi_d \simeq - \mbox{\boldmath $\sigma$}  \psi^{\dagger}_u \psi_d$
in a non-relativistic limit.  Here {\boldmath $\sigma$}
is the Pauli spin matrix, which in its application here flips the
quark spin.  Thus \cite{Ho89}

\be
\langle p \mid A^h_3 \mid n \rangle = - \frac{5}{3}
\label{qac}
\ee

\noindent and the normalization of the axial current is indeed modified in a
nucleon.  The naive quark model value of -5/3 is larger than the experimental
value of -1.26 obtained from the beta decay of the neutron.  However, the
nonrelativisitic treatment given here is not justifiable for quarks as light
as 10 MeV.  Typical relativistic quark calculations give values
in the range -1.1 to
-1.3 depending on the form of the quark wave functions used and the assumed
masses of the quarks.
\par
Since the quark model lacks the predictive precision required, we will follow
the historic and more phenomenological approach of writing down a general
Lagrangian using axioms and symmetry principles to limit the number of
terms.  For example, for the weak vector current, $V^h_{\mu}$, the most
general form is constructed by our writing down all possible ways of making a
Lorentz vector out of the vectors at our disposal:  $p_{\mu}$, the incoming
momentum of the nucleon; $p^\prime_{\mu}$, the outgoing momentum of the
nucleon; and $\gamma_{\mu}$, the spin matrices.  If it is further assumed that
a nucleon in a nucleus undergoing a weak interaction can be treated as a free
nucleon, which for the purposes of constructing
interaction operators satisfies the Dirac equation,
then the number of terms in the current is limited to
three.  This latter assumption is known as the impulse approximation.
Similar arguments also limit the number of terms in the axial current to
three.  Both currents are written as

\bea
V^h_{\mu} & = & i \overline{\psi}_p [\gV(k^2)\gamma_{\mu}
+ {\gM(k^2) \over 2M} \sigma_{\mu \nu} k_{\nu}
+i \gS(k^2)k_{\mu}]\psi_n
\nonumber \\
A^h_{\mu} & = & i \overline{\psi}_p [\gA(k^2) \gamma_{\mu} \gamma_5
+ {\gT(k^2) \over 2M} \sigma_{\mu \nu} k_{\nu} \gamma_5
+i \gP(k^2)k_{\mu} \gamma_5]\psi_n,
\label{VAc}
\eea

\noindent where $k_{\mu} = (p - p^{\prime})_{\mu}$ is the momentum transferred
from hadrons to leptons and $\overline{\psi}_p$ and
$\psi_n$ are field operators representing the proton and neutron respectively.
The form factors $\gV$, $\gM$, $\gS$, $\gA$, $\gT$, $\gP$ are arbitrary
functions of a Lorentz scalar constructed from the momenta $p^\prime$ and
$p$.  It is usual to choose this scalar to be $k^2$; any other scalar can be
expressed in terms of $k^2$ and the nucleon mass, $M$.  The values of these
form factors in the zero-momentum transfer limit, $k^2 \rightarrow 0$, are
known as the vector, weak magnetism, induced scalar, axial vector, induced
tensor and induced pseudoscalar coupling constants respectively.
\par
The number of arbitrary functions in $V^h_{\mu}$ and $A^h_{\mu}$ can be
reduced further by requiring that the currents be Hermitian
and invariant
under time reversal.  These conditions \cite{Be68} lead to the requirement
that $\gV(k^2)$, $\gM(k^2)$, $\gA(k^2)$ and $\gP(k^2)$ be real functions and
that $\gS(k^2) = \gT(k^2) = 0$.  Further, the statement above that the
normalization of the vector current is unchanged in going from the quark to
the nucleon Lagrangian requires $\gV(k^2 \rightarrow 0) = 1$.  Additional
constraints on the form factors follow from the conserved vector current
(CVC) and the partially conserved axial-vector current (PCAC) hypotheses, which
we discuss in later sections.
\par
It is convenient in nuclear physics to make no distinction
between protons and neutrons, but instead to use an isospin formalism.  Let
$\psi = (\psi_p, \psi_n)$ be an eight-component spinor operator representing
the nucleon field, with $\psi_p$ and $\psi_n$ each being four-component field
operators for the proton and neutron respectively.  Further, introduce Pauli
spin matrices, $\tau^1$, $\tau^2$ and $\tau^3$, forming the Cartesian
components
of a vector, {\boldmath $\tau$},
that are generalized in block form to act on $\psi$.
For example, acting with $\tau^3$ on a pure proton state or a pure neutron
state leads to

\be
\tau^3  \left( \begin{array}{c}
\psi_p \\ 0
\end{array} \right)
=  \left( \begin{array}{c}
\psi_p \\ 0
\end{array} \right) \  ; \
\tau^3  \left( \begin{array}{c}
0 \\ \psi_n
\end{array} \right)
=  - \left( \begin{array}{c}
0 \\ \psi_n
\end{array} \right).
\label{tau3}
\ee

\noindent These expressions serve to specify our isospin conventions.  The
currents in Eq.\,\ref{VAc} in an isospin formulation are now written

\bea
V^h_{\mu} & = & i \overline{\psi} \Bigl[ \gV(k^2)\gamma_{\mu}
+ {\gM(k^2) \over 2M} \sigma_{\mu \nu} k_{\nu}
+i \gS(k^2)k_{\mu} \Bigr] \sfrac{1}{2} \tau^a \psi
\nonumber \\
A^h_{\mu} & = & i \overline{\psi} \Bigl[ \gA(k^2)\gamma_{\mu} \gamma_5
+ {\gT(k^2) \over 2M} \sigma_{\mu \nu} k_{\nu} \gamma_5
+i \gP(k^2)k_{\mu} \gamma_5 \Bigr] \sfrac{1}{2} \tau^a \psi,
\label{VAct}
\eea

\noindent where $\tau^a$ stands for $\tau^1 \pm i \tau^2$
with a plus sign for $n \rightarrow p e^- \overline{\nu}_e$ and a minus sign
for $p \rightarrow ne^+ \nu_e$ charge-changing transitions.
\par
Another symmetry of these currents to be investigated is their
behaviour under
charge conjugation, which interchanges particles and
antiparticles, coupled with a 180$^\circ$ rotation about the 2-axis in
isospin space.  This combination is called G-parity and is a
symmetry of the strong interactions \cite{Lee81}:

\be
G = C \exp(i\pi T_2).
\label{Gpar}
\ee

\noindent Now, under charge conjugation, we have \cite{Lee81}

\be
C \psi_i C^{\dagger} = \psi^{\cc}_i \equiv \gamma_2 \psi_i^{\dagger},
\label{Cconj}
\ee

\noindent where $i = p,n$ and the representation of $C$ in terms of the Dirac
matrix, $\gamma_2$, is specific to the metric we are using.  Further, under
an isospin rotation,

\be
\exp(i \pi T_2) \left( \begin{array}{c}
\psi_p \\ \psi_n
\end{array} \right)
= -i \tau^2 \left( \begin{array}{c}
\psi_p \\ \psi_n
\end{array} \right)
= \left( \begin{array}{c}
-\psi_n \\ \psi_p
\end{array} \right).
\label{Irot}
\ee

\noindent Thus, combining these two operations, we have

\be
G \psi_p G^{\dagger} = -\psi^{\cc}_n \ ;
\qquad G \psi_n G^{\dagger} = \psi^{\cc}_p.
\label{Grot}
\ee

\noindent
Applying these steps to the first term in the vector current, Eq.\,
\ref{VAc}, we get

\bea
G V^h_{\mu} G^{\dagger} & = &
G \overline{\psi}_p G^{\dagger} \gamma_{\mu} G \psi_n G^{\dagger}
\nonumber \\
& = &
-i \psi_n^{\dagger \cc} \gamma_4 \gamma_{\mu} \psi^{\cc}_p
\nonumber \\
& = &
-i (\psi^{\cc}_n)^{\dagger}_{\alpha} (\gamma_4 \gamma_{\mu})_{\alpha \beta}
(\psi^{\cc}_p)_\beta
\nonumber \\
& = &
+i (\psi^{\cc}_p)_{\beta} (\gamma_4 \gamma_{\mu})^T_{\beta \alpha}
(\psi^{\cc}_n)^{\dagger}
\nonumber \\
& = &
+ i \psi^{\dagger}_p \gamma_2 \gamma^T_{\mu} \gamma^T_4 \gamma_2 \psi_n
\nonumber \\
& = &
+ i \overline{\psi}_p \gamma_{\mu} \psi_n
\nonumber \\
& = &
V^h_{\mu},
\label{GV}
\eea

\noindent where, in the third line, the expression is written out in
component form and reordered in the fourth line, introducing a minus sign,
because the fermion field operators anticommute.  Thus, the first term in the
vector current is invariant under G-parity transformations.  Applying the
same steps to the first term of the axial-current in Eq.\,\ref{VAc},
we get

\bea
G A^h_{\mu} G^{\dagger} & = &
i G \overline{\psi}_p G^{\dagger}\gamma_{\mu} \gamma_5 G \psi_n G^{\dagger}
\nonumber \\
& = &
- A^h_{\mu}.
\label{GGA}
\eea

\noindent We considered just the first terms of $V^h_{\mu}$ and $A^h_{\mu}$
because these are the only terms in the quark currents of the Standard Model.
Weinberg \cite{We58} has termed currents with the standard quark-model
transformation properties under G-parity as ``first class", while currents
with the opposite properties,

\bea
G V^{h(II)}_{\mu} G^{\dagger} & = & - V^{h(II)}_{\mu}
\nonumber \\
G A^{h(II)}_{\mu} G^{\dagger} & = & A^{h(II)}_{\mu},
\label{GII}
\eea

\noindent are termed ``second class".  Under this classification, the vector
and weak magnetism terms of $V^h_{\mu}$ and the axial-vector and pseudoscalar
terms of $A^h_{\mu}$ are first-class currents, while the induced scalar term
of $V^h_{\mu}$ and the induced tensor term of $A^h_{\mu}$ are second-class
currents.  To date, there is no unambiguous evidence for the presence of
second-class currents in nuclear beta decay, but it is nonetheless an
interesting experimental challenge to place limits on the possible presence
of these second-class currents.

\section{Vector Interaction} \label{VI}
\subsection{Conserved Vector Current Hypothesis}  \label{CVCh}

Briefly, let us return to the quark Lagrangian of Eq.\,\ref{Lint} for the
electromagnetic interaction and, in analogy with Eq.\,\ref{Jmuh}, write down
the electromagnetic current for the first family of quarks,

\be
V^{em}_{\mu} = i \sfrac{2}{3} \overline{\psi}_u \gamma_{\mu} \psi_u
- i \sfrac{1}{3} \overline{\psi}_d \gamma_{\mu} \psi_d,
\label{Jemq}
\ee

\noindent in units of the coupling constant, $e$, which is the
positron electric
charge.  Again, we can introduce isospin by defining a quark doublet spinor
$\psi_q = (\psi_u, \psi_d)$ such that

\be
V_{\mu}^{em} = i Q_{+} \overline{\psi}_q \gamma_{\mu} \psi_q I
+ i Q_{-} \overline{\psi}_q \gamma_{\mu} \psi_q \tau^3,
\label{Jemqi}
\ee

\noindent where $I$ is a unit matrix in isospin space;
$Q_{+}$ is the average quark charge,
$Q_{+} = \sfrac{1}{2} ( \sfrac{2}{3} - \sfrac{1}{3} )$,
and $Q_{-}$ the average difference,
$Q_{-} = \sfrac{1}{2} ( \sfrac{2}{3} + \sfrac{1}{3} )$.
Note that the
electromagnetic current has both isoscalar and isovector components.
When the steps outlined in sect.\,\ref{lforn} are repeated,
the electromagnetic current
between nucleons likewise exhibits isoscalar and isovector components and is
written

\bea
V_{\mu}^{em} & = & i \overline{\psi} \Bigl[ F_1^S(k^2)\gamma_{\mu}
+ {F_2^S(k^2) \over 2M} \sigma_{\mu \nu} k_{\nu}
+i F_3^S (k^2) k_{\mu} \Bigr] Q_{+} I \psi
\nonumber \\
& + &
i \overline{\psi} \Bigl[ F_1^V(k^2)\gamma_{\mu}
+ {F_2^V(k^2) \over 2M} \sigma_{\mu \nu} k_{\nu}
+i F_3^V (k^2) k_{\mu} \Bigr] Q_{-} \tau^3 \psi ,
\label{Jemni}
\eea

\noindent where now $Q_{+}$ is the average {\it nucleon} charge and
$Q_{-}$ the average charge difference: $Q_{+} = Q_{-} = \sfrac{1}{2}$.
Again, symmetry principles limit the number of terms to
three for both the isoscalar and isovector currents.  In terms of proton and
neutron form factors, we identify

\bea
F_i^S(k^2) & = & F_i^p(k^2) + F_i^n(k^2),
\nonumber  \\
F_i^V(k^2) & = & F_i^p(k^2) - F_i^n(k^2), \qquad i=1,2,3 .
\label{FSV}
\eea

\noindent Further, in going from quark to nucleon Lagrangians, the
normalization of the vector current remains unchanged so that the first form
factor $F_1$ in the $k^2 \rightarrow 0$ limit simply specifies the charge of
the nucleon in units of $e$: \viz $F_1^p(0) = 1$, $F_1^n(0) = 0$, or
equivalently $F_1^S(0) = 1$, $F_1^V(0) = 1$.  The second form factor relates
to the anomalous magnetic moment of the nucleon (in units of the nuclear
magneton, $e/(2M)$, with $\hbar = c = 1$): \viz $F_2^p(0) = \mu_p - 1 =
1.79$ and $F_2^n(0) = \mu_n = -1.91$, or equivalently $F_2^S(0) = -0.12$ and
$F_2^V(0) = 3.70$.  Finally, the electromagnetic current, $V_{\mu}^{em}$ is
a conserved current: that is,  $\partial_{\mu} V_{\mu}^{em} = 0$ or, as written
in
momentum space, $k_{\mu} V_{\mu}^{em} = 0$.  This follows from the fact
that the electromagnetic Lagrangian is invariant under the phase
transformation $\psi \rightarrow \psi^\prime = e^{ie \Lambda} \psi$ and
$\overline{\psi} \rightarrow \overline{\psi}^\prime = e^{-ie \Lambda} \psi$,
where $\Lambda$ is a real constant.  From Noether's
theorem \cite{deWS86}\nl ,
for every
transformation that leaves the Lagrangian invariant there is a corresponding
conserved current, in this case the electromagnetic current.  Evaluating
$k_{\mu} V_{\mu}^{em}$ term by term we have:

\be
\overline{\psi}(p^\prime)
k_{\mu} \gamma_{\mu} \psi(p) = \overline{\psi}(p^\prime) \gamma_{\mu}
(p_{\mu} -  p^{\prime}_{\mu})
\psi(p) = i \overline{\psi}(p^\prime)(M-M)\psi(p) = 0
\label{CVCpf}
\ee

\noindent from the Dirac
equation of a free nucleon; and $\sigma_{\mu \nu} k_{\mu} k_{\nu}$ summed
over repeated indices is zero from the antisymmetry of $\sigma_{\mu \nu}$;
while $k^2_{\mu} = k^2 \not= 0$.  Thus, current conservation requires that

\be
F_3(k^2) = 0
\label{F3}
\ee

\noindent for both isoscalar and isovector currents.
\par
It is noted that the isovector electromagnetic current, Eq.\,\ref{Jemni},
and the weak vector current, Eq.\,\ref{VAct}, are identical in structure
with the isospin matrix $\tau^3$ in the former case replaced by the matrix
$\tau^a$ in the latter case.  Feynman and Gell-Mann
\cite{FG58} postulated that these currents are members of an isotriplet vector
of
current operators.  A consequence of this hypothesis is that the weak vector
current is also a conserved current. This is called the conserved
vector current (CVC) hypothesis, from which it follows that

\bea
\gV(k^2) & = & F_1^V(k^2) \rightarrow 1 \qquad {\rm as}
\quad k^2 \rightarrow 0
\nonumber    \\
\gM(k^2) & = & F_2^V(k^2)
\nonumber    \\
\gS(k^2) & = & F_3^V(k^2) ~~ = ~~ 0 .
\label{CVCcc}
\eea

\noindent Note that $\gV(0) = 1$ always, irrespective of the nucleus in
which it might be experimentally measured.  That is, the vector coupling
constant is not renormalized in the nuclear medium.  Likewise, the
weak-magnetism
form factor is uniquely related to the corresponding
electromagnetic form factor:  for beta decays between isospin analogues it
relates to the anomalous isovector magnetic moment of the nuclear states in
question, while for non-analogue transitions it relates to the corresponding
isovector M1 gamma-decay rate.

So far, experimental tests of CVC have taken
two forms:  {\it (i)} showing $\gV(0)$ is a constant irrespective of the
nucleus in which it is measured, and {\it (ii)} showing $\gM(0) = F_2^V(0)$
by measuring $\gM(0)$ from either the lepton energy spectrum or an
angular correlation in a beta-decay experiment and
comparing it with $F_2^V(0)$  obtained from the
lifetime and branching ratio for the analogous $\gamma$-transition.  We
discuss both tests in the following sections.

\subsection{Superallowed Fermi Transitions}  \label{SFT}

Precise measurements of the $\beta$-decay between nuclear states with
$(J^{\pi},T) = (0^{+},1)$ provide the most straightforward yet
demanding probe of the vector interaction in nuclei.  Because
the axial current cannot contribute in first order to transitions
between spin-0 states, the experimental $ft$-value is related
directly to the vector coupling constant, $\GV = \GF V_{ud}$:

\be
ft(1+\delta_R) = \frac{K}{\GV^2 (1+\DRV) {\langle \MV \rangle}^2 },
\label{ftF}
\ee

\noindent with

\bea
{\langle \MV \rangle}^2 & = & 2(1-\delta_C) ~~ {\rm for}~T=1
\nonumber  \\
K/(\hbar c)^6 & = & 2 \pi^3 \ln 2 \hbar /(m_e c^2)^5 ~ = ~
(8120.271 \pm 0.012) \times 10^{-10} {\rm GeV}^{-4} {\rm s},
\label{consts}
\eea

\noindent where the physical constants used to evaluate $K$ were
taken from ref.\,\cite{PDG94}\nl .  Here $f$ is the statistical rate
function, $t$ is the partial half-life for the transition,
and $\langle \MV \rangle$ is the Fermi matrix element.
The radiative corrections,
$\delta_R$ and $\DRV$, and the charge correction, $\delta_C$,
are each of order 0.01, with $\delta_R$ and particularly $\delta_C$ being
dependent on nuclear structure.  There are theoretical
uncertainties introduced via these correction terms but they are
at least another order of magnitude smaller, so Eq.\,\ref{ftF}
provides an experimental technique for determining $\GV$ to
better than a part in a thousand.

To date, nine $0^{+} \rightarrow 0^{+}$ transitions have been
studied to this level of precision; they cover a wide range of
nuclear masses from $^{10}$C, the lightest parent, to $^{54}$Co,
the heaviest.  The results have been used to test the constancy of
$\GV$, which is a prediction of CVC, and to derive the value of
$V_{ud}$, the up-down element in the Cabibbo-Kobayashi-Maskawa
quark-mixing matrix.  A unitarity test on the matrix, made possible
by these results, is a fundamental measure of the adequacy of
the three-generation Standard Model.

\subsubsection{Experiments} \label{Expts}

The $ft$ value that characterizes a particular
$\beta$-transition is determined by three measured parameters: the transition
energy, $Q_{EC}$, which is used in calculating the statistical rate
function, $f$; the half-life, $t_{1/2}$, of the $\beta$-emitter and
the branching ratio, $R$, for the transition of interest, which
together yield the partial half-life, $t$.  World data have been surveyed
recently \cite{Ha90} for eight of the best-known superallowed emitters:
$^{14}$O, $^{26m}$Al, $^{34}$Cl, $^{38m}$K, $^{42}$Sc,
$^{46}$V, $^{50}$Mn and $^{54}$Co.  The results given in Table \ref{Exres}
are drawn from that survey, augmented by data on $^{10}$C decay
\cite{Ba84,Ba88,Ba89,Ba90,Az74,Kr91,Na91,Ro72,Sa95} and by several
other recent measurements of significance
\cite{Ki91,Wa92,Ko94a,Ko94b,Br94,Li94,Am94,Ha94}.

{\footnotesize
\begin{table} [t]
\begin{center}
\caption{Experimental results ($Q_{EC}$, $t_{1/2}$ and branching
ratio, $R$) and calculated corrections ($P_{EC}$, $\delta_C$ and $\delta_R$)
for $0^{+} \rightarrow 0^{+}$ transitions.
\label{Exres} }
\vskip 1mm
\begin{tabular}{ccccccccc}
\hline \\[-3mm]
 & $Q_{EC}$ & $t_{1/2}$ & $R$ & $P_{EC}$ & $ft$ & $\delta_C$ &
$\delta_R$ & $\F t$ \\
 & (keV) & (ms) & (\%) & (\%) & (s) & (\%) & (\%) & (s) \\
\hline \\[-3mm]
$^{10}$C & 1907.77(9) & 19290(12) & 1.4638(22) & 0.296 & 3040.1(51) &
0.18(4) & 1.30(4) & 3074.0(54) \\
$^{14}$O & 2830.51(22) & 70603(18) & 99.336(10) & 0.087 & 3038.1(18) &
0.23(4) & 1.26(5) & 3069.2(28) \\
$^{26m}$Al & 4232.42(35) & 6344.9(19) & $\geq$ 99.97 & 0.083
& 3035.8(17) &
0.31(4) & 1.45(2) & 3070.2(23) \\
$^{34}$Cl & 5491.71(22) & 1525.76(88) & $\geq$ 99.988 & 0.078
& 3048.4(19) &
0.54(4) & 1.33(3) & 3072.1(25) \\
$^{38m}$K & 6043.76(56) & 923.95(64) & $\geq$ 99.998 & 0.082
& 3047.9(26) &
0.56(4) & 1.33(4) & 3071.1(32) \\
$^{42}$Sc & 6425.58(28) & 680.72(26) & 99.9941(14) & 0.095
& 3045.1(14) &
0.36(4) & 1.47(5) & 3078.7(24) \\
$^{46}$V & 7050.63(69) & 422.51(11) & 99.9848(13) & 0.096 & 3044.6(18) &
0.37(4) & 1.40(6) & 3075.8(28) \\
$^{50}$Mn & 7632.39(28) & 283.25(14) & 99.942(3) & 0.100 & 3043.7(16) &
0.41(4) & 1.40(7) & 3073.8(28) \\
$^{54}$Co & 8242.56(28) & 193.270(63) & 99.9955(6) & 0.104
& 3045.8(11) &
0.50(4) & 1.39(7) & 3073.1(28) \\
 & & & & & & \multicolumn{2}{c}{Average, $\overline{\F t}$} & 3073.1(11) \\
 & & & & & & \multicolumn{2}{c}{$\chi^2/(N-1)$}  & 1.34 \\
\hline
\end{tabular}
\end{center}
\end{table}
}

The experimental uncertainties quoted for the measured decay energies
 (typically 1 part in 20,000), half-lives (1 part in 3,000) and
branching ratios (1 part in 70,000) are at the limits of what is currently
possible.  Even though such quantities are frequently measured
in nuclear physics, the extreme demands on precision in these studies
have led to experimental techniques highly refined for the purpose.
Several examples, briefly described, will illustrate the point.

{\em 2.2.1.1 Decay energies.}  The nine superallowed $\beta$-emitters
listed
in Table \ref{Exres} populate daughters with stable ground states.
As a result, reaction $Q$-value measurements, which lend themselves
to experimental precision, can be used to determine decay energies
instead of direct measurements of positron end-point energies, which
are frought with precision-limiting difficulties.  For example, in
the case of $^{26m}$Al
$ \rightarrow ^{26}$Mg, the decay energy,
$Q_{EC}$, has been measured by three different routes:

{\it (i)} $^{26}$Mg($p,n)^{26m}$Al.  The threshold energy for this
reaction was determined \cite{Br94} by detection of the
$^{26m}$Al activity
produced as a function of proton energy, with a $^{26}$Mg target
located in the focal plane of a magnetic spectrometer.
The energy of the proton beam was calibrated by comparison with a
$^{133}$Cs$^{+}$ beam accelerated through a voltage difference $V$
to pass along the same path in the spectrometer.  The value of $V$
was then compared via a resistive divider with a one-volt standard.
The measured $Q$ value, $Q_{pn}$, derived from the observed threshold
energy, is related to $Q_{EC}$ by:

\be
Q_{EC} = -Q_{pn} - 782.354~~{\rm keV}.
\label{pn}
\ee

{\it (ii)} $^{25}$Mg$(p,\gamma )^{26m}$Al and $^{25}$Mg$(n,\gamma )
^{26}$Mg.
For the first reaction, proton energies at four resonances were
established \cite{Ki91} by comparison with known
resonances in $^{27}$Al$
(p,\gamma )^{28}$Si.  For the second, thermal neutrons were
used, with $\gamma$-ray energies being determined \cite{Ki91} by
comparison with known calibration lines.  The measured $Q$ values
yield $Q_{EC}$ through the relation:

\be
Q_{EC} = Q_{n \gamma } - Q_{p \gamma } -782.354~~{\rm keV}.
\label{png}
\ee

\noindent Though comparable uncertainties are claimed as for
method {\it (i)}, it is evident that this method is less direct and involves
a variety of secondary calibration standards.

{\it (iii)} $^{26}$Mg$(^{3}He,t)^{26}$Al (1058 keV).  This reaction
was observed \cite{Ko87a} in conjunction with
$^{14}$N$(^{3}He,t)^{14}$O from a composite
target in which $^{14}$N had been implanted into $^{26}$Mg.
The target assembly was biased periodically to $\sim $80 kV, thus
permitting the two triton groups corresponding to the $^{14}$O ground
state and the 1058 keV state of $^{26}$Al alternately to follow
the same path through a magnetic spectrometer.  The voltage at which
they perfectly overlapped corresponds to the $Q$-value difference,
$\Delta Q$, between the two reactions.  The energy difference between
the 1058 keV state and the $^{26}$Al $0^{+}$ isomer was
determined \cite{Ko87b} from the relevant $\gamma$-ray energy,
$E_{\gamma_1}$.  Similarly, the ground state of $^{14}$N can be related
to the $0^{+}(T=1)$ excited state by a $\gamma$ ray of energy,
$E_{\gamma_2}$.  Thus,

\be
Q_{EC}({\sf A=26}) - Q_{EC}({\sf A=14}) = E_{\gamma_2} - E_{\gamma_1} - \Delta
Q.
\label{Het}
\ee

\noindent This method, too, depends on secondary calibration standards
but it provides valuable interconnections betwen pairs of
superallowed $0^{+} \rightarrow 0^{+}$ transition $Q_{EC}$ values.

{\em 2.2.1.2 Half-lives.}  To achieve the required precision, isotopic
purity
of the sample must be assured.  Currently, all emitters except
$^{14}$O and $^{10}$C have been measured \cite{Ko94a} with an
on-line isotope
separator.  Typically, thousands of multi-scaled
spectra are acquired from successive samples observed in a
large-solid-angle, high-efficiency, low-background gas proportional
counter; the counting electronics have a well-defined, non-extendable
dead time.  The best analysis procedure uses modified Poisson
variances and is based on a simultaneous fit to up to 500 decay
curves with the same half-life but with individual intensity parameters
for each decay curve.  This procedure is checked \cite{Ko94a} with
hypothetical data, computer-generated to simulate closely the
experimental counting conditions, and has been proven correct
at the 0.01\% level.

{\em 2.2.1.3 Branching ratios.}  In most cases, the superallowed
transition
dominates ($\geq$ 99\% ) the decay of the parent nucleus and is best
determined
via measurement of the other decay branches, their branching ratios
being subtracted from 100\% .  For these other branches, precision is
not as important as completeness; therefore special techniques have been
developed (\eg \, ref.\cite{Ha94}) to observe $\beta$-delayed $\gamma$ rays
from branches as weak as a few parts per million.  For the only case
in which the superallowed transition is itself weak -- the decay of
$^{10}$C -- the technique used \cite{Kr91} is to compare
the intensity of $\beta$-delayed $\gamma$ rays in $^{10}$B, produced
via the reaction $^{10}$B$(p,n)^{10}$C $(\beta^{+}) ^{10}$B$^\ast$,
with the same $\gamma$ rays produced via $^{10}$B$(p,p^{\prime})^{10}$B$^\ast$;
for added precision, a large $\gamma$-ray array \cite{Sa95} has now
been used to observe these $\gamma$ rays in alternating off-line,
on-line measurements.

\subsubsection{The $ft$ values} \label{ftvalues}

The statistical rate function, $f$, is defined
by

\be
f = \int_{1}^{W_0} p W (W_0 - W)^2 F(Z,W) C(W) {\rm d}W,
\label{fint}
\ee

\noindent where $W$ and $p$ are the electron energy and momentum,
respectively, in electron rest-mass units, and $W_0$ is the
maximum $\beta$-energy, which is related directly to the measured
$Q_{EC}$ via the equation $W_0 + 1 = Q_{EC}/(m_e c^2)$. The quantity
$F(Z,W)$ is known as the Fermi function and $C(W)$ the shape
correction factor \cite{Sc66}\nl .  The latter has a value close to unity
for most values of $W$, its departure from unity reflecting the
influence of screening by atomic electrons, the dependence of the
nuclear matrix elements on W, and the presence of second-forbidden
matrix elements.  Values of the statistical rate function have been
computed for each superallowed transition following the techniques
described in refs.\,\cite{Ha90,To73}\nl .  Each result has an associated
uncertainty reflecting the experimental uncertainty in the $Q_{EC}$
value used in the calculation.

Next, the partial half-life for each superallowed emitter must be
obtained from its total half-life and branching ratio according to
the formula

\be
t = t_{1/2} (1 + P_{EC})/R ,
\label{partt}
\ee

\noindent where $P_{EC}$ is the calculated electron-capture probability.
Table \ref{Exres} lists the $P_{EC}$ values calculated by the method
described in ref.\,\cite{Ha90}\nl , which employs exact bound-state electron
wave functions.  Though the calculated values of $P_{EC}$ are small in all
cases, the correction is comparable to the experimental uncertainties
on $t_{1/2}$ and $R$, and must be properly accounted for.

Finally, the $ft$-values that incorporate all experimental uncertainties
are listed for the nine emitters in Table \ref{Exres}.  They
incorporate the complete set of world data and form the experimental
basis for the fundamental tests that follow.

\subsubsection{Theoretical corrections} \label{Theocor}

In a nucleus, the charged particles involved in
$\beta$ decay interact with an external electromagnetic field, which
leads to effects of order $(Z\alpha )^m$, $m$=1,2..., where $Z$ is
the atomic number of the daughter nucleus and $\alpha$ is the
fine-structure constant.  For the electron (or positron), these
effects are accounted for in the calculation of $f$ (see Eq.\,\ref{ftF}),
which uses \cite{Ha90,To73} exact solutions of the Dirac
equation for electrons in the
field of a modelled nuclear charge distribution.
For the proton, these effects are reflected in the breaking of nuclear
isospin symmetry and a resulting correction, $\delta_C$, to the
Fermi matrix element.

There are also effects that arise from interactions between the charged
particles themselves.  These are of order $Z^m \alpha^n$, with $m<n$,
and lead to the radiative corrections, $\delta_R$ and $\DRV$.
By convention, the former incorporates those corrections that are
nuclear-structure dependent, the latter, those that are not.

{\em 2.2.3.1 Charge correction,} $\delta_C$.  Both Coulomb and
charge-dependent
nuclear forces destroy isospin symmetry between the initial and final
states in superallowed $\beta$-decay.  The consequences are twofold:
  there are different degrees of configuration mixing in the two states,
and, because their binding energies are not identical, their radial
wave functions differ.  Calculations \cite{Ha90,Ha94,To73,Or89}
accommodate both effects by splitting $\delta_C$ into two components,
$\delta_{C_1}$ and $\delta_{C_2}$, to account for configuration mixing
and radial mismatch, respectively.  In all cases $\delta_{C_2}>\delta_{C_1}$.

To establish $\delta_{C_1}$, shell-model wave functions were calculated
with appropriate nucleon-nucleon interactions modified to include:
$(i)$ two-body Coulomb terms in the proton-proton part of the
Hamiltonian, $(ii)$ a 2\% increase in proton-neutron elements (justified
by the charge dependence observed in nucleon-nucleon scattering data),
and $(iii)$ one-body elements determined from the single-particle energies of
closed-core-plus-proton and -neutron nuclei.  These basic elements were
then adjusted to reproduce the $b$- and $c$-coefficients of
the isobaric-multiplet mass equation that describes the multiplet
involved in the superallowed transition.  The results have also been tested
independently by measurements \cite{Ha94}\nl , in some of the same nuclei,
of weak non-analogue $0^{+} \rightarrow 0^{+}$ transitions, which
are sensitive to the same effects.

The larger of the two components, $\delta_{C_2}$, has been calculated
from two different approaches.  One \cite{To77} describes
the proton in the parent
nucleus and the neutron in the daughter nucleus by full parentage expansions
in terms of Woods-Saxon wave functions suitably matched to the
experimentally known separation energies.  The other
\cite{Or89} employs a self-consistent Hartree-Fock calculation
in which the overall strength of the central potential was adjusted
so that the single-particle eigenvalues agree with the experimental
separation energies.  The former approach leads to values for $\delta_{C_2}$
that are consistently larger than those from the latter.

In Table \ref{Exres}, we list $\delta_C$ (=$\delta_{C_1} + \delta_{C_2}$)
for all nine transitions.  The values for $\delta_{C_1}$ were taken
from ref.\,\cite{Ha90} updated by new results in ref.\,\cite{Ha94};
those for $\delta_{C_2}$ are the unweighted averages from the
results of refs.\,\cite{Or89,To77}\nl .  ``Statistical" uncertainties,
reflecting
the scatter of the calculated values, were derived in ref.\,\cite{Ha90}
and have been adopted here.

{\em 2.2.3.2 Radiative corrections,} $\delta_R$ {\em and} $\DRV$.
In the context
of Standard-Model tests, we need consider only radiative-correction
terms that distinguish semi-leptonic nuclear $\beta$-decay from
purely leptonic $\mu$-decay \cite{Ma86}\nl .  By grouping them according
to whether they are nuclear-structure dependent ($\delta_R$) or not
($\DRV$), we write \cite{To92}:

\bea
\delta_R & = & \frac{\alpha}{2 \pi} \left[ \,\overline{g}(E_m)
+ \delta_2 + \delta_3 + 2 \,C_{NS} \,\right]
\label{dR} \\
\DRV & = & \frac{\alpha}{2 \pi} \left[ \,
4\ln(\mZ/m_p) + \ln(m_p/\mA)
+ 2 \,C_{\rm{Born}} \,\right] + \ldots ,
\label{DR}
\eea

\noindent where, for clarity, only leading-order terms have been presented
for $\DRV$.  In these equations,
$E_m$ is the maximum electron energy in the $\beta$-decay,
$m_p$ and $\mZ$ are the masses of the proton and $Z$-boson, and $\mA$
is a mass parameter associated with the $A_1$-meson but, in this application,
is allowed the range $400 \leq \mA \leq 1600$ MeV.  The function $g(E,E_m)$
was first defined by Sirlin \cite{Si67} as part of the order-$\alpha$ universal
photonic contribution arising from the weak vector current;
it is here averaged over the
electron spectrum to give $\overline{g}(E_m)$.  The effects of
order-$Z\alpha^2$ and -$Z^2\alpha^3$ have
recently been definitively calculated \cite{Si87} and are
incorporated as $\delta_2$ and $\delta_3$, respectively.  Finally,
the terms $C_{\rm{Born}}$ and $C_{NS}$ arise from the order-$\alpha$
axial-vector photonic contributions: the former accounts for
single-nucleon contributions, while the latter covers two-nucleon
contributions and is consequently dependent on nuclear structure.
We adopt here the calculations of Towner \cite{To94}\nl .

Built from these components (see Eq.\,\ref{dR}), calculated values
for $\delta_R$ are listed in Table \ref{Exres}.  The largest component,
$\overline{g}(E_m)$, is responsible for approximately 1\% in each
case, $\delta_2$ and $\delta_3$ together account for less than 0.5\%,
and the effects of $C_{NS}$ range from $-0.3\%$ to +0.1\%.  The
calculated value for $\DRV$ is

\be
\DRV = (2.46 \pm 0.09) \ \%.
\label{DRvalu}
\ee

\noindent In this case, the leading logarithic term dominates, with a
contribution of 2.1\%.  The remainder arises from $C_{\rm{Born}}$ (0.20\%)
and higher order terms not explicitly listed in Eq.\,\ref{DR}.  The
second term in the equation effectively establishes the quoted uncertainty,
which reflects the range of values allowed for $\mA$.

\subsubsection{Implications for the Standard Model} \label{Impsm}

It is now convenient to define a ``corrected" $\F t$
value, which is related to the conventional $ft$ value via:

\be
\F t = ft (1 + \delta_R) (1 - \delta_C).
\label{scrptft}
\ee

\noindent As a comparison with Eq.\,\ref{ftF} makes clear, all
nuclear-structure-dependent effects should have been removed from the
$\F t$ values.  It is the consistency of the determined $\F t$ values
over a wide variety of superallowed $0^{+} \rightarrow 0^{+}$
transitions that constitutes a demanding test of CVC.

\begin{figure}[t]
\begin{center}
\setlength{\unitlength}{0.240900pt}
\ifx\plotpoint\undefined\newsavebox{\plotpoint}\fi
\sbox{\plotpoint}{\rule[-0.175pt]{0.350pt}{0.350pt}}%
\begin{picture}(1500,900)(0,0)
\tenrm
\sbox{\plotpoint}{\rule[-0.175pt]{0.350pt}{0.350pt}}%
\put(264,315){\rule[-0.175pt]{4.818pt}{0.350pt}}
\put(242,315){\makebox(0,0)[r]{3070}}
\put(1416,315){\rule[-0.175pt]{4.818pt}{0.350pt}}
\put(264,473){\rule[-0.175pt]{4.818pt}{0.350pt}}
\put(242,473){\makebox(0,0)[r]{3075}}
\put(1416,473){\rule[-0.175pt]{4.818pt}{0.350pt}}
\put(264,630){\rule[-0.175pt]{4.818pt}{0.350pt}}
\put(242,630){\makebox(0,0)[r]{3080}}
\put(1416,630){\rule[-0.175pt]{4.818pt}{0.350pt}}
\put(354,158){\rule[-0.175pt]{0.350pt}{4.818pt}}
\put(354,113){\makebox(0,0){5}}
\put(354,767){\rule[-0.175pt]{0.350pt}{4.818pt}}
\put(580,158){\rule[-0.175pt]{0.350pt}{4.818pt}}
\put(580,113){\makebox(0,0){10}}
\put(580,767){\rule[-0.175pt]{0.350pt}{4.818pt}}
\put(805,158){\rule[-0.175pt]{0.350pt}{4.818pt}}
\put(805,113){\makebox(0,0){15}}
\put(805,767){\rule[-0.175pt]{0.350pt}{4.818pt}}
\put(1030,158){\rule[-0.175pt]{0.350pt}{4.818pt}}
\put(1030,113){\makebox(0,0){20}}
\put(1030,767){\rule[-0.175pt]{0.350pt}{4.818pt}}
\put(1256,158){\rule[-0.175pt]{0.350pt}{4.818pt}}
\put(1256,113){\makebox(0,0){25}}
\put(1256,767){\rule[-0.175pt]{0.350pt}{4.818pt}}
\put(264,158){\rule[-0.175pt]{282.335pt}{0.350pt}}
\put(1436,158){\rule[-0.175pt]{0.350pt}{151.526pt}}
\put(264,787){\rule[-0.175pt]{282.335pt}{0.350pt}}
\put(45,562){\makebox(0,0)[l]{\shortstack{$\F t~(s)$}}}
\put(850,23){\makebox(0,0){$Z$ of daughter}}
\put(264,158){\rule[-0.175pt]{0.350pt}{151.526pt}}
\put(354,441){\circle*{24}}
\put(444,290){\circle*{24}}
\put(670,322){\circle*{24}}
\put(850,381){\circle*{24}}
\put(940,350){\circle*{24}}
\put(1030,589){\circle*{24}}
\put(1120,498){\circle*{24}}
\put(1211,435){\circle*{24}}
\put(1301,413){\circle*{24}}
\put(354,271){\rule[-0.175pt]{0.350pt}{81.906pt}}
\put(344,271){\rule[-0.175pt]{4.818pt}{0.350pt}}
\put(344,611){\rule[-0.175pt]{4.818pt}{0.350pt}}
\put(444,202){\rule[-0.175pt]{0.350pt}{42.398pt}}
\put(434,202){\rule[-0.175pt]{4.818pt}{0.350pt}}
\put(434,378){\rule[-0.175pt]{4.818pt}{0.350pt}}
\put(670,249){\rule[-0.175pt]{0.350pt}{34.930pt}}
\put(660,249){\rule[-0.175pt]{4.818pt}{0.350pt}}
\put(660,394){\rule[-0.175pt]{4.818pt}{0.350pt}}
\put(850,303){\rule[-0.175pt]{0.350pt}{37.821pt}}
\put(840,303){\rule[-0.175pt]{4.818pt}{0.350pt}}
\put(840,460){\rule[-0.175pt]{4.818pt}{0.350pt}}
\put(940,249){\rule[-0.175pt]{0.350pt}{48.421pt}}
\put(930,249){\rule[-0.175pt]{4.818pt}{0.350pt}}
\put(930,450){\rule[-0.175pt]{4.818pt}{0.350pt}}
\put(1030,513){\rule[-0.175pt]{0.350pt}{36.376pt}}
\put(1020,513){\rule[-0.175pt]{4.818pt}{0.350pt}}
\put(1020,664){\rule[-0.175pt]{4.818pt}{0.350pt}}
\put(1120,410){\rule[-0.175pt]{0.350pt}{42.398pt}}
\put(1110,410){\rule[-0.175pt]{4.818pt}{0.350pt}}
\put(1110,586){\rule[-0.175pt]{4.818pt}{0.350pt}}
\put(1211,347){\rule[-0.175pt]{0.350pt}{42.398pt}}
\put(1201,347){\rule[-0.175pt]{4.818pt}{0.350pt}}
\put(1201,523){\rule[-0.175pt]{4.818pt}{0.350pt}}
\put(1301,325){\rule[-0.175pt]{0.350pt}{42.398pt}}
\put(1291,325){\rule[-0.175pt]{4.818pt}{0.350pt}}
\put(1291,501){\rule[-0.175pt]{4.818pt}{0.350pt}}
\sbox{\plotpoint}{\rule[-0.350pt]{0.700pt}{0.700pt}}%
\put(264,413){\usebox{\plotpoint}}
\put(264,413){\rule[-0.350pt]{282.335pt}{0.700pt}}
\end{picture}
\vskip 1mm
\caption{$\F t$ values for the nine precision data and the best
least squares one-parameter fit  \label{fig1}}
\end{center}
\end{figure}
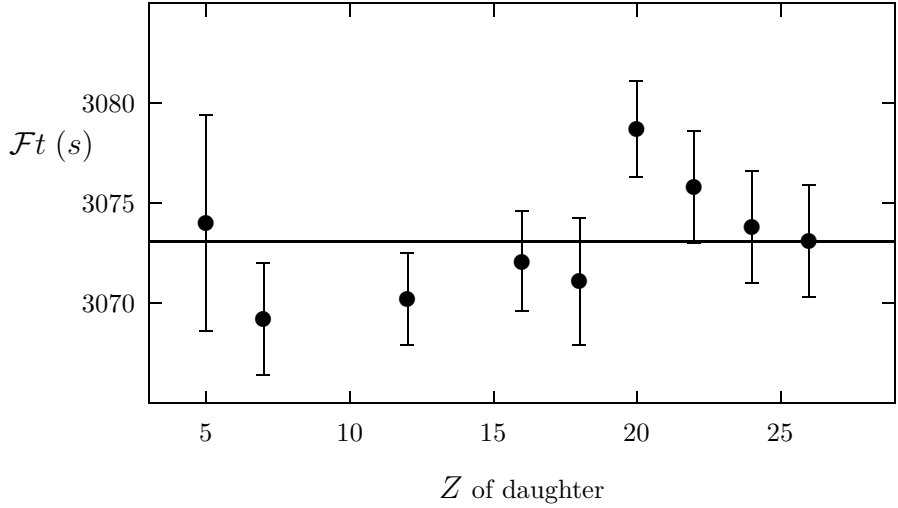

Table \ref{Exres} contains all requisite experimental and calculated
information for the determination of $\F t$ values, and the results
for all nine transitions appear in the last column of that table
together with their weighted average, $\overline{\F t}$, and scaled
``statistical" uncertainty.  The
same results are displayed in Fig.\,\ref{fig1}.  There is no
statistically significant evidence of inconsistencies in the data,
thus verifying the expectation of CVC at the level of $4 \times 10^{-4}$,
the fractional uncertainty quoted on the average $\F t$-value.

With the consistency of the measured $\F t$-values established, we turn
to the evaluation of $V_{ud}$ and the unitarity test of the CKM matrix.
The up-down quark mixing matrix element is given by

\bea
V_{ud}^2 & = & \GV^2 / \GF^2
\nonumber  \\
& = & \frac{K}{2 \GF^2 (1+\DRV) \overline{\F t} }.
\label{Vud2}
\eea

\noindent In using the average $\overline{\F t}$ value from
Table \ref{Exres} in
Eq.\,\ref{Vud2}, it is
important now for us to include in the
$\overline{\F t}$-value uncertainty the
effects of a ``systematic" uncertainty in the charge correction,
$\delta_C$, as well as the ``statistical" uncertainty already described.
The ``systematic" uncertainty of $\pm 0.08$\% arises from systematic
differences between two independent model calculations \cite{To77,Or89}
of $\delta_{C_2}$, and has been elaborated on in ref.\,\cite{Ha90}\nl .
The result is

\be
\overline{\F t} = 3073.1 \pm 3.0 ~~{\rm s}
\label{Ftbar}
\ee

\noindent Thus, with

\be
\GF /(\hbar c)^3 = (1.16639 \pm 0.00002) \times 10^{-5} \ {\rm GeV}^{-2}
\label{GF}
\ee

\noindent from ref.\,\cite{PDG94} and $\DRV$ from Eq.\,\ref{DRvalu},
we obtain

\be
V_{ud} = 0.9736 \pm 0.0006 .
\label{Vudvalu}
\ee

\noindent The quoted uncertainty is dominated by uncertainties in the
theoretical corrections, $\DRV$ and $\delta_C$; experimental
errors on the input data for the $\overline{\F t}$-value contribute
less than $\pm 0.0002$ to the uncertainty in Eq.\,\ref{Vudvalu}.

This result for $V_{ud}$ leads to the most demanding test yet available
of the unitarity of the Cabibbo-Kobayashi-Maskawa matrix, in which

\be
\mids V_{ud} \mids^2 +
\mids V_{us} \mids^2 +
\mids V_{ub} \mids^2 = 1
\label{unit}
\ee

\noindent must be satisfied if there are three generations of quarks
prescribed by the Standard Model.  Evidently, $V_{ud}$ is by far the
largest of the three matrix elements and its uncertainty the most
critical.  The value of $V_{us}$ is taken to be \cite{PDG94}
$\mids V_{us} \mids \ = 0.2205 \pm 0.0018$, the average of two
results, one from the analysis of $K_{e3}$ decays, the other
from hyperon decays.  Finally, $V_{ub}$ can be derived, with substantial
uncertainty, from the semi-leptonic decay of $B$ mesons, but is
sufficiently small \cite{PDG94}\nl , at $0.0032 \pm 0.0009$, as to be
negligible in the unitarity test.  The result,

\be
\mids V_{ud} \mids^2 +
\mids V_{us} \mids^2 +
\mids V_{ub} \mids^2 = 0.9965 \pm 0.0015,
\label{unitvalu}
\ee

\noindent differs from unitarity at the 98\% confidence level.

The significance of this apparent non-unitarity is still being debated.
Taken at face value it indicates the need for some extension to the
three-generation Standard Model (see sect.\,\ref{BSM}).  However,
it may also reflect some undiagnosed inadequacy in the evaluation
of $\mids V_{us} \mids $  \footnote { A larger value is recommended
in the analysis of ref.\cite{GHK92} }
or $\mids V_{ud} \mids $\, .  For example, it has been proposed \cite{Wi90}
that the nuclear-model-dependent calculations of $\delta_C$ should
be replaced by an empirical approach in which the uncorrected data
are fitted by a smooth $Z$-dependent function.  Extrapolation of this
function
to $Z=0$ would yield the recommended value of $\F t$.  With the data
in Table \ref{Exres}, a variation of this approach is to allow a
$Z$-dependent fit of the nine $\F t$-values: doing so with linear
$Z$ dependence does lower the reduced $\chi^2$ somewhat (from 1.34 to
1.04).  The result is $\F t({\sf Z=0}) = 3068.1 \pm 3.1$, which yields a
new unitarity sum (Eq.\,\ref{unitvalu}) of $0.9980 \pm 0.0015$.
It must be emphasized, however, that although this approach does yield
a less provocative result, there is no substantive explanation for any
correction not already accounted for in $\delta_C$ and $\delta_R$,
and there is no statistically significant indication from the data
that one is required.

\subsection{Experimental Tests of CVC in Weak Magnetism Terms}
\label{ExWM}

The conserved vector current hypothesis, besides requiring that
$\gV (0) $ is constant ($ = 1$),
as verified in superallowed transitions,
also predicts that the weak magnetism form factor is uniquely related
to the corresponding electromagnetic form factor.  In this section
we will discuss experimental tests of this latter prediction,
concentrating our attention principally on the
$( J^{\pi}, T = 1^{+},1) \rightarrow
(0^{+},0)$ allowed axial-vector mirror $\beta$-transitions in the $A=12$
nuclei: $^{12}$B $\rightarrow ~^{12}$C $+ e^{-} +
\overline{\nu}_e$ and
$^{12}$N $\rightarrow ~^{12}$C $+ e^{+} + \nu_e$.
For oriented parent nuclei, the beta-decay
rates (upper sign for $^{12}$N decay and lower sign for $^{12}$B
decay) are given by \cite{HP77}

\bea
d^3 \Gamma (e^{\pm}) & = & \frac{1}{\hbar} \frac{\GF^2}{4 \pi^4}
V_{ud}^2 \mid \! F^{\pm}_A (0) \! \mid ^{2}  F_{\pm}(Z,W) p W (W_0 - W)^2
(1 + \eta_{\pm} + a_{\pm} W) dW d\Omega_e
\nonumber \\[1mm]
& & \times [1 \pm (h_1 - h_{-1}) (1 + \alpha_{\pm}W) P_1(\cos \theta_e)
+ (1-3h_0) \alpha_{\pm} W P_2(\cos \theta_e)],
\label{WMrate}
\eea

\noindent where $F_{\pm}(Z,W)$ is the Fermi function for the $e^{\pm}$
decays, $p$ and $W$ are the electron momentum and total energy
in electron rest-mass units; and  $W_0$ is the maximum electron energy.
The shape correction factor, $C(W)$, has been explicitly written
as
$(1 + \eta_{\pm} + a_{\pm} W)$. In addition,
$h_1,h_{-1},h_0$ are the populations of the $m = 1,-1,0$ states
of $^{12}$N or $^{12}$B, normalized such that $h_1 + h_{-1} + h_0
= 1$; $\cos \theta_e = \hat{p} \cdot \hat{z}$, and $\Gamma = \ln 2
/t$, where $t$ is the decay half-life.  The coefficients
$\eta_{\pm}$, $a_{\pm}$ and $\alpha_{\pm}$ are expressed, in
the formalism of Hwang and Primakoff \cite{HP77}\nl , in terms of
{\em nuclear} form factors, $F$:

\bea
\eta_{\pm} & = & \frac{W^{\pm}_0}{3M} \left( \pm 2
\frac{F^{\pm}_M (0)}{F^{\pm}_A (0)}
+ \frac{F^{\pm}_E (0)}{F^{\pm}_A (0)} \right)
\nonumber  \\
a_{\pm} & = & \mp \frac{4}{3M}
\frac{F^{\pm}_M (0)}{F^{\pm}_A (0)} + \delta_{\pm}(\alpha Z)
\nonumber  \\
\alpha_{\pm} & = & \frac{1}{3M} \left(
\mp \frac{F^{\pm}_M (0)}{F^{\pm}_A (0)}
- \frac{F^{\pm}_E (0)}{F^{\pm}_A (0)} \right),
\label{epff}
\eea

\noindent where $M$ is the nucleon mass and $\delta_{\pm}(\alpha Z)$
are radiative corrections calculated to be
$\delta_{-} = 1.31~{\rm GeV}^{-1}$ and
$\delta_{+} = 2.69~{\rm GeV}^{-1}$ by Behrens and Szybisz \cite{BS75}\nl .
These nuclear form factors
originate in an elementary-particle treatment \cite{HP77}
of the decays in which no
explicit consideration is given to the dependence of the wavefunctions
on the coordinates of the constituent nucleons, nor is the weak current
itself decomposed into contributions from individual nucleons.
Rather, the details of nuclear structure are buried in the
nuclear form factors, which have then to be determined from
experiment.

With sufficient data, this approach
provides a treatment that is essentially independent of nuclear models
and permits analyses to explore the consequences of symmetry
properties and conservation hypotheses.  But with insufficient data,
some of the form factors have to be obtained from other
sources, the most common being to appeal to the impulse approximation.
In the latter case, the starting point is the single-nucleon
currents of Eq.\,\ref{VAct} reduced to their nonrelativistic form
to yield one-body operators, the matrix elements of which are evaluated
with many-body wavefunctions.  The coupling constants of
Eq.\,\ref{VAct},
$g(k^2) \simeq g(0)$, are determined from data on the
free proton and neutron and treated as constants, unchanged
from nucleus to nucleus.  The connection with the nuclear
form factors of the elementary-particle approach is given by
\cite{HP77,De70}

\bea
\sqrt{2} F^{\pm}_M (0) & \rightarrow & - (1 + \gM)
\langle \mbox{\boldmath $\sigma$} \rangle
-  \langle {\bf L} \rangle
\nonumber  \\
\sqrt{2} F^{\pm}_A (0) & \rightarrow & - \left (
\gA \mp  \frac{W^{\pm}_0}{2M} \gT \right )
\langle \mbox{\boldmath $\sigma$} \rangle
\nonumber  \\
\sqrt{2} F^{\pm}_E (0) & \rightarrow & - (\gA \pm  \gT)
\langle \mbox{\boldmath $\sigma$} \rangle
- 2 \gA \langle i{\bf r}(\mbox{\boldmath $\sigma \cdot p$})  \rangle,
\label{ffIA}
\eea

\noindent where, for example, $
\langle \mbox{\boldmath $\sigma$} \rangle
$ is the Gamow-Teller matrix element defined more completely below
in Eq.\,\ref{MGT}.

The transition rate for the analogous gamma decay, $^{12}$C$^{\ast}
(1^{+},1) \rightarrow ^{12}$C$(0^{+},0)$, is specified by

\be
\Gamma(^{12}{\rm C}^{\ast} \rightarrow ~^{12}{\rm C} ) =
\frac{\alpha}{3} \frac{E^{3}_{\gamma}}{M^2} \mid \!
\sqrt{2} \mu (0) \! \mid ^{2},
\label{gamwdth}
\ee

\noindent where $\alpha$ is the fine-structure constant, $E_{\gamma}$
the photon decay energy and $\mu (0)$ the isovector magnetic
{\em nuclear} form factor, which, in the nonrelativistic limit,
connects to impulse-approximation quantities
\cite{HP77,De70} via

\be
\sqrt{2} \mu (0)   \rightarrow   - (\mu_p - \mu_n +1)
\langle \mbox{\boldmath $\sigma$} \rangle
-  \langle {\bf L} \rangle ,
\label{M1ffIA}
\ee

\noindent with $\mu_p$ and $\mu_n$ the proton and neutron {\em nucleon}
magnetic moments respectively.  The test of CVC consists of the
verification of the relation

\be
F^{\pm}_M (0) = \sqrt{2} \mu (0).
\label{CVCWM}
\ee

\noindent  The $\sqrt{2}$ displayed in Eq.\,\ref{CVCWM}
occurs because the isospin operator for beta decay is $(\tau^1 \pm
i \tau^2)$, whereas the isospin spherical tensor operator that relates
to the $\tau_3$ operator of gamma decay through an isospin rotation
is $(\tau^1 \pm i \tau^2)/\sqrt{2}$.

The weighted average of six
measurements \cite{Ch73} of $\Gamma_{\gamma}$ that use the technique
of nuclear resonance fluorescence (after reevaluation with
currently accepted branching ratios) is $\Gamma_{\gamma} = 40.4
\pm 2.0$ eV.  The three most recent measurements of $\Gamma_{\gamma}$
that use $^{12}{\rm C}(e,e^{\prime})$ yield $\Gamma_{\gamma} =
37.0 \pm 1.1$ eV (ref.\cite{Ch73}), $35.74 \pm 0.86$ eV (ref.\cite{Sp72})
and $38.5 \pm 0.8$ eV (ref.\cite{De83}).
Following de Braeckeleer \cite{De92}\nl , we adopt
the value $\Gamma_{\gamma} = 38.8 \pm 1.6$ eV, which is an average
of the combined data with an error chosen to make the two data
sets consistent with each other.  There is a small
correction \cite{CH76} due to isospin mixing of the
$^{12}{\rm C}^{\ast}(1^{+},1)$
state at 15.11 MeV with the
$(1^{+}, 0)$ state at 12.71 MeV, but its magnitude is
smaller than the stated error on $\Gamma_{\gamma}$ so we consider
it to be included in that estimate.  Thus, from Eqs.\,\ref{gamwdth}
and \ref{CVCWM} we obtain as a prediction of CVC:

\be
F^{\pm}_M (0) = 2.02 \pm 0.04.
\label{FMCVC}
\ee

Note that the form factors $F_A$ and $F_E$ both contain second-class
current terms through the impulse-approximation coupling constant, $
\gT$.  Thus, the analysis proceeds in one of two ways:  either one
accepts the CVC prediction, Eq.\,\ref{FMCVC}, and uses additional
data to set limits on second-class currents, or one sets $\gT = 0$
and uses additional data to verify the CVC hypothesis.  Following
the latter choice, and deducing  axial form factors from the
experimental half-lives \cite{Aj85} to be $F^{-}_A (0) = 0.512 \pm 0.001$
and $F^{+}_A (0) = 0.481 \pm 0.002$, the slope parameters, $a_{\pm}$,
(see Eq.\,\ref{epff}) are predicted from CVC to be

\bea
a_{-} & = & 5.61 + 1.31 ~~ = ~~6.92 ~ {\rm GeV}^{-1}
\nonumber  \\
a_{+} & = & -5.97 + 2.69 ~~ = ~~-3.28 ~ {\rm GeV}^{-1}
\nonumber  \\
\Delta a  & = & a_{-} - a_{+}
{}~~ = ~~10.20 \pm 0.16 ~ {\rm GeV}^{-1}.
\label{aCVC}
\eea

\noindent  The order-$\alpha Z$ corrections \cite{BS75}\nl ,
$\delta_{\pm}(
\alpha Z)$,
which appear explicitly as the second terms in the equations for
$a_{\pm}$,
are significant and shift the prediction
away from the symmetric result of $a_{-} \simeq -a_{+}$ expected
in the absence of any isospin-symmetry breaking.  Behrens and
Syzbisz \cite{BS75} also predict a sizeable quadratic term in the
shape correction factor, $b_{\pm} W^2$, in Eq.\,\ref{WMrate},
which should be included in the
analysis.  However the contribution from this term is roughly the
same in $e^{-}$ and $e^{+}$ transitions, so it can be omitted
providing we restrict our attention to $\Delta a$, which
is more reliable
experimentally and theoretically since many
systematic uncertainties cancel out.

The CVC hypothesis alone cannot predict values for the correlation
coefficients, $\alpha_{\pm}$, because of the presence of the $F_E$
form factor in Eq.\,\ref{epff}.  However a prediction can
be given for the difference \cite{HP77}:

\be
\Delta \alpha = \alpha_{-} - \alpha_{+}
 ~~ = ~~2.89 \pm 0.04 ~ {\rm GeV}^{-1},
\label{alpCVC}
\ee

\noindent where, as in Eq.\,\ref{aCVC}, the quoted
error reflects only the uncertainty in the $M1$ $\gamma$-decay
width, $\Gamma_{\gamma}$.  The experimental test of the CVC
hypothesis, therefore, is to measure the difference in slope
parameters, $\Delta a$, or the correlation coefficients, $\Delta
\alpha$, in the mirror mass-12 transitions and compare with the predictions
in Eqs.\,\ref{aCVC} and \ref{alpCVC}.  We consider some examples
in the following sections.

\subsubsection{The total decay rate}
\label{ttdr}

For non-aligned $^{12}$B and $^{12}$N nuclei, we can integrate over
electron directions and energies in Eq.\,\ref{WMrate},
giving for the ratio of decay rates
in the mirror transitions the result

\bea
\frac{ft^{+}}{ft^{-}} & = & \frac
{\mid \! F^{-}_A (0) \! \mid ^{2} }
{\mid \! F^{+}_A (0) \! \mid ^{2} } \frac
{ 1 + \eta_{-} + a_{-} W^{-}_0 /2 }
{ 1 + \eta_{+} + a_{+} W^{+}_0 /2 }
\nonumber  \\
& \rightarrow & 1 - \frac{4}{3} \frac{W^{+}_0 + W^{-}_0 }{2M}
\frac{\gT}{\gA} + \delta_{\rm nucl},
\label{ftratio}
\eea

\noindent where $f$ is the integrated Fermi function
(see Eq.\,\ref{fint}).  The
energy-dependent term, $aW$, has been replaced by its average value,
$a W_0 /2$,
and terms proportional to $(W^{+}_0 - W^{-}_0 )$ have been neglected.
The second line represents the impulse-approximation limit.  It
would seem that a comparison of $ft$ values in mirror transitions would
be a profitable way to search for second-class currents.  This,
indeed, was proposed by Wilkinson \cite{Wi70} and first results
appeared encouraging.  However, a definitive experiment was performed
by Wilkinson and Alburger \cite{WA71} in which the asymmetry of
$ft$ values was measured as a function of $(W^{+}_0 + W^{-}_0 )$
in the beta decays of $^{8}$Li and $^{8}$B.  This is a unique case
in which the daughter state is broad, decaying itself by $\alpha$
emission.  This effectively permits the beta end-point energy to be
varied within a single decay.  Their results showed the ratio of
$ft$ values to be almost independent of $(W^{+}_0 + W^{-}_0 )$,
implying little or no evidence for second-class currents.

Another difficulty with this type of analysis concerns the existence
of the additional term $\delta_{\rm nucl}$ in Eq.\,\ref{ftratio}.
It is given by
$\mid \!
\langle \mbox{\boldmath $\sigma$} \rangle ^{-} /
\langle \mbox{\boldmath $\sigma$} \rangle ^{+} \! \mid ^{2} =
1 + \delta_{\rm nucl}$ and
arises from the fact that
$\langle \mbox{\boldmath $\sigma$} \rangle ^{+} \neq
\langle \mbox{\boldmath $\sigma$} \rangle ^{-}$
since the wavefunctions for $^{12}$B
and $^{12}$N are not exact isospin eigenstates: the isospin symmetry
is broken by
the internucleon electromagnetic interaction and the neutron-proton
mass difference.  Nuclear-structure
calculations \cite{Wi71,To73b} indeed compute values for
$\delta_{\rm nucl}$ that are comparable with
the total experimental asymmetry, and thus mask
the evidence, if any, for the presence of second-class
currents.

\subsubsection{Electron-spectrum shape}
\label{Ees}

To study the electron-energy spectrum for
non-aligned $^{12}$B and $^{12}$N nuclei, we integrate only
over electron directions in Eq.\,(\ref{WMrate}).
A measurement of the spectrum
makes it possible to determine the coefficient of the energy-dependent
term $a_{\pm} W$.  This is the classic experiment of Lee \etal \,
\cite{LMW63}\nl , which first claimed to verify CVC.  However,
a re-appraisal by Calaprice and Holstein \cite{CH76}
questioned the original analysis of the experiment and weakened the
experimental support for the validity of CVC.
The critique has been discussed in a new analysis by the original group
of experimenters \cite{WLM77} using the data from the original
experiment, with the conclusion that their data still support CVC.
At the same time, a
new measurement of $a_{\pm}$ was performed by Kaina \etal \,
\cite{Ka77} using a NaI crystal as a spectrum analyzer instead of
the magnetic spectrometer employed in the earlier measurements.
The results for $a_{\pm}$ were not in
agreement with previous
(revised or not) determinations.

\begin{table}[t]
\begin{center}
\caption{Difference of slope parameters, $\Delta a$, for $^{12}$B
and $^{12}$N mirror beta decays.
\label{adiff} }
\vskip 1mm
\begin{tabular}{rcl}
\hline  \\[-3mm]
\multicolumn{1}{c}{$\Delta a$}  & $b^{+~a}_1$  & Remarks  \\
 ( GeV$^{-1}$) &  (\%) &   \\
\hline  \\[-3mm]
$9.2 \pm 1.2$ & 2.4 & ref.\,\cite{MM62}\nl , as revised by \cite{CH76}
and $a_{+}$ increased by 0.7  \\
$15.3 \pm 2.8$ & 2.4 & ref.\,\cite{GP63}\nl , as revised by \cite{CH76}
and $a_{+}$ increased by 0.7  \\
$8.6 \pm 2.4$ & 2.1 & ref.\,\cite{WLM77}  \\
$9.1 \pm 0.9$ & 1.8 & ref.\,\cite{Ka77}\nl , with $a_{+}$ decreased by 0.7  \\
$10.5 \pm 2.1$ & & ref.\,\cite{Mi89}\nl , branching ratio, $b^{+}_1$, not
given \\
\hline \\
\multicolumn{3}{l}{{\footnotesize $^{a}$Branching ratio assumed
in original reference.}}
\end{tabular}
\end{center}
\end{table}

One cause of the trouble lies in the unresolved decay branch from $^{12}$N
to the first excited state in $^{12}$C.
The conclusion derived from the experiment depends strongly upon
the knowledge of the branching ratio, $b_1^{+}$, for the decay to that
state, which must
be subtracted from the
observed result.  For example \cite{WLM77}\nl , a change from 1.8\% to 2.1\% in
$b^{+}_1$ produces a variation in the extracted $a_{+}$ of
$-0.7 ~{\rm GeV}^{-1}$, a 25\%
change.  A second cause of trouble lies in the systematic uncertainties
surrounding the experimental energy-dependent response functions.
So far, these seem insurmountable in the determination of $a_{\pm}$,
but are deemed largely to cancel in $\Delta a$; as a result,
it is the latter quantity that we quote here.
In Table \ref{adiff} are the results from
four experiments as corrected for a common $2.1 \%$ branching ratio
as reviewed by Grenacs \cite{Gr85}\nl , and
a preliminary result \cite{Mi89} from a fifth experiment.
The weighted average is $\Delta a = 9.5 \pm 0.7$.  By combining
in quadrature  the effects of
an uncertainty of 0.3\% in $b^{+}_1$,
we produce a recommended value of

\be
\Delta a = 9.5 \pm 1.0 ~{\rm GeV}^{-1} ,
\label{adexpt}
\ee

\noindent  which agrees with the CVC prediction given in Eq.\,\ref
{aCVC}.

\subsubsection{Polarization and alignment measurements}
\label{paam}

We now turn to experiments with oriented nuclei and measurements of the
decay electron's angular distribution as given in Eq.\,\ref{WMrate}:

\be
W(\theta_{e}) = 1 \pm (h_1 - h_{-1}) (1 + \alpha_{\pm}W) P_1(\cos \theta_e)
+ (1-3h_0) \alpha_{\pm} W P_2(\cos \theta_e) .
\label{angdis}
\ee

\noindent  Typical experimental arrangements are described by
Minamisono \cite{Mi73}\nl , Haskell and Mad\-ansky \cite{HM73} and
Haskell \etal \, \cite{HCM75}\nl .  Polarized $^{12}$B and $^{12}$N
recoils are produced through the reactions $^{10}{\rm B}(^{3}{\rm He},
n)^{12}{\rm N}$ and $^{11}{\rm B}(d,p)^{12}{\rm B}$, with the
precise magnetic substate populations of the product nuclei depending
on the incident beam energy and the recoil angle.  The polarized
recoiling nuclei are implanted into monocrystalline metal foils
in a large magnetic holding field in
which the degeneracy of the Zeeman levels is broken by the
quadrupole couplings.  The polarization
is detected  through
an asymmetry in the angular distribution of the electrons
in the subsequent $\beta$ decays.

\begin{table}[t]
\begin{center}
\caption{Beta-decay asymmetry coefficients, $\alpha_{\pm}$ in
GeV$^{-1}$.  \label{alpexpt} }
\vskip 1mm
\begin{tabular}{ccl}
\hline  \\[-3mm]
$\alpha_{-} (^{12}{\rm B})$ &
$\alpha_{+} (^{12}{\rm N})$ & Reference  \\
\hline  \\[-3mm]
$-(0.07 \pm 0.20)$ & & Lebrun \etal \, \cite{Le78} \\
$+(0.24 \pm 0.44)$ & & Br\"{a}ndle \etal \, \cite{Br78a} \\
$+(0.25 \pm 0.34)$ & $-(2.77 \pm 0.52)$ & Sugimoto \etal \,
\cite{Su78}  \\
$+(0.10 \pm 0.30)$ & $-(2.73 \pm 0.39)$ & Br\"{a}ndle \etal \,
\cite{Br78b}  \\
$-(0.01 \pm 0.21)$ & $-(2.67 \pm 0.56)$ & Masuda \etal \,
\cite{Ma79}  \\
$+(0.046 \pm 0.053)$ & $-(2.795 \pm 0.100)$ & Minamisono \etal \,
\cite{Mi86}  \\
$-(0.174 \pm 0.059)$ & $-(2.774 \pm 0.086)$ & Minamisono \etal \,
\cite{Mi93}  \\[3mm]
$-(0.043 \pm 0.046)$ & $-(2.780 \pm 0.063)$ & Average values
\\
\hline
\end{tabular}
\end{center}
\end{table}

The first determinations \cite{STG75} of $\alpha_{\pm}$ from the
$\beta$-energy dependence of the $P_{1}(\cos \theta_{e})$
term in the decay electron's
angular distribution (see Eq.\,\ref{angdis}) turned out to be fraught
with experimental difficulties (primarily from backscattering),
since a small effect is to be isolated from a large leading
term (\ie \ $\alpha W \ll 1$).  More reliable data have come from later
experiments that concentrated on the $P_{2}(\cos \theta_{e})$ term
in Eq.\,\ref{angdis}.  A summary of experimental results
is given in Table \ref{alpexpt}.  The data are all in accord
and the average value for the difference in the mirror correlation
coefficients is

\be
\Delta \alpha = \alpha_{-} - \alpha_{+} = 2.737 \pm 0.078 ~{\rm GeV}^{-1}
\label{aldiexp}
\ee

\noindent  This result is in good agreement with the CVC expectation
given in Eq.\,\ref{alpCVC}.

\subsubsection{$\beta$-$\alpha$ correlations}
\label{bacorr}

A similar test of CVC can be mounted in the mass-8 mirror transitions
$^{8}{\rm Li}(2^{+},1) \rightarrow \ ^{8}$Be$(2^{+},0) +
e^{-} + \overline{\nu}_e$ and
$^{8}{\rm B}(2^{+},1) \rightarrow \ ^{8}$Be$(2^{+},0) +
e^{+} + \nu_e$.  The broad daughter state, which becomes aligned
with respect to the electron direction in
the process, subsequently breaks up into a pair of $\alpha$ particles.
The degree of alignment is measured from the
$\cos^2 \theta$ distribution of $\alpha$-particles, where $\theta$
is the angle between the electron and $\alpha$-particle directions.
The $\beta$-$\alpha$ directional correlation in the A=8 nuclei
and the electron angular distribution from aligned A=12 nuclei
are therefore essentially equivalent phenomena \cite{Gr85} with the
role of the initial and final states inverted.

The CVC prediction for this case requires the analogous M1
$\gamma$-transition to be measured, but this poses more problems than in the
mass-12 case.  Firstly, the $2^{+} \rightarrow 2^{+}$ sequence implies that the
electromagnetic transition is a mix of M1 and E2 multipoles, so a
mixing ratio has to be measured to determine the M1 fraction.
Secondly, the analogue state of the decay parent in $^{8}{\rm Be}$ is
a nearly degenerate doublet (16.6 and 16.9 MeV) of which both
members are roughly equal mixes of T=0 and T=1 eigenstates.
De Braeckeleer \cite{De94} recently reported on a determination of
the M1 and E2 widths needed for the CVC test in which the
$^{4}{\rm He}(\alpha , \gamma )$ cross section was measured as a function of
energy across the 16.6 and 16.9 MeV levels and photon
angular distributions were obtained for both resonances.  The analysis
leads to an isovector M1 width of $2.80 \pm 0.18$ eV, in serious
disagreement
with previous determinations by Bowles and Garvey \cite{BG78,BG82}\nl .

The $\beta$-$\alpha$ correlations
in the two mirror decays
have been measured by Tribble and
Garvey \cite{TG75} and the results are in
agreement with the CVC prediction based on the earlier M1 width,
although there are large uncertainties.  A subsequent experiment
\cite{MGG80} detecting both alpha particles enabled the correlation
to be determined as a function of the final-state energy.  The results
show a deviation from the CVC prediction, which is dependent
on the final-state energy.
A repeat experiment \cite{De92} is
currently underway, in which it is proposed to
measure the triple $\beta$-$\nu$-$\alpha$ angular correlation.
The neutrino direction will be inferred from the recoil of the second
$\alpha$ particle.  Furthermore, by an appropriate choice of geometry,
it may be possible to separate the weak-magnetism
and the second-class-current contributions to the alignment.
The experiment \cite{De92} is expected to improve present limits
on second-class currents by a factor of five.

\subsubsection{$\beta$-$\gamma$ correlations}
\label{bgcorr}

Another similar test of CVC is based on the mass-20 mirror transitions
$^{20}{\rm F}(2^{+},1) \linebreak[1]
\rightarrow \ ^{20}$Ne$(2^{+},0) + e^{-} +
\overline{\nu}_e$ and $^{20}{\rm Na}(2^{+},1) \rightarrow \
^{20}$Ne$(2^{+},0) + e^{+} + \nu_e$.  In this case, the daughter
nucleus de-excites by $\gamma$-ray emission rather than $\alpha$
emission as in the mass-8 case.  The $\beta$-$\gamma$ angular
correlation measurement is equivalent to the $\beta$-$\alpha$
correlation experiment and the analysis therefore follows
analogously.  The results \cite{DR78,TM78,TMT81} confirm the CVC
hypothesis, but only to
an accuracy of $\pm 25\%$.

\section{Axial-vector Interaction in the Nucleon} \label{AVI}
\subsection{Partially Conserved Axial-vector Current Hypothesis}  \label{PCACh}

We start with a small aside and consider the decay of the pion:
$\pi^{-} \rightarrow \ell^{-} \overline{\nu}_{\ell}$,
where $\ell$ represents either the first generation,
$e^{-} \overline{\nu}_e$,
or second generation, $\mu^{-} \overline{\nu}_{\mu}$, leptons.
Since the pion is a bound state of $d$ and $\overline{u}$ quarks,
its decay is a straightforward example of a semi-leptonic charged-current
weak interaction.  The $T$-matrix for the decay process is given by
Eq.\,\ref{Tfi0}, and
is easily computed for free quarks.  However, for pion
decay, the quarks are bound within the pion and, as
mentioned in sect.\,\ref{lforn}, the required quark density functions
are not reliably computed.  Thus, a more phenomenological approach
is followed in which we parameterize the matrix element of the quark hadronic
current, $J_\mu^h$, evaluated between the pion and the vacuum,
$ \langle 0|J_\mu^h|\pi^{-}\rangle $.  Since the pion
is spinless,
the only available four-vector is the momentum transfer between
the hadron and leptons, $k_\mu$.  Thus, the current is
parameterized as

\be
J_\mu^h = if_\pi k_\mu \pi^a,
\label{fkmu}
\ee

\noindent where $\pi^a$ is the field operator for the pion,
with the superscript
$a$ an isospin index, and $f_\pi$ is some real numerical constant,
which parameterizes our inability to evaluate the hadron matrix
element.  In principle, $f_\pi$ should be calculable in QCD.  Note that
since the pion is a pseudoscalar particle and $k_\mu$ a polar vector,
only the axial portion of $J_\mu^h$, namely $A_\mu^h$, is operative
in Eq.\,\ref{fkmu}.  The lifetime for the decay is evaluated
in a standard way \cite{CB83,Re90} to give:

\be
\tau^{-1}(\pi^{-} \rightarrow \ell^{-} \overline{\nu}_{\ell}) =
\frac{\GF^2 f_\pi^2}{4\pi} V_{ud}^2 m_{\ell}^2 m_\pi
\left(1 - \frac{m_{\ell}^2}{m_{\pi}^2} \right)^2.
\label{pitau}
\ee

\noindent The decay is predominately to muons.  Thus, with the measured pion
lifetime, corrected for other small branches and adjusted by
a few percent for radiative corrections, a value for $f_\pi$ is
determined from Eq.\,\ref{pitau} with the known values of
$m_\mu, m_\pi, \GF$ and $V_{ud}$.  The result \cite{PDG94} is
\footnote{Note our definition of $f_\pi$ differs from the
Particle Data group \cite{PDG94} by $\sqrt{2}$}

\be
f_\pi = 0.663 m_\pi = 92.42 \pm 0.26 \, {\rm MeV}.
\label{fpi}
\ee

\noindent The constant, $f_\pi$, is called the pion-decay constant.

The axial-vector current has no electromagnetic analogue.
Furthermore,  it is not a conserved current since
the pion does decay.  The axial current and its
derivative for pion decay are therefore

\bea
A_\mu^h & = & if_\pi k_\mu \pi^a
\nonumber  \\
\del_\mu A_\mu^h & = & ik_\mu A_\mu^h~~=~~-f_\pi k_\mu^2 \pi^a
{}~~=~~m_\pi^2 f_\pi \pi^a,
\label{delpi}
\eea

\noindent where $k_\mu^2=-m_\pi^2$ for free pions.  Its divergence
is non-zero because $f_\pi \neq 0$ and $m_\pi \neq 0$.
However the pion is light on the hadronic mass scale
and its mass can be considered small.  Gell-Mann and Levy \cite{GL60}
introduced the concept of a partially conserved axial current (PCAC)
by postulating that conservation is achieved in the ``soft-pion limit",
in which the pion mass tends to zero.  Thus, the divergence of the
axial current appears to be related to the creation operator for a
pion.  If
this principle is also applied to the axial current operative in the
decay of the neutron, it can be assumed
that the current
is dominated by the pion-pole graph.  That is, a pion is created
at the proton-neutron vertex and then subsequently decays.
Schematically, the axial current is \cite{CB83}

\bea
A_\mu^h & = & (n \rightarrow p \pi^{-}\, \rm{vertex} ) \times
( \pi^{-}\, \rm{propagator} ) \times (\pi^{-} \rightarrow e^{-}
\overline{\nu}_e
\, \rm{vertex})
\nonumber  \\
& = & i \gpNN \FpNN (k^2) \overline{\psi} \gamma_5 \psi \tau^a ~~\times ~~
(k^2+m_\pi^2)^{-1} ~~\times ~~ if_\pi k_\mu
\nonumber  \\
& = & i\, \frac{2 \gpNN \FpNN (k^2) f_\pi}{k^2+m_\pi^2} \,\overline{\psi}
[i k_\mu \gamma_5 ] \psi \, \sfrac{1}{2} \tau^a,
\label{Amuh1}
\eea

\noindent where $\gpNN$ is the pion-nucleon coupling constant.
Here $\FpNN (k^2)$ is a vertex form factor, taken to be a smooth
function of $k^2$ with $\FpNN (k^2=-m_\pi^2) = 1$ so that
$\gpNN$ becomes the physically-measured coupling constant.
Thus, the derivative of this current in the soft-pion limit
(neglecting the derivative of the vertex form factor) is

\be
\lim_{m_\pi^2 \rightarrow 0} \del_\mu A_\mu^h =
-i\, 2\gpNN \FpNN (k^2) f_\pi \, \overline{\psi} [\gamma_5] \psi
\, \sfrac{1}{2} \tau^a.
\label{dAmuh1}
\ee

Another possible pole graph involves the $A_1$-meson, a meson of
the same spin-parity quantum numbers as the axial current.  Then
the axial current is

\bea
A_\mu^h & = & (n \rightarrow p A_1 \,\rm{vertex}) \times
(A_1 \,\rm{propagator}) \times (A_1 \rightarrow e^{-}
\overline{\nu}_e \,\rm{vertex})
\nonumber  \\
& = & (-i \gA g_\rho \FANN (k^2) \overline{\psi} \gamma_\mu \gamma_5
\tau^a \psi)
{}~~ \times ~~ \left( \frac{\delta_{\mu \nu} + k_\mu k_\nu /\mA^2}
{k^2 + \mA^2} \right) ~~\times~~ \left( -\frac{\mA^2}{2g_\rho} \right)
\nonumber \\
& = & i \gA \FANN (k^2) \frac{\mA^2}{k^2+\mA^2} \overline{\psi} \left[
\gamma_\mu \gamma_5 + \frac{k_\mu \not{\!k} \gamma_5}{\mA^2} \right]
\psi \,\sfrac{1}{2} \tau^a ,
\label{Amuh2}
\eea

\noindent where $\not{\!k} = k_{\alpha} \gamma_{\alpha}$.
The $A_1NN$ coupling constant is extracted from a phenomenological
chirally-invariant Lagrangian \cite{LN68} in which the $A_1$-meson
is treated as the chiral partner of the $\rho$-meson, and its mass
related to that of the $\rho$-meson via the Weinberg
relation \cite{We67}: $\mA = \sqrt{2} m_{\rho}$.
In particular,
we quote from the work of Ivanov and Truhlik \cite{IT79} where
$g_{A_1NN} = \gA g_\rho$ with $g_\rho$ the coupling constant for
the decay of the rho-meson to two pions.  Similarly, in the same Lagrangian,
the decay of the $A_1$-meson to leptons is given in terms of the
same coupling constant, $g_\rho$, which then cancels out in forming
the axial current in Eq.\,\ref{Amuh2}. This is the feature of
chiral Lagrangians that all vertices involving $\rho$ and $A_1$ mesons
are expressed in terms of a common coupling constant, here $g_\rho$.
Again, a smooth vertex function has been introduced with
$\FANN (k^2 = -\mA ^2) = 1$.
The derivative of this current is

\bea
\del_\mu A_\mu^h & = & -\gA \FANN (k^2) \frac{\mA^2}{k^2+\mA^2} \overline{\psi}
\left[ \not{\!k} \gamma_5 + \frac{k_\mu \not{\!k} \gamma_5}{\mA^2}
\right] \psi \,\sfrac{1}{2} \tau^a
\nonumber  \\
& = & -\gA \FANN (k^2) \overline{\psi} \left[ \not{\!k} \gamma_5 \right]
\psi \sfrac{1}{2} \tau^a
\nonumber  \\
& = & 2i \gA \FANN (k^2) M \overline{\psi} \left[ \gamma_5 \right]
\psi \sfrac{1}{2} \tau^a,
\label{dAmuh2}
\eea

\noindent with $M$ the nucleon mass.
Then, if the pion- and $A_1$-meson pole graphs,
Eqs.\,\ref{dAmuh1} and \ref{dAmuh2}, are added together,
the derivative of the
axial current in the soft-pion limit becomes

\be
\lim_{m_\pi^2 \rightarrow 0} \del_\mu A_\mu^h =
2i(\gA \FANN (k^2) M - \gpNN \FpNN (k^2) f_\pi) \overline{\psi} [\gamma_5] \psi
\,\sfrac{1}{2} \tau^a,
\label{dAmuh3}
\ee

\noindent and the requirement that this derivative vanish leads to
the condition

\be
\gA \FANN (k^2) M = \gpNN \FpNN (k^2) f_\pi.
\label{GTRp}
\ee

\noindent If, in addition, it is assumed that the vertex form factors
are slowly varying functions of $k^2$, such that they can be replaced
by their limiting values of unity, then

\be
\gA M = \gpNN f_\pi.
\label{GTR}
\ee

\noindent This result is known as
the Goldberger-Treiman relation.

The accepted value for $\gpNN$ for many years was that determined
from $\pi$-nucleon scattering data
by Bugg, Carter and Carter \cite{BCC73} of $13.40 \pm 0.08$ and
confirmed by the Karlsruhe-Helsinki \cite{KH80} group.  Recently,
this value has been challenged, notably by the Nijmegen group
\cite{Nij93}\nl , and
smaller values are currently recommended.  Three recent determinations
are: $13.03 \pm 0.03$ (ref.\cite{Nij93}),
$13.14 \pm 0.07$ (ref.\cite{AWP94}),
and $13.12 \pm 0.13$ (ref.\cite{BM94}).  Since this issue is still under
discussion we will use the last value as a conservative estimate
for what follows.  With this choice for $\gpNN$, the pion-decay
constant from Eq.\,\ref{fpi}, and the average of the
proton and neutron masses, $M = 938.919$ MeV, the Goldberger-Treiman
relation predicts for the axial-vector constant
$\mids \gA \mids \ = 1.29 \pm 0.01$.
This is reasonably close to the value 1.26 obtained from neutron
beta decay (see next section).  The difference of only 2\%
is remarkable testimony to the
accuracy of the PCAC hypothesis.

Finally, the sum of Eqs.\,\ref{Amuh1} and \ref{Amuh2} yield for the
total axial current

\bea
A_\mu^h & = & i\gA \FANN (k^2) \overline{\psi} \left[ \frac{\mA^2}{k^2+\mA^2}
\gamma_\mu \gamma_5 - \frac{(\mA^2-m_\pi^2)}{(k^2+\mA^2)
(k^2+m_\pi^2)} k_\mu \not{\!k} \gamma_5 \right] \psi
\,\sfrac{1}{2} \tau^a
\nonumber  \\
& = & i\gA \FANN (k^2) \frac{\mA^2}{k^2+\mA^2} \overline{\psi} \left[
\gamma_\mu \gamma_5 + \frac{2Mi(1-m_\pi^2/\mA^2)}{k^2+m_\pi^2}
k_\mu \gamma_5 \right] \psi
\,\sfrac{1}{2} \tau^a,
\label{Amuh3}
\eea

\noindent which can be compared with the phenomenological expression
of Eq.\,\ref{VAct}
to obtain expressions for the form factors $\gA(k^2)$ and $\gP(k^2)$.
The results are

\be
\gA(k^2) = \gA \FANN (k^2) \frac{\mA^2}{k^2+\mA^2}
\label{gAk2}
\ee

\noindent and

\be
\gP(k^2) = \frac{2M\gA(k^2)}{k^2+m_\pi^2} \left( 1-\frac{m_\pi^2}{\mA^2}
\right),
\label{gPk2}
\ee

\noindent or

\be
\gP (k^2) = \frac{2 \gpNN f_\pi }{k^2 + m_{\pi}^2} \FpNN (k^2).
\label{gPk2p}
\ee

\noindent Since these derivations are based on meson-pole graphs, the
form factors deduced are of monopole form, with the range given by
the mass of the appropriate meson.
For $\gA (k^2)$, use of a phenomenological chiral Lagrangian has
fixed $\mA = \sqrt{2} m_{\rho} = 1086~{\rm MeV}$, a value not too
far from the physical $A_1$ meson mass \cite{PDG94} of
$1230 \pm 40~{\rm MeV}$.
Experimental information has also been obtained
from neutrino-nucleon scattering \cite{Ah88} at the Brookhaven AGS,
where the antineutrino quasielastic reaction
$ \overline{\nu}_{\mu} p \rightarrow \mu^{+} n$
was studied in the momentum range up to $k^2 \sim 1.0~{\rm GeV}^2$.
A value of an axial-vector mass, $\mD$, in a {\it dipole}
parameterization of the form factor was determined from the shape
of the momentum distribution to be $1.09 \pm 0.04~{\rm GeV}$.
This agrees with the monopole value, $\mA$, obtained from the
phenomenological chiral Lagrangian and suggests the vertex form
factor $\FANN (k^2)$ is likewise a monopole function of
similar range.

In Eqs.\,\ref{gPk2} and \ref{gPk2p} we gave two expressions for the
pseudoscalar coupling constant.  The first uses the Goldberger-Treiman
relation and includes a factor $(1-m_{\pi}^2 / \mA ^2 )$, which
originates in the propagator for the $A_1$-meson.
The second expression assumes that the
pseudoscalar coupling constant is entirely due to the pion-pole graph,
and is the expression we will use here.  In semi-leptonic decays, the
pseudoscalar coupling constant is multiplied
by the mass of the lepton, $m_\ell$, in the
expression for the transition rate.  Thus, for processes involving first
generation leptons, such as nuclear beta decay, the pseudoscalar
term gives a negligible contribution:
see sect.\,\ref{RMCH}.
In contrast, for processes such as muon capture, involving
second-generation leptons, the contribution is significant
and characterized by the dimensionless quantity

\be
m_\mu \frac{\gP(k^2)}{\gA(0)} = \frac{2 \gpNN f_{\pi} m_\mu}
{\gA (k^2+m_\pi^2)}
\FpNN (k^2) = 7.04 \pm 0.07,
\label{gpga}
\ee

\noindent in which $\gpNN = 13.12 \pm 0.13$, $f_{\pi} = 92.42 \pm
0.26~{\rm MeV}$, $\gA = 1.2599 \pm 0.0025$ from neutron decay and
$\FpNN (k^2) \simeq 1$.  A correction to this expression has recently
been given by Bernard, Kaiser and Meissner \cite{BKM94}\nl .  It is based
on the chiral Ward identity of QCD and exploits heavy-meson chiral
perturbation theory to find an expression to order $k^2$.  The
result is

\be
m_\mu \frac{\gP(k^2)}{\gA(0)} = \frac{2 \gpNN f_{\pi} m_\mu}
{\gA (k^2+m_\pi^2)}
- \frac{1}{3} m_{\mu} M \rA^2,
\label{gpgac}
\ee

\noindent where $\rA^2$ is the mean square axial radius of the nucleon
obtained from the axial-vector form factor: $\gA (k^2) = \gA (0)
( 1 - \sfrac{1}{6} k^2 \rA^2 )$.  In dipole fits \cite{Ah88}\nl ,
$\rA = \sqrt{12} /\mD = 0.63 \pm 0.02~{\rm fm.}$ corresponding
to $\mD = 1.09 \pm 0.04~{\rm GeV}$.  The prediction for the dimensionless
pseudoscalar quantity then becomes

\be
m_\mu \frac{\gP(k^2)}{\gA(0)} = (7.04 \pm 0.07)
- (0.33 \pm 0.02) = 6.71 \pm 0.08.
\label{gpgap}
\ee

\noindent In both Eqs.\,\ref{gpga} and \ref{gpgap}, the
numerical result is obtained for a value of $k^2$ appropriate
for muon capture ($k^2=-m_\mu^2+2m_\mu Q \simeq 0.88 m_\mu^2$) where $Q$
is the reaction Q-value.  A measurement of $\gP$ in muon capture
by hydrogen, in addition to providing a test of PCAC and
the dominance of the pion pole, could, if sufficient accuracy were
achieved, confirm or deny the correction term predicted \cite{BKM94}
from heavy-meson chiral perturbation theory.

\subsection{Neutron Decay}  \label{NeutD}

The decay of the neutron is the simplest $\beta$-decay mode that
involves both the vector and the axial-vector interactions.  In
principle, its study should provide us with the most direct view
of $\gA$ and $\gV$, uncluttered by complex nuclear wave functions or
by renormalization effects of the nuclear medium.  In practice,
however, the experimental uncertainties have, until
recently, been large.
The value of
$\gV$, as determined from neutron decay, is a factor of five less
precise than the value obtained from the $0^{+} \rightarrow 0^{+}$
nuclear $\beta$-decays (see sect.\,\ref{SFT}).  Ultimately, though,
its potential is greater since the nuclear result has reached a plateau
established by theoretical uncertainties that are not likely to be
soon reduced.

In the same notation used to describe the superallowed nuclear decays
(see Eq.\,\ref{ftF}), the neutron decay rate can be written:

\be
ft(1 + \delta_R ) = \frac{K}{(\GV^{\prime 2} \langle \MV \rangle ^{2}
+ \GA^{\prime 2} \langle \MA \rangle ^{2} )} ,
\label{ftneut}
\ee

\noindent with

\bea
\GV^{\prime 2} = (1 + \DRV ) \GV^2
\nonumber \\
\GA^{\prime 2} = (1 + \DRA ) \GA^2 .
\label{GVpGAp}
\eea

\noindent Here $\GV = \GF V_{ud}$ and $\GA = \GF V_{ud} \gA$ with
$\GF$ the fundamental weak-interaction coupling constant determined
from muon decay, to which all other coupling constants are referenced.
For neutron decay, $\langle \MV \rangle^{2} = 1$ and $\langle \MA
\rangle^{2}
= 3$.

Although the calculation of the matrix elements, $\MV$ and $\MA$,
does not offer any
theoretical challenge in the case of neutron decay, it is evident
that a measurement of the neutron's $ft$-value must be complemented
by one additional measurement before useful information can be
extracted on the individual coupling constants, $\GV^{\prime}$
and $\GA^{\prime}$.
To appreciate what other measurements are possible, we write the
differential of the probability distribution for neutron $\beta$ decay,
$n \rightarrow p e^{-}\overline{ \nu}_{e}$, in the following form
\cite{Du91}:

\be
{\rm d}^5 \Gamma \propto
N(W) \left [ 1 + a \frac{v}{c} \hat{{\bf p}}_e \cdot \hat{{\bf p}}
_{\overline{\nu}}
 + \langle \mbox{\boldmath $\sigma$}_n \rangle
\cdot \left ( A  \frac{v}{c} \hat{{\bf p}}_e + B  \hat{{\bf p}}
_{\overline{\nu}}
 \right ) \right ]
{\rm d}\Omega_e {\rm d}\Omega_{\overline{\nu}} {\rm d}W ,
\label{Wneut}
\ee

\noindent in which $N(W)$ is the usual allowed electron energy
spectrum and
$ \langle \mbox{\boldmath $\sigma$}_n \rangle $
is the average neutron polarization vector.  The three coefficients,
$a,A$ and $B$, known respectively as the $e$-$\overline{\nu}$
angular-correlation
coefficient, the $\beta$-asymmetry parameter and the
$\overline{\nu}$-asymmetry
parameter, are each related to the ratio of coupling constants:
\viz ,

\be
a = \frac{1 - \lambda^2}{1 + 3 \lambda^2} ~~; ~~~~
A = -2 \frac{\lambda (\lambda +1)}{1 + 3\lambda^2} ~~; ~~~~
B = 2 \frac{\lambda (\lambda -1)}{1 + 3\lambda^2},
\label{aAB}
\ee

\noindent with

\be
\lambda = \GA^{\prime} / \GV^{\prime} \simeq \gA .
\label{Lambda}
\ee

\noindent All three coefficients have
been measured and the first two, being more sensitive to $\lambda$,
have been combined with the neutron's $ft$ value to determine
$\GV^{\prime}$ and $\GA^{\prime}$.

\subsubsection{$ft$-value measurements} \label{ftneuts}

As with nuclear decay, the $ft$ value for
neutron decay depends upon its $Q_{EC}$ value, half-life and
branching ratio.  The $Q_{EC}$ value is derived from the
evaluated masses \cite{Au93} of the neutron and proton, which
are known to high precision via a combination of Penning-trap
mass measurements (see, for example, ref.\,\cite{Va93}) and
measurements of the neutron binding energy in deuterium
(\eg \  ref.\,\cite{Gr86}).  The result is $Q_{EC} =
1293.3392(22)$ keV.

Measurement of the neutron lifetime has defied such high levels of precision.
Its relatively long, 10.2 min half-life, together with the difficulty
in containing a well-defined source, have conspired to
restrict precision to a half percent or so -- more than an
order-of-magnitude worse than the best nuclear-lifetime measurements
-- in spite of heroic efforts.  The measurements (for a recent
review, see ref.\cite{Sc92}) can be separated into two distinct
categories: those in which the decay products are counted from
a defined segment of a slow-neutron beam, and those in which cold
neutrons are stored and their number measured after a well-determined
time.  These categories have been described \cite{Du91} as ``counting
the dead" and ``counting the survivors", respectively.

{\it 3.2.1.1  Neutron-beam experiments}.  These experiments depend upon
an accurate knowledge of the efficiency, $\epsilon_d$, with which
decay products can be detected, and upon the total number of neutrons, $N$,
being viewed by the decay-product detectors.  If $C_d$ is the
observed decay rate, then

\be
C_d = \frac{N \epsilon_d}{\tau} = \frac {\rho_N V_d \epsilon_d}
{\tau} .
\label{taubeam}
\ee

\noindent  Here $\tau$ is the neutron mean-life (conventionally
used in this field instead of the half-life: $t = \tau \ln 2$), $
\rho_N$ is the neutron density and $V_d$ is the sensitive decay
volume.  If a continuous neutron beam is used, special attention is
required to define the boundaries of $V_d$; with a pulsed beam, $V_d$
is defined by the dimensions of the neutron packet, but
at the expense of a reduced counting rate.

{\footnotesize
\begin{table}[t]
\begin{center}
\caption{Experimental results$^{a}$ for neutron decay \label{exptneut}}
\vskip 1mm
\begin{tabular}{cclllll}
\hline  \\[-3mm]
& Method & \multicolumn{2}{c}{Measured values } &~&
\multicolumn{2}{c}{Average values} \\
\cline{3-4}
\cline{6-7}
& & ~~~~~~~~1 & ~~~~~~~~2  & & \multicolumn{1}{c}{subsets} &
\multicolumn{1}{c}{overall}  \\
\hline  \\[-3mm]
$\tau$(s) & $n$ beam & $~~918 \pm 14$ \cite{Ch72} & $~~891 \pm 9$ \cite{Sp88}
& & &  \\
& & $~~876 \pm 21$ \cite{La88} & $~~878 \pm 31$ \cite{Ko89} & & & \\
& & $~~893.6 \pm 5.3$ \cite{By90} & & &
$~~894.2 \pm 4.7$ &   \\
& $n$ trap & $~~903 \pm 13$ \cite{Ko86,Mo89} & $~~893 \pm 20$ \cite{Mo89}
& & & \\
& & $~~877 \pm 10$ \cite{Pa89}
& $~~887.6 \pm 3.0$ \cite{Ma89}
& & & \\
& & $~~888.4 \pm 3.3$ \cite{Ne92} &
$~~882.6 \pm 2.7$ \cite{Ma93} & & $~~885.9 \pm 1.7$ &  \\
& & & & & & $~~887.0 \pm 2.0$  \\
$\lambda$ & $\beta$-asym. & $-1.254 \pm 0.015^b$ \cite{Kr75} &
$-1.257 \pm 0.012^b$ \cite{Er79}
& & & \\
& & $-1.262 \pm 0.005$ \cite{Bo86} &
$-1.2544 \pm 0.0036$ \cite{Er91}  & & & \\
& & $-1.266 \pm 0.004$ \cite{Sc95} &
& & $-1.2599 \pm 0.0026$ & \\
& $e$-$\overline{\nu}$ corr. & $-1.259 \pm 0.017$ \cite{St78} & & &
$-1.259 \pm 0.017$ & \\
& & & & & & $-1.2599 \pm 0.0025$ \\
\hline \\
\multicolumn{7}{l}{\footnotesize
{$^{a}$ Following the practice used for
 superallowed decay, we retain only those measurements with}} \\
\multicolumn{7}{l}{\footnotesize
{uncertainties
that are within a factor of ten of the most precise
measurement for each quantity.}} \\
\multicolumn{7}{l}{\footnotesize
{All such measurements not withdrawn
by their authors, of which we are aware in early 1995,}} \\
\multicolumn{7}{l}{\footnotesize
{are listed.}} \\
\multicolumn{7}{l}{\small {$^b$ Corrected for weak magnetism and recoil
following ref.\,\cite{Wi82} }}
\end{tabular}
\end{center}
\end{table}
}

Results for the five most-precise beam measurements appear in
Table \ref{exptneut}.  Of the three continuous-beam experiments,
one \cite{Ch72} detected the decay electrons, the other two
\cite{Sp88,By90} detected protons.  The most recent, by Byrne \etal
\  \cite{By90}\nl , stored the decay protons in a variable-length Penning
trap, the magnetic axis of which coincided with the neutron-beam
axis.  To suppress background, the protons were released periodically
in a brief burst and accelerated into a solid-state detector.  The
two remaining beam experiments used pulsed cold-neutron beams:
the first \cite{La88} observed decay electrons with an in-line
superconducting spectrometer, PERKEO; the other \cite{Ko89}
used a helium-filled time-projection chamber to observe the
decay electrons and measure the neutron density in the same apparatus.

{\it 3.2.1.2 Stored-neutron experiments}.  Neutrons of sufficiently low
energy ($\leq 10^{-7}$ eV) can be successfully confined by material
walls, from which they are coherently scattered, by gravity or by
a magnetic-field gradient acting on their magnetic moment.  In
experiments of this type, a reproducible but not necessarily well
known number of neutrons is stored and then the number that remains
after a storage time, $t$, is counted.  The result is related to
the neutron mean-life by

\be
N(t) = N(0) \exp \left [ -t \left ( \frac{1}{\tau} + \frac{1}{
\tau_{\,{\rm loss}}} \right ) \right ],
\label{taustore}
\ee

\noindent in which $\tau_{\,{\rm loss}}$ reflects the various neutron-loss
processes except $\beta$ decay.  The measurement is then repeated
for many different storage times and under different conditions
that affect the loss processes.  Because the loss effects are deemed to
be reasonably well understood, these results can be extrapolated
to give a zero-loss limit, thus yielding the neutron mean-life.

Results for six measurements with neutron traps are listed in
Table \ref{exptneut}: five of the experiments used ultra-cold neutrons
in gravitational traps with diverse wall materials; one employed a
magnetic storage ring.  The earliest trap \cite{Ko86} had aluminum
walls, adjustable in temperature from 80 to 750 K, and movable
plates to vary the neutron collision frequency.  This same
apparatus was later used \cite{Mo89} with a thin layer of heavy-water
ice on the walls to reduce losses substantially.  A wall
coating of hydrogen-free diffusion-pump oil was used by Mampe \etal
\cite{Ma89} to minimize loss in their variable-volume chamber,
while Nesvizhevskii \etal \  \cite{Ne92} chose frozen oxygen on cooled
beryllium walls.  The most recent refinement \cite{Ma93} combined
oil-coated walls with on-line detection of neutrons inelastically
scattered by the vessel wall, one of the most important contributions
to neutron loss.  Even the magnetic storage ring of Paul \etal
\  \cite{Pa89} suffered losses because
some injected neutrons were captured in unstable orbits; it is
believed that those neutrons disappeared rapidly and that decay
at later times was entirely due to $\beta$ decay.

The lifetime measurements within each category are internally consistent
but when they are considered as a single body of data the agreement
is less satisfactory though acceptable:  the confidence level is
$\sim 25 \%$.  It has been suggested \cite{Gr90} that the difference
between the average lifetimes for beam and trap measurements
could reflect the decay $n \rightarrow H \overline{\nu}_e$, which
would be unobserved in the former.  If one takes this approach,
the averages quoted in Table \ref{exptneut} yield a branching-ratio
limit for the hydrogen-atom branch of $\leq 2 \%$ at the $95 \%$
confidence limit.  Calculations \cite{So87} suggest that the
branch is actually much smaller ($4 \times 10^{-4} \%$), and in
deriving the $ft$-value for the neutron $\beta$-decay we have
neglected it.

The statistical rate function with a radiative correction for
neutron $\beta$ decay has been calculated by Wilkinson \cite{Wi82}\nl .
Correcting his result for the current best $Q_{EC}$ value, and
combining it with the overall average mean-life quoted in
Table \ref{exptneut}, we obtain

\be
ft(1 + \delta_{R}) = 1054.4 \pm 2.4 ~s .
\label{ftrneut}
\ee

\subsubsection{Determination of $\lambda$} \label{lamneut}

To date, the most effective measure of $\lambda$ has proven to be
the $\beta$ asymmetry from the decay of polarized
neutrons.  Such experiments are designed with a $\beta$ detector
on the axis of neutron polarization; then spectra are recorded for both
polarization directions to yield the asymmetry (see Eq.\,\ref{Wneut}):

\be
\frac {N^{+} - N^{-}} {N^{+} + N^{-}} = \frac{v}{c} P A \cos \theta ,
\label{NpNm}
\ee

\noindent in which $N^{\pm}$ is the electron count rate for each
direction of neutron polarization, $P$ is the magnitude of the
polarization, $\theta$ is the angle between the electron momentum and
the polarization direction, and $v$ is the electron velocity.  In
practice, $A$ must also be corrected for the effects ($\sim 1 \%$)
of nuclear recoil and weak magnetism \cite{Wi82} before it can be used
to extract a value for $\lambda$ via Eq.\,\ref{aAB}.

The results of five such experiments are listed in Table \ref{exptneut}.
Three
\cite{Kr75,Er79,Er91}
required electrons
to be observed
in coincidence with protons in order to reduce
background events, while
the other two obviated the need for proton detection by relying in one
case \cite{Bo86} on the large
volume and high efficiency of the in-line PERKEO spectrometer
and in the other \cite{Sc95} on a time-projection chamber operated in
coincidence with plastic scintillators to determine the
electrons' track and energy, respectively.
Since the mid 1980's, the
availability of high-flux beams of polarized cold neutrons has
helped improve the precision of such measurements.

Only one experiment to determine the $e$-$\overline{\nu}$
angular-correlation
coefficient, $a$, has achieved sufficient precision to contribute
to the determination of $\lambda$ (see Table \ref{exptneut}).
Stratowa \etal \,\cite{St78} measured the energy spectrum of protons
recoiling down a tangential reactor beam tube from neutrons decaying
in the centre of the reactor.  The shape of this spectrum
characterizes the correlation between the unobserved decay
products: the electron and the antineutrino.

The five measurements of $\lambda$ based on the observed
$\beta$-asymmetry are consistent with one another, and with the
measurement from the $e$-$\overline{\nu}
$ correlation.  The resultant
world average is given in Table \ref{exptneut}.

\subsubsection{Implications for the Standard Model} \label{ImpSM}

We are now in a position to extract experimental
values for the vector and axial-vector coupling constants from
neutron decay, compare the former with the result from $0^{+}
\rightarrow 0^{+}$ nuclear beta decays, and examine the implications
for the unitarity test of the CKM matrix (see sect.\,\ref{Impsm}).
The current experimental status from neutron decay can be summarized in two
expressions,

\bea
\lambda = \GA^{\prime} / \GV^{\prime} & = & -1.2599 \pm 0.0025
\nonumber  \\
(\GV^{\prime 2} + 3 \GA^{\prime 2}) /(\hbar c)^6 & = & (7.701 \pm 0.018)
\times 10^{-10} ~~{\rm GeV}^{-4},
\label{GAGV}
\eea

\noindent where the latter follows from combining Eqs.\,\ref{ftneut}
and \ref{ftrneut}.  The solution of these equations, shown
graphically in figure 2, gives the following results for the
effective coupling constants:

\bea
\GA^{\prime} /(\hbar c)^3 & = & - (1.4566 \pm 0.0018) \times 10^{-5}
{}~~ {\rm GeV}^{-2}
\label{gGA}  \\
\GV^{\prime} /(\hbar c)^3 & = &  (1.1561 \pm 0.0023) \times 10^{-5}
{}~~ {\rm GeV}^{-2}  ~~~~ [{\rm neutron}] .
\label{GVneut}
\eea

\begin{figure}[t]
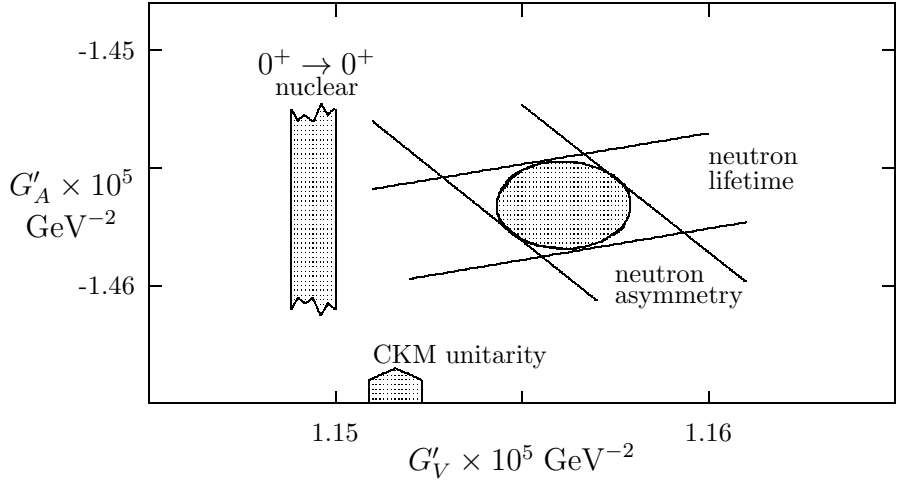

\begin{center}
\setlength{\unitlength}{0.240900pt}
\ifx\plotpoint\undefined\newsavebox{\plotpoint}\fi
\sbox{\plotpoint}{\rule[-0.175pt]{0.350pt}{0.350pt}}%

\vskip 1mm
\caption{Allowed regions of $G_A ^{\prime}$
and
$G_V^{\prime} $ from neutron decay, nuclear
superallowed decays and CKM unitarity  \label{fig2}}
\end{center}
\end{figure}

In sect.\,\ref{SFT} we described the study of $0^{+} \rightarrow
0^{+}$ $\beta$-decay transitions in nine nuclei.  The results,
shown in Table \ref{Exres}, led to an average $\overline{\F t}$ value which,
with the help of Eqs.\,\ref{ftF} and \ref{scrptft},
becomes:

\be
\GV^{\prime} /(\hbar c)^3  =   (1.1494 \pm 0.0006) \times 10^{-5}
{}~~ {\rm GeV}^{-2}  ~~~~ [0^{+} \rightarrow 0^{+}] ,
\label{GVnucl}
\ee

\noindent where the quoted error bar includes provision for a
``systematic" uncertainty in calculating the charge corrections
(see sects. 2.2.3.1 and \ref{Impsm}).  The neutron
result is considerably less precise but, even so, it is in
serious disagreement with nuclear decays.  The discrepancy is
not understood.  The analysis of the nuclear results suffers from
uncertainties in the effects of charge dependence and nuclear
structure; the neutron work is faced with more formidable
experimental challenges.  There is no reason to suppose, at this time,
that either result is in doubt beyond its quoted uncertainty.

Following the approach taken in sect.\,\ref{Impsm}, the neutron result
for the vector coupling constant can be used to test the unitarity
of the CKM matrix, yielding

\be
\mids V_{ud} \mids ^{2} +
\mids V_{us} \mids ^{2} +
\mids V_{ub} \mids ^{2} = 1.0077 \pm 0.0040,
\label{unitn}
\ee

\noindent which differs from unitarity at a confidence level of
$\sim 99 \%$, but in the opposite sense to the result from
nuclear $\beta$ decay.  In the absence of any viable explanation
within the Standard Model for the discrepancy between the neutron and
nuclear results, nor any reason to prefer one over the other, we
can do no better than take the weighted average of both results
for $\GV^{\prime}$, suitably inflating the resultant uncertainty.
This leads to

\bea
\GV^{\prime}/(\hbar c)^3   & = &  (1.1498 \pm 0.0015) \times 10^{-5}
{}~~ {\rm GeV}^{-2}  ~~~~ [{\rm average}]
\nonumber  \\
\mids V_{ud} \mids & = & 0.9741 \pm 0.0017  ~~~~ [{\rm average}]
\label{GVavg}
\eea

\noindent and

\be
\mids V_{ud} \mids ^{2} +
\mids V_{us} \mids ^{2} +
\mids V_{ub} \mids ^{2}  =  0.9978 \pm 0.0037 ~~~~ [{\rm average}],
\label{unitavg}
\ee

\noindent which agrees with unitarity.

\subsection{Muon Capture in Hydrogen} \label{RMCH}

The induced pseudoscalar coupling constant, $\gP$ of Eq.\,\ref{VAc},
is the most poorly determined of all the nucleon
weak-interaction coupling constants.  This is because
in semi-leptonic weak processes involving first-generation
leptons, such as beta decay or electron capture, the
pseudoscalar term gives negligible contribution to the
transition rate.  Consider the T-matrix involving
just the pseudoscalar term:

\bea
T_{fi} & = & - \frac{\GF}{\sqrt{2}} V_{ud} \, i \overline{\psi}_p
( i \gP k_\mu \gamma_5 ) \psi_n \, i \overline{\psi}_e
\gamma_\mu ( 1 + \gamma_5 ) \psi_{\overline{\nu}_e}
\nonumber \\
& = & - \frac{\GF}{\sqrt{2}} V_{ud} \, i \overline{\psi}_p
( - \frac{\gP}{2M} k_\mu \not{\! k} \gamma_5 ) \psi_n
\, i \overline{\psi}_e
\gamma_\mu ( 1 + \gamma_5 ) \psi_{\overline{\nu}_e}
\nonumber \\
& = & - \frac{\GF}{\sqrt{2}} V_{ud} \,i \overline{\psi}_p
( - \frac{\gP}{2M}  \not{\! k} \gamma_5 ) \psi_n
\, i \overline{\psi}_e
\not{\! k} ( 1 + \gamma_5 ) \psi_{\overline{\nu}_e}
\nonumber \\
& = & - \frac{\GF}{\sqrt{2}} V_{ud} \,i \overline{\psi}_p
( \frac{\gP m_e}{2M}  \not{\! k} \gamma_5 ) \psi_n
\,  \overline{\psi}_e
( 1 + \gamma_5 ) \psi_{\overline{\nu}_e} ,
\label{pssmall}
\eea

\noindent  where
$k = p_e+p_{\nu}$.
Here pseudoscalar coupling is transformed to pseudovector
coupling in line two, a step valid only for on-shell nucleons.
The factor $k_\mu$ is moved into the lepton matrix element in line
three, where application of the Dirac
equation, $\overline{\psi}_e \not{\! p}_e = i m_e \overline{\psi}_e$,
shows immediately that the pseudoscalar term is of size
$\gP m_e /(2M)$ and is very small.  However, for
second-generation leptons participating in such processes as muon
capture, the factor becomes $\gP m_\mu /(2M)$ and is much larger.

Consequently, experiments to determine $\gP$ are based on muon capture
in hydrogen.  The two main processes that contribute to the
disappearance of a negative muon stopped in pure hydrogen are
muon decay,
$ \mu^{-} \rightarrow e^{-} + \overline{\nu}_e + \nu_\mu $,
and nuclear capture,
$ \mu^{-} + p \rightarrow n + \nu_\mu $.
In the latter process, a 5.2 MeV neutron is emitted.
Because the weak interaction is strongly spin dependent, an
accurate specification of the initial muon-proton state
is vital to the interpretation of the experimental results.
The history of a negative muon, stopping in a
hydrogen target, is as follows.
At low velocity, the muon is captured by a proton and cascades to
the $1S$ state in less than $10^{-9}$s; this neutral $(\mu p)$
atomic system is formed in a statistical mixture of triplet
and singlet states.
In general, the triplet state converts to the singlet state
with a rate depending upon the hydrogen density.  Even in gaseous
hydrogen, for pressures above five atmospheres, the conversion
rate is large compared to the natural decay rate, so nuclear
capture occurs predominantly from the singlet state.

In liquid hydrogen,
the singlet $(\mu p)$ system further converts rapidly into
a $(p \mu p)^{+}$ molecular ion if the muon
has not decayed first. Consequently, it is from
the stable molecular ion that the muon is most likely to
be captured or to decay.  Most experiments are able to measure the
time of each capture and thus the configuration from which
capture takes place is known.  Usually only capture events
that have originated in a $(p \mu p)$ system are included
in the analysis.  The $(p \mu p)^{+}$ molecular ion is almost
invariably formed in the ortho state  for which the capture
rate, $\Lambda_{om}$, is given by $2 \gamma_{o} ( \sfrac{3}{4}
\Lambda_s + \sfrac{1}{4} \Lambda_t )$, where $\Lambda$ refers
the the capture in the triplet and singlet atoms, and $2 \gamma_{o}$
is a wave function overlap factor of the muon
with the proton, calculated to be
$2 \gamma_{o} = 1.01 \pm 0.01$.  There is a
small probability that the ortho state will decay to the
lower-energy para-molecular state before the capture occurs.
The capture rate from the para state is given by $\Lambda_{pm} =
2 \gamma_{p} ( \sfrac{1}{4} \Lambda_s + \sfrac{3}{4} \Lambda_t ) $,
with $2 \gamma_{p} = 1.15 \pm 0.01$, and is smaller than
the ortho rate because $\Lambda_s \gg \Lambda_t$.  Nevertheless,
a correction to the measured capture rate has to be applied
for this ortho-to-para transition and this correction
significantly influences the extraction of the $\gP$
coupling constant.

Early experiments \cite{Bl62,Ro63,Al69,By74} obtained the
capture rate by observing the outgoing 5.2 MeV neutrons.
The principal limitation was the calibration of the
absolute efficiency of the neutron detectors.
A more recent Saclay experiment
\cite{Ba81a} obtained the capture rate by comparing the
lifetime $\tau_{\mu^{-}}$ of negative muons stopped in
liquid hydrogen with the lifetime $\tau_{\mu^{+}}$ of
positive muons.  Both $\tau_{\mu^{-}}$ and $\tau_{\mu^{+}}$
were measured by detection of the delayed muon-decay electrons.
The capture rate was then obtained from the difference
in these two lifetimes.
In all cases,
the hydrogen in the target had to be highly
purified and isotopically pure.
If small amounts of deuterium or other atoms
are present, then an irreversible transfer to the deuteron
(or heavier atom) is likely. Use of pure hydrogen, therefore,
is very important to minimize this loss.
The Columbia and Saclay
experiments \cite{Bl62,Ro63,Ba81a} used liquid hydrogen
targets; the capture rates were corrected
for the ortho-to-para transition, which was explicitly
measured in a subsequent Saclay experiment \cite{Ba81b}\nl .
The Dubna and CERN experiments \cite{Al69,By74} used gas
targets.

The resulting values of the dimensionless quantity $m_\mu \gP /\gA $
from these five experiments are: $4.8 \pm 6.3$ (ref.\cite{Bl62}),
$8.7 \pm 3.7$ (ref.\cite{Ro63}), $8.2 \pm 3.1$ (ref.\cite{Al69}),
$6.3 \pm 4.7$ (ref.\cite{By74}) and
$5.6 \pm 2.6$ (refs.\cite{Ba81a,Ba81b}).
The average is $6.9 \pm 1.5$ in excellent agreement with
the PCAC value of 6.7 from Eq.\,\ref{gpgap}, evaluated
with $k^2 = 0.88 m_{\mu}^2$, which is appropriate for hydrogen. Were
it not for the ortho-to-para correction, the value from
the Saclay experiment would have been $10.6 \pm 1.6$
rather than the quoted $5.6 \pm 2.6$!  A new measurement
of the ortho-to-para transition rate with
improved accuracy would be very desirable.

Despite the good agreement between the ordinary muon capture (OMC)
experiments and the PCAC prediction, it should be noted that the
OMC rate is insensitive to the exact value of $\gP$ since
an individual measurement, based on a 4\% OMC rate determination
at Saclay \cite{Ba81a}\nl , gives a 42\% uncertainty for $\gP$.
This is because, in liquid hydrogen, 80\% of the capture takes place
from the ortho-molecular state, which is predominantly the atomic
singlet state.  In contrast, in radiative muon capture (RMC),
in which OMC is accompanied by a bremsstrahlung photon, the
photon carries away one unit of angular momentum and the
capture is mainly from the triplet atomic state.  The triplet
capture rate is 15 times more sensitive than the singlet rate
to the value of $\gP$.  Beder and Fearing \cite{BF87} have
estimated that an 8\% measurement of the RMC rate could provide a
10\% value for $\gP$.

The experimental problems in measuring RMC in hydrogen are
formidable.  The partial RMC branching ratio (for $E_\gamma
\geq 57$ MeV) is only $1.6 \times 10^{-8}$ and the bremsstrah\-lung
from radiative muon decay or from the Michel electrons is
nearly five orders of magnitude larger.  Consequently, the
apparatus requires good energy resolution and an accurate
knowledge of the high-energy tail of the response function
in order to guarantee that the RMC signal region is not
contaminated by events from the bremsstrahlung region.
The low branching ratio also
requires an extremely high purity level in the hydrogen liquid
itself; otherwise the $\mu^{-}$ will be lost from the proton
before the capture can take place.  Impurity levels of
$\leq 10^{-9}$ for all elements other than hydrogen are
necessary since the capture probability increases as $Z^4$.
An experiment \cite{Ha93} presently underway at TRIUMF hopes
to achieve an 8\% measurement of the RMC capture rate.

\section{Axial-vector Interaction in Nuclei} \label{AVInuc}
\subsection{Quenching of the axial-vector current in nuclei}  \label{qavcn}

Since the vector current is a conserved current, its normalization is
unchanged from one environment to another: $\gV(0)$ is unity at
the quark level, at the single-nucleon level and in finite nuclei.
In contrast to this, the axial current is not conserved and its
normalization therefore becomes environment dependent: $\gA(0)$ is
unity at the quark level, is greater than unity (of order 1.26) at
the single-nucleon level and reduced relative to 1.26 in finite
nuclei.  The degree of reduction, however, is nucleus dependent.
Since pions cause the failure of the axial current to be conserved,
as evidenced in the PCAC hypothesis (see Eq.\,\ref{delpi}),
and since nucleons in nuclei are interacting by the exchange of pions,
it is not surprising that the axial current in nuclei is
modified.  One mechanism frequently discussed in this context
involves the $\Delta(1232, s=3/2, t=3/2)$-isobar, which can be
viewed as an internal spin-isospin-flip excitation of a single
nucleon.  A nucleon, for example, in emitting a pion can be
raised to its $\Delta$-excited state.  The $\Delta$ is subsequently
de-excited back to the nucleon through an interaction with
the axial current, while the pion is absorbed by a second nucleon.
This and related processes are calculable in perturbation theory
but the results are strongly model dependent, relying on
assumptions made concerning the isobar's interactions.  It is
probably worth repeating here the cautionary remarks of
Lipkin and Lee \cite{LL87}: ``There has been considerable debate
of the possibility that isobars may be present in nuclei.  However, it
is not clear exactly what this means.  The isobar is a highly
unstable particle with a natural width much larger than the
spacing between nuclear levels.  It is, therefore, unclear whether
an isobar present in a nucleus can be considered as an elementary
fermion in some approximation, as a three-quark composite, or as
a pion-nucleon resonance.  Furthermore, an isobar present in a
normal nuclear ground state is very far off shell, and it is
not clear how to extrapolate the properties of such a complicated
object far off shell.  All this confuses the issue and makes it
difficult to test experimentally for the presence of such isobars."

The idea that the axial-vector coupling constant might be
renormalized for nucleons embedded in a nucleus was first discussed
by Ericson \etal \,\cite{EFT73}\nl .  They pointed out that
the pion field within the nucleus might be suppressed at small
momenta because of the strong short-range repulsion between
nucleons in the medium.  Thus, an axial-vector $\beta$-decay in
nuclei would, using the arguments of PCAC, be similarly
suppressed.  This was worked out in more detail by Rho \cite{Rho74}
and Ohta and Wakamatsu \cite{OW74}\nl , who showed that the
$\Delta$-isobar would be the main mediator in effecting such a
suppression.  Experimental data, as we will shortly discuss,
also give evidence for this suppression, but whether or not
it is due to the influences of $\Delta$-isobars is not so
clear.  There are other many-body corrections to be applied
to the nuclear-structure calculations, which arise from
short-range central and tensor correlations between nucleons
and which also lead to a suppression of axial-vector rates
\cite{SIA78,BH82}\nl .

One of the primary sources of experimental information on
axial-vector transitions in nuclei comes from nuclear beta decay.
Referring to the axial current given in Eq.\,\ref{VAct}, we
ignore second-class currents and note that the pseudoscalar
coupling term gives a negligible contribution.  Thus, only the
$i \overline{\psi} \gA \gamma_\mu \gamma_5 \sfrac{1}{2}
\tau^a \psi $ term is operable.  In nuclear physics it is
sufficiently accurate to consider a nonrelativistic limit
of this current in which only leading terms in $p/M$,
where $p$ is the nucleon momentum and $M$ its mass, are
retained and four-component Dirac spinors, $\psi$, are
replaced by two-component Pauli spinors, $\chi$.  The leading
term arises when the Lorentz index, $\mu$, is space-like:

\be
{\bf A}^h = - \gA \chi^\dagger \mbox{\boldmath $\sigma$}
\sfrac{1}{2} \tau^a \chi ,
\label{GTop}
\ee

\noindent where {\boldmath $\sigma$} is the Pauli spin operator
and $\tau^a=\tau^1 \pm i\tau^2$, with the upper sign for $\beta^{-}$
transitions and the lower sign for $\beta^{+}$ transitions.
The quantity between $\chi^\dagger \chi$ is taken as a
one-body operator, the matrix elements of which are evaluated in the
many-body nucleus with shell-model wavefunctions.  The
experimentally determined quantity, the $ft$-value,
for a pure axial-vector transition is
related to this matrix element via ({\it cf.} Eq.\,\ref{ftneut})

\be
ft (1 + \delta_R )
= \frac{6146 \pm 6}{\gA^2 B(GT)}~~ {\rm s},
\label{ftGT}
\ee

\noindent where

\bea
B(GT) & = & \langle \MA \rangle^2
\nonumber \\
\langle \MA \rangle  & = & \frac{\hat{J_f}}{\sqrt{2} \hat{J_i}}
\langle T_i \, T_{zi} \, 1 \, \mp 1 \mid T_f \, T_{zf} \rangle
\langle J_f \, T_f \tbar \mbox{\boldmath $\sigma \tau$}
\tbar J_i T_i \rangle .
\label{MGT}
\eea

\noindent Here the spin, isospin and $z$-projection of isospin
are denoted $J_i, T_i$ and $T_{zi}$ for the initial state and
$J_f, T_f$ and $T_{zf}$ for the final state.  The triple-barred
matrix element is a reduced matrix element in both spin and
isospin spaces in the notation of Brink and Satchler \cite{BS68}
and $\hat{J} = (2J+1)^{1/2}$.  The constant, $6146 \pm 6$ s,
in Eq.\,\ref{ftGT} is given by $2\,\overline{\F t}$, where
$\overline{\F t}$ is the average
electromagnetically-corrected $ft$ value for the superallowed
Fermi $0^{+} \rightarrow 0^{+}$ beta transitions (see Eq.\,\ref{Ftbar}).

The operator {\boldmath $\sigma \tau $}, known as the
Gamow-Teller (GT) operator, has no dependence on orbital
angular momentum and hence has no non-zero matrix elements
between states of different orbital symmetry.  In particular,
in a shell-model calculation carried out in a complete basis
of harmonic oscillator states (\eg \ a $(0p_{3/2},0p_{1/2})$-basis
for $p$-shell nuclei or a $(0d_{5/2},1s_{1/2},0d_{3/2})$-basis
for $sd$-shell nuclei) the GT-operator acting on
any member of the basis connects only to other states within
the basis and not to any state with occupancy in a different
major oscillator shell.  Complete untruncated shell-model
calculations within a single major oscillator shell are
called $0 \hbar \omega $ calculations.  A working definition
for the term {\em quenching} is as follows:  for beta decays
populating well-defined, isolated states in the daughter
nucleus, the ratio of the experimentally measured rate
to the calculated rate in a full $0 \hbar \omega $
calculation is called the {\em quenching factor}.  An average
quenching factor implies an average over many transitions.  The definition
is model dependent in the sense that the shell-model
calculation is dependent on the choice of residual interaction,
but for an average value it is hoped that this dependence
is not severe. \footnote{ However, see some contrary remarks
by Brown \cite{Br92} in a discussion of the $\beta$-decay
of $^{37}$Ca.}  Equivalently, one can discuss a quenching in
terms of a renormalization of the axial-vector coupling
constant.  We define

\be
\gAeff = \gA + \delta \gA ,
\label{gAeff}
\ee

\noindent where $\gA$ is the free-nucleon value of 1.26 and
$\delta \gA $ the correction to it; thus, the quenching factor
is $(\gAeff / \gA )^2 $.

The first systematic attempt to determine an average quenching
factor, $q$, was by Wilkinson \cite{Wi74} from a study of
$\beta$-transitions in the $p$-shell and the lower $sd$-shell
nuclei ($A$=17-21), in which he obtained $q = 0.80 \pm 0.06$.  A more
comprehensive analysis by Brown and Wildenthal \cite{BW85}
using full $0 \hbar \omega $ calculations throughout the
$sd$-shell for 189 data, for which the experimental error
on the deduced GT matrix element was 10$\%$ or less, found an
average quenching factor of $q = 0.63 \pm 0.01$;  the quoted
error here is limited to the statistical error on the fit.  Significant
quenching seems to be evident.
One configuration has a special role to play here, for example in the
ground states of $A=15,17,39$ and $41$ nuclei: it is
a closed-LS-shell-plus
(or minus)-one nucleon. In these cases,
the full $0 \hbar \omega $ calculation
involves just a single-particle matrix element, so
microscopic calculations can
seek to evaluate corrections to the single-particle matrix
element and identify the origin of the quenching phenomenon.
The Tokyo \cite{SIA78,ASBH87} and Chalk River \cite{TK79,To87}
groups have attempted this with similar results.

Generally, the quenching stems either from the inadequacy
of the single-particle description of the nuclear state
or from the inadequacy of the one-body GT operator.  Calculations
of the first effect, frequently called core polarization, estimate
through perturbation theory the admixtures of $2p \hyphen 1h$ and
$3p \hyphen 2h$
configurations in the basically single-particle wave function
and evaluate the impact the admixtures have on the calculated
matrix element.  At LS closed shells there is no contribution to the
core polarization in first order because
the one-body GT operator has no spatial dependence and thus cannot excite
a $1p \hyphen 1h$ state from a closed LS shell.  Therefore, calculations
must be taken to second order in perturbation theory.
This leads to a time-consuming calculation,
as there is no selection rule to limit the intermediate-state
summation and the convergence is slow.  This is particularly
true with tensor forces in the residual interaction as
first stressed by Shimizu \etal \, \cite{SIA78}\nl .  This propensity
of the tensor force to couple strongly to high-lying states
has been called `tensor correlations' and the phenomenon
leads to a reduction in the GT matrix element.

The other inadequacy concerns the use of one-body operators.
Corrections arise because nucleons in nuclei interact through
the exchange of mesons, and this exchange can be perturbed
by the action of the weak axial current.  Since this perturbation
requires at least two nucleons to be involved, corrections
of this type lead to two-body GT operators.  The possible
pion-exchange processes can be classified into two types:
Born graphs, involving only nucleons and pions, and non-Born
graphs in which the axial current either excites a nucleon to its
isobar $\Delta $ state, or converts
the pion into a heavy meson, such as the $\rho $. The important
point is that for axial currents the pion Born graphs are
identically zero \cite{CR71}\nl .  For the non-Born graphs, we
can distinguish between diagrams involving nucleon
excitations -- referring to these as isobar graphs -- and
those in which a pion is converted to a $\rho $ meson -- referring
to these as $\rho \pi $ or meson-exchange-current (MEC)
graphs.  In general, the MEC graphs give a small contribution
and are not an important ingredient in the quenching of the
GT matrix element.

There are two further points to consider.  The core-polarization
calculation corrects the matrix element of a one-body operator
evaluated in the closed-shell-plus-one configuration for the
presence of $2p \hyphen 1h$ and $3p \hyphen 2h$ admixtures in the
single-particle
wave function.  The perturbation calculation is carried out
to second order in the residual interaction (or to the fourth
power in the meson-nucleon coupling constants).  It is
logical, therefore, that the matrix elements of two-body
operators should likewise be corrected for $2p \hyphen 1h$ and
$3p \hyphen 2h$ admixtures.  Since the two-body operator itself involves
the meson-nucleon coupling constants to the second power, it
is sufficient to estimate this correction to first order in the
residual interaction.  These terms have been called `crossing
terms' in the work of the Tokyo group \cite{ASBH87}\nl .  The
second point concerns the one-body GT operator, which is
obtained as a leading term in the nonrelativistic reduction
of a relativistic axial-vector current.  There is a few per cent
correction coming from the next-order terms in the
reduction.

In summary, then, corrections to lowest-order shell-model estimates
of the GT matrix element come from: core polarization, isobar
currents, MEC currents, crossing terms and relativistic corrections.
All these ingredients for closed-LS-shell-plus-one nuclei have
been calculated by Arima \etal \, \cite{ASBH87} and by Towner
and Khanna \cite{TK79}\nl .  We quote in Table \ref{dMGT} some results from
a review by Towner \cite{To87}\nl , in which more details can be found.

{\small
\begin{table}[t]
\begin {center}
\caption {Corrections to the ground-state diagonal and spin-flip
off-diagonal matrix elements, $\delta \langle M_A \rangle $,
expressed as a percentage
of the single-particle value, $\langle M_A \rangle _{sp}$,
for closed-shell-plus-one
configurations at $A=17$ and $A=41$.  Quenching factors, $q$, are also
given.
\label{dMGT}  }
\vskip 1mm
\begin{tabular} {lrrrr}
\hline \\[-3mm]
&  \multicolumn{4}{c}{$\delta \langle M_A \rangle /
\langle M_A \rangle _{sp}~{\rm  in}~ \%$}  \\
\cline{2-5}
& $d_{5/2} \rightarrow d_{5/2}$
& $d_{5/2} \rightarrow d_{3/2}$
& $f_{7/2} \rightarrow f_{7/2}$
& $f_{7/2} \rightarrow f_{5/2}$   \\[3mm]
\hline  \\[-3mm]
Core polarization & -9.0~~~~ & -11.3~~~~ & -11.8~~~~ & -14.1~~~~ \\
Isobars           & -0.9~~~~ & -5.9~~~~ &  -2.1~~~~ &  -6.2~~~~ \\
MEC               & -1.2~~~~ &   0.3~~~~ &  -1.4~~~~ &   0.2~~~~ \\
Crossing terms    &  3.0~~~~ &   2.3~~~~ &   3.4~~~~ &   2.6~~~~ \\
Relativistic corr & -2.1~~~~ &  -1.2~~~~ &  -2.3~~~~ &  -1.3~~~~ \\[3mm]
Total             & -10.2~~~~ & -15.8~~~~ & -14.2~~~~ & -18.8~~~~ \\[3mm]
Quenching factor  &  0.81~~~~ &  0.71~~~~ & 0.74~~~~ &  0.66~~~~  \\
Experiment, $q$   & 0.74 $\pm$ 0.01 & & 0.54 $\pm$ 0.01 & \\
\hline
\end{tabular}
\end{center}
\end{table}
}

Note that the calculated quenching factors, although in
the expected range of 0.66 to 0.81, are higher than the experimental
values for the diagonal matrix elements and, for
the $0d$ orbits, are also higher than the average quenching factor of
$0.63 \pm 0.01$ determined by Brown and Wildenthal \cite{BW85}\nl .
Furthermore, there is more quenching evident in spin-flip
than in diagonal matrix elements,
with approximately two-thirds of the quenching coming from
core polarization (notably tensor correlations) and one-third
from isobars.

\subsection{$(p,n)$ and $(n,p)$ reactions}  \label{pnnp}

The reactions $(p,n)$ and $(n,p)$ are comparable to $\beta^{-}$
and $\beta^{+}$ decay, respectively, but are not limited by the
energetics of a radioactive decay.  In studying these reactions,
it is useful to define a Gamow-Teller strength function by the
relation:

\be
S(\omega ) = \sum_f \langle \MA (i \rightarrow f) \rangle ^2
\delta (E_f-E_i-\omega ) ,
\label{Sw}
\ee

\noindent where $\omega$ is excitation energy,
$\delta$ is a resolution function,
and $i$ and $f$ represent the initial and final nuclear eigenstates
of energies $E_i$ and $E_f$.  The major peaks in $S(\omega )$ are
identified with a collective state: the Gamow-Teller giant resonance
(GTR).  Unfortunately, beta decay has access to nuclear states in
a very limited energy window and misses the strongest states
in the strength function.  Therefore, to map out the complete
strength function one needs probes that allow the independent
variation of both energy transfer and momentum transfer to the
target.  This is possible with hadronic probes such as the
$(p,n)$ and $(n,p)$ charge-exchange reactions.

The Gamow-Teller resonance has quantum numbers $J^\pi=1^{+}, L=0,
S=1, T=1$ indicating it is populated through spin- and
isospin-excitations.  In the $(p,n)$ reaction this mode is reached
via the $V_{\sigma \tau }
\mbox{\boldmath $\sigma $}_p \cdot
\mbox{\boldmath $\sigma $}_j
\mbox{\boldmath $\tau $}_p \cdot
\mbox{\boldmath $\tau $}_j $
component of the interaction between the projectile and target
nucleons.  The labels
$p$ and $j$ refer to the projectile and the struck nucleon
respectively.  With respect to the target, the $(p,n)$ probe has
a similar spin-isospin operator structure as the GT operator
of $\beta$-decay.  However, the $(p,n)$ reaction simulates the
$\beta$-decay conditions only if the cross-section is
measured at very small momentum transfer.
In the $(p,n)$ reaction, this condition
can be met only for $0^{\circ }$ scattering and at high bombarding
energies.  As there is no orbital excitation in a GT transition,
the angular distribution for the $(p,n)$ reaction to the GTR will
have a characteristic $L=0$ shape, with the maximum cross-section
at zero degrees.

The GTR occurs at a higher excitation energy than the isobaric
analogue of the target ground state (IAS), which carries the
quantum numbers $J^\pi = 0^{+}, L=0, S=0, T=1$.  The $(p,n)$
reaction excites the IAS through the force component $V_{\tau}
\mbox{\boldmath $\tau $}_p \cdot
\mbox{\boldmath $\tau $}_j $,
which requires no spin flip $(S=0)$.  While the IAS is the
dominant peak in the spectrum at low incident energies
$(E_p \leq 100$ MeV), it is completely swamped \cite{Os92} by the large
GT cross-section at $E_p = 200$ MeV.  The large difference
in the excitation strengths of the IAS and GTR as a function
of the incident projectile energy is due to the strong
energy dependence of the $V_{\tau}$-term in the
projectile-target-nucleon interaction.  The force strength
$V_{\tau}$ is greatly reduced with increasing projectile
energy, while the force strength $V_{\sigma \tau}$ is nearly
independent of incident energy.

The $(p,n)$ reaction is a powerful spectroscopic tool for
measuring the GT strength functions of nuclei.  The zero-degree
$(p,n)$ cross-section can be calculated in a distorted-wave theory
and compared to experiment to extract the spectroscopic
information: namely $B(GT)$ of Eq.\,\ref{MGT}.  However, there
is sufficient uncertainty in the distorted-wave theory,
especially with regard to its absolute normalization, that it
is usual to calibrate the theory by measuring the $(p,n)$
cross-sections of specific transitions for which the $B(GT)$
values are already known from $\beta $-decay \cite{Ta87}\nl .
Once this is done, then the $(p,n)$ reaction can be used to
determine the GT matrix elements of states that cannot be
populated by $\beta $-decay.  In particular, it is possible
to map out the whole GT response function in the zero-degree
spectrum.  There are, however, other states with multipolarities
$L > 0$, which also contribute to the $0^{\circ}$ $(p,n)$
cross-section.  In general, these states receive their
maximum cross-sections at larger scattering angles, but their
angular distributions extend forward to $0^{\circ}$.  These
states lead to a background contribution that has to be subtracted
from the data.
A discussion of the theoretical techniques for doing so can
be found in the papers of Osterfeld \cite{Os82,OCS85}\nl .

The breakthrough in the use of the $(p,n)$ reaction to study
GT resonances came in 1980 from experiments at the Indiana
University Cyclotron Facility.  These experiments
\cite{Ba80,Go80,Ho80,Ga81,An80} demonstrated the existence
of very collective spin-isospin modes in nuclei.  An excellent
review of these and the many subsequent experiments is to be
found in ref.\,\cite{Os92}\nl .  The spectra of nuclei with neutron
excess are seen to be dominated by one prominent peak, which
is interpreted as the giant Gamow-Teller resonance.  This
collective mode was already predicted by Ikeda, Fujii and
Fujita \cite{IFF63}\nl , who inferred the existence of the GTR
from the absence of spin-isospin strength at low excitation
energies.  Their argument is based on a sum rule.  Let us
integrate the strength function of Eq.\,\ref{Sw}
to define a sum:

\be
S = \int S(\omega ) d\omega = \sum_f \langle \MA (i \rightarrow f)
\rangle ^2 .
\label{sumr}
\ee

\noindent Then, from the commutator algebra of isospin,
$ [\tau^x,\tau^y ] = 2i \tau^z $, we obtain for the difference
in the sum rule for $\beta^{-}$ and $\beta^{+}$ emissions from
a given nucleus the model-independent result \cite{IFF63}:

\be
S(\beta^{-}) - S(\beta^{+}) = 3(N-Z) .
\label{sumdif}
\ee

\noindent That is, the difference in the sums for the $(p,n)$
and $(n,p)$ reactions on a particular target is quite independent
of nuclear structure.  In heavy nuclei
with large neutron excess, $S(\beta^{+}) \simeq 0$, because
all states into which a proton could be transferred within the
same major shell are already fully occupied: the GT transition
is Pauli blocked.  Setting $S(\beta^{+}) = 0$ then gives a lower
limit for the $\beta^{-}$ sum: \viz

\be
S(\beta^{-}) \geq 3(N-Z) .
\label{sumlim}
\ee

\noindent At intermediate energies $(E_p \sim 100-200$ MeV), the
selective excitation of the GT giant resonance by the $(p,n)$
reaction has been observed in almost the entire periodic
table.  The striking result \cite{Ga81} is that experimentally only
$(60 \pm 10) \%$ of the expected sum-rule limit, $3(N-Z)$, can be
located in the excitation region where the major GT peaks occur.
This means that at least $40 \%$ of the GT strength
is unobserved in the spectra.  As described in the previous section
for beta decay,
this missing GT strength has
been interpreted as evidence for isobar admixtures in the nucleonic
wave function \cite{EFT73,Rho74,BM81} or, alternatively, as indicating
\cite{BH82} that high-lying
$2p \hyphen 2h$ states mix with the low-lying $1p \hyphen 1h$ GT
states and shift
GT strength above and below the GTR.  If this latter
mechanism is the right one, then the so-called missing GT strength
would actually be located in the background.
The detection of this strength
would not be an easy task
because the spectrum on the high-energy side of the GTR
is continuous and structureless, making it difficult to
extract information.  Calculations by Smith and Wambach
\cite{SW87} indicate that significant strength from $2p \hyphen 2h$ mixing
resides in the tails of the resonances, suggesting that
the quenching due to $\Delta$-isobars is
$10-20 \%$.  This is a further indication that quenching
due to $\Delta$'s is small.

In conclusion we note two recent experiments that have
elicited an exchange of comments in the literature:

\bi

\item A group \cite{Ve89} at TRIUMF has measured angular distributions
of the $^{54}$Fe$(p,n)^{54}$Co and $^{54}$Fe$(n,p)^{54}$Mn
cross-sections at incident beam energies of 300 MeV to test
the GT sum rule, Eq.\,\ref{sumdif}, for a case in which the GT
strength is strong in both channels.  The $(p,n)$
cross-section
at $0^{\circ}$ for $ L=0$
was summed up to an excitation energy of $E_x = 15$
MeV.  The strength above this excitation energy was not included
because of increasing uncertainties in removing dipole
and higher multipole strength at higher energies.  The integrated
strength below $E_x = 15$ MeV is $S(\beta^{-}) = 7.5 \pm 1.2$
in agreement with earlier results \cite{Ra83} at 160 MeV of
$S(\beta^{-}) = 7.8 \pm 1.9$.  The $0^{\circ}$ $(n,p)$
cross-section was summed up to $E_x = 10$ MeV, the same criteria
being used to decide the upper limit of the integration.  The
integrated strength is $S(\beta^{+}) = 3.1 \pm 0.6$.  A more
recent experiment \cite{Ro93} at a lower energy of 97 MeV
finds $3.5 \pm 0.3 \pm 0.4$, the first uncertainty being due to
statistics and multipole decomposition, the second due to
uncertainties in normalization.  The sum rule difference
in the TRIUMF data thus yields

\be
S(\beta^{-}) - S(\beta^{+}) = (7.5 \pm 1.2) - (3.1 \pm 0.6)
= 4.4 \pm 1.3,
\label{sumtest}
\ee

\noindent to be compared with the model-independent value of
$3(N-Z) = 6$ in this case.  The result is consistent
with the quenching factor of 0.6 seen throughout the periodic
table in $(p,n)$ reactions where the $(n,p)$ strength is
strongly Pauli blocked.  However, Goodman \etal \,\cite{Go90} have commented
that this work does not test the model-independent
sum rule because both measured strength functions are incomplete,
there being additional strength undetected either because it lies
outside the measured energy domain or because it is hidden
in the background.

\item A recent beta-decay experiment by Garc\'{\i}a \etal \,
\cite{Ga91} has extended the measured GT strength function in the
$\beta^{+}$ decay of $^{37}$Ca up to 8 MeV excitation in the
daughter nucleus.  Delayed protons were detected in coincidence with
gamma rays to identify proton decays of $^{37}$K daughter levels
to excited states in $^{36}$Ar.  Some 43 previously undetected
proton groups were observed yielding an integrated $B(GT)$
twice as large as that observed previously.  In terms of the
sum-rule limit, the experiment observed a summed GT strength
of 1.95 compared to the sum-rule requirement of at least 9 units.
Because $^{37}$Ca is the isospin mirror of $^{37}$Cl, it is
possible to compare the strength seen in this decay with that
measured in the $^{37}$Cl$ (p,n) ^{37}$Ar reaction studied by
Rapaport \etal \, \cite{Ra81}\nl .  Garc\'{\i}a \etal \, made
such a comparison and concluded that the $\beta^{+}$ decay
contains $50 \%$ more GT strength in this energy region than
the $(p,n)$ experiment.
\par
This conclusion led to a flurry of correspondence
\cite{Au92,Ra92}\nl .
Two issues were raised.  First, in the analysis of the $(p,n)$ data,
insufficient Gamow-Teller strength was assigned near
the isobaric analogue state, which in the reaction is
populated by both Fermi and GT operators.  Second, the beta-decay
data for the low-lying states have to take account of both
delayed-gamma and delayed-proton emission \cite{Ga95}\nl .
\par
The current situation is that the ratio of $B(GT)$ values determined
from beta-decay and $(p,n)$ experiments show a state-by-state
scatter with discrepancies in the $B(GT)$ values that can vary by
as much as a factor of two.  However, these states
are below 8 MeV excitation and their
$B(GT)$ values are very small.  Discrepancies of this size,
therefore, are not necessarily surprising because, for weak
transitions with $B(GT) \ll 1$, one can expect inadequacies in
the reaction model used to extract $B(GT)$ values
from forward-angle charge-exchange cross-sections.  The
importance of the $A=37$ example is that it allows a
comparison of GT strength inferred from $(p,n)$ experiments
and beta-decay experiments over a significantly larger
excitation energy range and for significantly more
transitions than any other system studied so far.
Further experiments are underway.
\par
Despite the large number of GT transitions identified in this
example, the $\beta^{+}$ experiment found only 2/9
of the sum-rule limit.  Brown \cite{Br92} points out that
if experiment and theory are compared over an energy region that
only exhausts a small fraction of the sum rule, the results are very
sensitive to
the choice of residual
interaction used in the shell-model calculations.  Some older
interactions reproduced the data better than the preferred
Wildenthal interaction \cite{Wi84}\nl . This, claims Brown, indicates
a need for further improvement in the $sd$-shell Hamiltonian.

\ei

\subsection{Radiative Muon Capture in Nuclei} \label{RMCnuc}

It was explained in sect.\,\ref{RMCH} that muon-capture data provide
the best opportunity to measure the induced pseudoscalar coupling
constant.  Data on ordinary muon capture (OMC) by the proton yield a
value consistent with the PCAC hypothesis, but the accuracy is not sufficient
to provide a demanding test of the prediction.
Radiative muon capture (RMC)
potentially provides greater sensitivity,  but the low total
capture probability in hydrogen ($\sim 10^{-3}$) combined with the
low expected branching ratio RMC/OMC ($\sim 10^{-5}$) has so far
precluded a measurement of the elementary radiative capture process,
although results from an experiment \cite{Ha93} at TRIUMF are
anticipated soon.

RMC measurements are easier for heavier nuclei, in which the muon
capture probability is greatly enhanced: roughly by a factor of $Z^4$,
where $Z$ is the charge number of the capturing nucleus.  Exclusive
nuclear RMC to a specific final nuclear state has not yet been attempted
and would pose formidable experimental challenges.  All measurements
of RMC to date have been inclusive ones, in which the final nuclear
states are not experimentally resolved.  These suffer the disadvantage
that the nuclear response has to be calculated from some theory
and this introduces considerable model dependence.

Most nuclear RMC calculations use the impulse approximation, in which
the amplitude for the elementary process is summed incoherently
over all the nucleons in the nucleus.  Even within the impulse approximation,
the calculation is non-trivial, since the RMC amplitudes must be
summed over all final states.  The earliest calculations adopted the
closure approximation \cite{RT65,Fe66,BD70} for this final-state
summation, in which a single average excitation energy
with a corresponding maximum photon energy, $k_{\rm{max}}$,
is assigned to the final nucleus.
The RMC photon spectrum then takes a polynomial form,

\be
\frac{d\Lambda}{dE_\gamma} \simeq (1 - 2x + 2x^2) x (1 - x)^2,
\label{poly}
\ee

\noindent where $x = E_\gamma/k_{\rm{max}}$ and $E_\gamma$ is the photon
energy.  The use of the closure approximation has been criticized
\cite{Ch81} and, indeed, such calculations typically overestimate the
RMC branching ratio by large factors \cite{GT87}\nl .  Nevertheless,
the polynomial form given in Eq.\,\ref{poly} usually reproduces
the data fairly well provided $k_{\rm{max}}$ is treated as a variable
parameter.  Christillin, Rosa-Clot and Servadio
\cite{CRS80} were able to avoid the closure approximation for
medium-heavy nuclei by integrating over the nuclear response
obtained in a nonrelativistic Fermi-gas model.  The only free
parameter was the effective nucleon mass, $M^\star$, which was
fixed at $M^\star = 0.5 M$ from a global fit to experimental OMC
rates.  The RMC rate was then calculated in a consistent manner.
A recent calculation by Fearing and Welsh \cite{FW92} cast some
doubt on the validity of the nonrelativistic Fermi-gas model
calculations.  They adopt a fully relativistic Fermi-gas model
along with the local-density approximation and realistic
nuclear-density distributions and a number of other refinements;
they find values for the OMC and RMC rates that are different by
factors of two
or more.  This issue remains unresolved.

In lighter nuclei, a more microscopic account of the response function
can be attempted.  Christillin \cite{Ch81} uses a phenomenological
nuclear excitation spectrum for $^{40}$Ca consisting of
Lorentzian giant dipole and quadrupole components.  The widths and
relative strengths are adjusted to reproduce existing photoabsorption
and OMC data.  A similar approach is used by Christillin and Gmitro
\cite{CG85} for $^{16}$O except explicit account is taken of
low-lying states in the $^{16}$N final nucleus in addition to the
giant dipole and quadrupole terms.  Lastly, Roig and Navarro \cite{RN90}
have invoked SU(4) symmetry and applied sum-rule techniques to
calculate RMC on $^{12}$C, $^{16}$O and $^{40}$Ca.  The branching
ratios predicted are substantially lower than those provided
by either the phenomenological or microscopic models.

{\small
\begin{table} [t]
\begin{center}
\caption{ Branching ratios and pseudoscalar coupling constants
extracted
from RMC measurements at PSI and at TRIUMF with various theoretical
calculations of the nuclear response function.
\label{RMCBR} }
\vskip 1mm
\begin{tabular}{rcccccr}
\hline \\[-3mm]
 & Branching & \multicolumn{4}{c}{$m_{\mu}\gP /\gA$} & \\
\cline{3-6}
Target & Ratio & & & & & ~~Reference \\
 & $\times 10^{-5}$ & CRS80$^a$ & CG85$^b$ &
GOT86$^c$ & RN90$^d$ & \\
\hline \\[-3mm]
$^{12}$C~~ & $2.7 \pm 1.8$ & $7.2^{+5.6}_{-11.8}$ & $9.5 \pm 7.2$ & & &
\cite{Do88}~~~ \\
 & $2.16 \pm 0.24$ & & $7.1 \pm 0.9$ & & & \cite{Ar91}~~~ \\
 & $2.13 \pm 0.24$ & & & $13.1^{+1.8}_{-2.0}$ & & \cite{Ar91}~~~ \\
$^{16}$O~~ & $2.44 \pm 0.47$ & $6.0^{+2.0}_{-3.0}$ & $8.4 \pm 1.9$ & & &
\cite{Do88}~~~ \\
 & $2.22 \pm 0.23$ & & $7.3 \pm 0.9$ & & & \cite{Ar91}~~~ \\
 & $2.18 \pm 0.21$ & & & $13.6^{+1.6}_{-1.9}$ & & \cite{Ar91}~~~ \\
$^{27}$Al~~ & $1.83^{+0.58}_{-0.25}$ & $2.5^{+3.9}_{-1.7}$ & & & &
\cite{Do88}~~~ \\
$^{40}$Ca~~ & $2.30 \pm 0.21$ & $6.3^{+1.0}_{-1.5}$ & & & &
\cite{Do88}~~~ \\
 & $2.07 \pm 0.14$ & & & $5.0 \pm 1.7$ & & \cite{Ar92}~~~ \\
 & $2.09 \pm 0.19$ & & & & $7.8 \pm 0.9$ & \cite{Ar92}~~~ \\
$^{\rm{nat}}$Fe~~ & $1.71 \pm 0.17$ & $3.0 \pm 1.3$ & & & &
\cite{Do88}~~~ \\
Mo~~ & $1.26 \pm 0.10$ & $0.0^{+1.6}_{-4.1}$ & & & & \cite{Ar92}~~~ \\
Sn~~ & $1.03 \pm 0.08$ & $0.1^{+1.4}_{-7.5}$ & & & & \cite{Ar92}~~~ \\
$^{165}$Ho~~ & $0.75 \pm 0.13$ & $-0.5 \pm 1.4$ & & & &
\cite{Do88}~~~ \\
Pb~~ & $0.60 \pm 0.05$ & $\leq 0.2$ & & & & \cite{Ar92}~~~ \\
$^{209}$Bi~~ & $0.62 \pm 0.08$ & $0.2 \pm 1.1$ & & & & \cite{Do88}~~~ \\
\hline \\[-3mm]
\multicolumn{7}{l}{$^a$ref.\,\cite{CRS80}; $^b$ref.\,\cite{CG85};
$^c$ref.\,\cite{GOT86};
$^d$ref.\,\cite{RN90} }
\end{tabular}
\end{center}
\end{table}
}

In Table \ref{RMCBR} we tabulate the branching ratio and $\gP$ values
extracted from RMC measurements at PSI \cite{Do88} and at TRIUMF
\cite{Ar91,Ar92}
with various theories for the nuclear response functions.  It is evident
that the observed branching ratio is falling for heavier nuclei.  It
also appears that $\gP$ is being quenched but this result is heavily
dependent on the theory of the nuclear response, which is
unsatisfactory at the present time.

In light nuclei, there are a few OMC experiments to explicit final
nuclear states that lead to a determination of the pseudoscalar coupling
constant.  To isolate the pseudoscalar term from a term involving
the regular axial-vector current requires either two complementary
measurements on the same transition or one measurement and a calculation.
Further, with the pseudoscalar term identified, a calculation of the
matrix element is required to determine the coupling constant.  We
cite three examples:

\bi

\item One favoured case is the
$\mu^{-} + ^{12}$C$(0^{+}) \rightarrow \,^{12}$B$(1^{+}) + \nu_\mu$
transition.  In a $0 \rightarrow 1$ capture, there are three
independent observables \cite{Gr85}: \viz \,the capture rate,
$\Gamma^{\rm{cap}}$, the polarization, $P_{\rm{av}}$, of the
recoil nucleus (here $^{12}$B) along the muon spin direction,
and the longitudinal polarization of
the recoil nucleus, $P_L$, where

\bea
P_{\rm{av}} & \equiv & \langle {\bf J} \cdot \mbox{\boldmath $\sigma$}
_{\mu} \rangle /J,
\nonumber  \\
P_L & \equiv & \langle {\bf J} \cdot \hat{{\bf v}}
\rangle /J ,
\label{PPP}
\eea

\noindent and $\langle \mbox{\boldmath $\sigma$}_{\mu}
\rangle = {\bf P}_{\mu}$ is the muon polarization, ${\bf J}$ the nuclear
spin, and $\hat{{\bf v}}$ the recoil direction.  Since the helicity of
the muon neutrino, $h_{\nu_{\mu}}$, is $-\sfrac{1}{2}$, $^{12}$B
can have only $m=-1$ or $m=0$ spin directions in the recoil
frame.  The helicity of the final state ($^{12}$B$+\nu_\mu$)
is thus $-\sfrac{1}{2}$ or $+\sfrac{1}{2}$, for which the amplitudes
are denoted by $M_{-1/2}$ and $M_{+1/2}$ respectively.  The observables
are expressed in terms of these amplitudes by

\bea
\Gamma^{\rm{cap}} & = & C (2+x^2)
\nonumber \\
P_{\rm{av}} & = & P_\mu \sfrac{2}{3} (1+2x)/(2+x^2)
\nonumber \\
P_L & = & -2/(2+x^2),
\label{xobser}
\eea

\noindent where $x=\sqrt{2}M_{+1/2}/M_{-1/2}$, and $C$ is a rate
constant depending on known quantites.  An interesting feature
is that $x$ can be
determined from the ratio of $P_{\rm{av}}$
to $P_L$:

\be
R = P_{\rm{av}}/P_L = - \sfrac{1}{3} (1+2x),
\label{Rratio}
\ee

\noindent when $P_{\mu} = 1$.
This ratio is not only more sensitive to $x$ than either
observable separately, but it is virtually immune to the systematic
uncertainities associated with absolute measurements of recoil
polarizations.  Roesch \etal \, \cite{Ro82} determined $x$ in a
novel recoil-implantation experiment that enabled $P_{\rm{av}}$
and $P_L$ to be measured {\em simultaneously}.  The parameter $x$
depends not only on the pseudoscalar coupling constant, but
also on nuclear matrix elements and other weak-interaction form
factors, such as weak magnetism.  With shell-model calculations
of nuclear matrix elements, and with beta-decay experiments to establish
the form factors (see sect.\,\ref{ExWM}), the dimensionless
quantity $m_\mu \gP / \gA$ is determined to be $9.0 \pm 1.7$.
This result is
slightly larger than the PCAC prediction, Eq.\,\ref{gpgap}, but is
consistent with the RMC results in Table \ref{RMCBR}.

\item Another interesting case is the
$\mu^{-} + ^{16}$O$(0^{+}) \rightarrow \,^{16}$N$(0^{-}) + \nu_\mu$
transition and its inverse, the beta decay
$^{16}$N$(0^{-}) \rightarrow \,^{16}$O$(0^{+}) + e^{-} +
\overline{\nu}_{e}$,
for which both transition rates have been measured \cite{Pa75,Ga83}\nl .
The analysis here depends on the shell-model calculations used to
compute matrix elements of parity-changing operators.
Gagliardi \etal \,\cite{Ga83} obtained $m_{\mu}\gP /\gA = 11 \pm 2$;
however, subsequent evaluations by Warburton \etal \,\cite{Wa94} and by
Haxton and Johnson \cite{HJ90}\nl , with much improved shell-model
wave functions, resulted in smaller values: in the first case,
$m_{\mu}\gP /\gA = 7.4 \pm 0.5$ and, in the second, it was in the
range 7 -- 9. Both are
consistent with PCAC.

\item Lastly, Gorringe \etal \,\cite{Go94} recently reported obtaining
a value of $m_{\mu} \gP /\gA$ from
measurements
of the rates and hyperfine dependencies of $\mu^{-}$ capture on
$^{23}$Na to explicit daughter states in $^{23}$Ne.
The $1S$ atomic ground state of a muonic
atom in the case of a $J_i \neq 0$ nucleus, is split into two
hyperfine (HF) states with angular momenta $F_{+} = J_i + 1/2$
and $F_{-} = J_i - 1/2$.  For examples where the transition
decreases the spin of the nucleus by one unit (\eg \ $\sfrac{3}{2}^{+}
\rightarrow \sfrac{1}{2}^{+}$ in $\mu^{-}$ $^{23}$Na), capture
from the $F_{-}$ state is generally dominated by $\gA$ and the
Gamow-Teller (GT) matrix element, whereas capture from the $F_{+}$
state is generally dominated by $\gP$ and the same GT matrix element.
Consequently, the ratio of $\mu$-capture
rates, $\Lambda_{+}/\Lambda_{-}$, is a strong function of
$\gP /\gA$ and, to the extent that the GT matrix element dominates other
terms, is independent of nuclear wave functions.  Hyperfine
transitions lead to the depopulation of the upper hyperfine state of
the $\mu^{-}$ atom ($F_{+}$ in $^{23}$Na) in favour of the lower
hyperfine state and yield a $\mu^{-}$ capture time dependence
that is sensitive to $\Lambda_{+}/\Lambda_{-}$.  Gorringe \etal
\cite{Go94} were able to measure this time dependence in the $\mu^{-}$
capture on $^{23}$Na to the 1017 keV $\sfrac{1}{2}^{+}$ and
1833 keV $\sfrac{3}{2}^{+}$ states in $^{23}$Ne to obtain
$m_{\mu}\gP /\gA = 7.6^{+2.1}_{-2.5}$ and $\leq 7.1$, respectively,
consistent with the PCAC prediction.

\ei

In summary, our knowledge of $\gP$ in nuclei is very sketchy
and depends heavily on nuclear-structure evaluations for its
determination.  There is a hint in light nuclei, $^{12}$C and
$^{16}$O, that $\gP$ might be slightly enhanced over the PCAC value, while
in heavy nuclei it seems almost completely quenched.
Some support for the first remark comes from
a calculation by Kolbe, Langanke and Vogel \cite{KLV94}\nl , who with
the continuum random phase approximation are able to reproduce the
total capture rates for ordinary muon capture on the
light nuclei, $^{12}$C, $^{16}$O and $^{40}$Ca,
without renormalization of $\gA$ or $\gP$.  The calculated
rates would be significantly lower if in-medium quenching of coupling
constants were introduced.

That $\gP$
might be significantly renormalized in heavy nuclei was first
discussed by Ericson \etal \,\cite{EFT73,Er71,DEFT76,Er78}\nl , who link
the pseudoscalar coupling term to the
pion-pole graph, described by Eq.\,\ref{Amuh1}.  In the
model of ref.\cite{Er78} the quenching is evaluated to be
$\gPeff = 0.3 \gP$ for infinite nuclear matter.  This work
is based on only $p$-wave pion-nucleon interactions.  Recently,
Delorme and Ericson \cite{DE94} pointed out that the repulsive
$s$-wave interaction significantly alters this estimate, and their
revised value is now $\gPeff = 0.6 \gP$.
Kirchbach and Riska \cite{KR94} also stress
that the effective induced pseudoscalar coupling constant in nuclei
originates from four different sources:  the space-like and
time-like parts of the axial current and the space-like and
time-like parts of the pion pole, each of which is renormalized
in a different way in the nuclear medium.
The net effect
for radiative muon capture in heavy nuclei
in their calculation
is a substantial
quenching of transition rates, which increases with mass number
as suggested by the experimental data.

\subsection{Meson-exchange currents in time-like axial-charge
transitions} \label{MECt}

In contrast to the quenching evident in allowed Gamow-Teller
transitions discussed in sect.\,\ref{qavcn}, transitions originating
in the time part of the axial-vector current appear {\em enhanced}.
Again there
are two sources of explanations, both of which contribute
to the final outcome.  First, despite use of the best available
shell-model calculations, the shell-model configuration space
has inevitably been truncated, and corrections for this truncation
are necessary.  These are the core-polarization corrections
discussed in sect.\,\ref{qavcn}.
Second, the one-body operator
from the impulse approximation is augmented by meson-exchange currents
in which the axial-vector current prompts a meson to be emitted
by one nucleon and reabsorbed on a second nucleon.  Such
processes lead to two-body operators.  The pion, as the lightest-mass
meson, produces a long-range operator and so it is the most
important meson to consider for exchange currents.  In any evaluation of
these currents, the pion-absorption amplitude is well known, being
determined from the long-range part of the nucleon-nucleon force,
but the pion-production amplitude is less certain.  In beta decay,
the momentum transfer is a few
MeV and is certainly small relative to the nucleon rest mass.
If, further, it is assumed that, in describing the pion-production
amplitude, it is sufficiently accurate to take the limit that the
pion momentum is small as well -- \ie \ the ``soft-pion
limit" -- then this vertex is determined in a model-independent
way through the exploitation of low-energy theorems \cite{AD66}\nl .
The net result is that the pion-production amplitude is given
by the commutator, $(1/f_\pi ) [Q_5,A_\mu ]$, where $A_\mu$ is
the hadronic axial current, $Q_5$ the axial-charge operator and $f_\pi$
the pion-decay constant.  Were we dealing with zero-mass quarks with
point couplings, then the evaluation of this commutator would be
trivial:  the charges would be generators of closed $SU(2) \times
SU(2)$ algebra and the commutator would be determined from the
group algebra.  These same commutators are nonetheless adopted
in this context, following the premise of current algebra
\cite{AD68,AFFR73} that the currents given by the quark model
exhibit algebraic properties that are of more general validity.
{}From the $SU(2) \times SU(2)$ group algebra, the commutator is

\be
[Q^a_5,A^b_\mu ] = i \epsilon_{abc} V^c_\mu,
\label{commu}
\ee

\noindent where the superscripts are isospin indices and
$\epsilon_{abc}$ the completely antisymmetric tensor in these
indices.  What this equation expresses is that the leading term in
the weak pion-production amplitude is a current proportional
to a vector current, $V_\mu $.

To continue, we evaluate this commutator between nucleon spinors
and take the nonrelativistic limit.  For the case in which the
Lorentz index, $\mu$, is time-like the commutator is of order
$\O (1)$, while when $\mu$ is space-like, it is of order
$\O (p/M)$.  Here $p$ is the momentum lost by the nucleon
making the transition, and is small by comparison with $M$, the
nucleon mass.  By contrast, an evaluation of the axial current,
$A_\mu$, itself between nucleon spinors -- the impulse
approximation -- leads, in the nonrelativistic limit, to terms
of order $\O (p/M)$ when $\mu$ is time-like, and of order
$\O (1)$ when $\mu$ is space-like.  Thus, from these limits,
it is evident that meson-exchange currents (MEC) will
be more important in axial-charge transitions (originating
in the time component of the axial current) than in Gamow-Teller
transitions (originating in the space component).  This
observation was first stressed by Kubodera, Delorme and Rho
\cite{KDR78}\nl .

In sect.\,\ref{qavcn}, we discussed corrections to some allowed
Gamow-Teller matrix elements and from Table \ref{dMGT} we see that
the MEC correction is small: of the order of one percent.
The isobar contribution is not part of the soft-pion MEC
correction.  This is because the pion-production amplitude at the
vertex where a nucleon transforms to an isobar is of order
$\O (q/M)$, where $q$ is the pion momentum, and vanishes in the
soft-pion limit.
However, for the space part of the axial current, the soft-pion terms are
suppressed -- of order $\O (p/M)$ --  and are comparable in size to the
correction terms.  In this situation the soft-pion theorems
by themselves are of little use.  Rather, one must start with a model
Lagrangian and explicitly evaluate the meson-exchange currents
it would generate.

For the rest of this section, we will only consider the time
part of the axial current, for which
the soft-pion terms are not suppressed:
this leads to what are called axial-charge
transitions.  The two-body operator, after the nonrelativistic
limit has been taken, is Fourier transformed to coordinate space and
the rank-0 multipole is projected out, giving \cite{To92b,To86}

\be
A_0({\rm soft}\hyphen \pi ) = - \frac{1}{2f_\pi} \frac{\gpNN}{2M}
\frac{m_\pi^2}{4 \pi} \, \mbox{\boldmath $\sigma$}_{+} \cdot
{\bf \hat{r}} \, Y_1(x_\pi) \,
i (\mbox{\boldmath $\tau$}_1 \times \mbox{\boldmath $\tau$}_2),
\label{A02b}
\ee

\noindent where $\mbox{\boldmath $\sigma$}_{+} = \mbox{\boldmath
$\sigma_1 + \sigma_2$}$, ${\bf r} = {\bf r}_1 - {\bf r}_2$, and
$Y_1(x) = (1 + 1/x) e^{-x}/x$, with $x = m_{\pi} r$.
The corresponding one-body operator from the impulse
approximation is

\be
A_0({\rm IA}) = \gA \frac{\mbox{\boldmath $\sigma \cdot \nabla$}}
{2M} \mbox{\boldmath $\tau$}.
\label{A01b}
\ee

\noindent The interesting question is how big are the matrix
elements of the pion-exchange terms relative to those from
the impulse approximation.  This, then, requires a nuclear-structure
calculation for the nucleus in question.  Delorme \cite{De82}
has given an estimate for nuclear matter and finds the ratio
$A_0({\rm soft} \hyphen \pi )/A_0({\rm IA})$ to be large, being of order
60\% at nuclear-matter densities.  Axial-charge transitions are
clearly heavily renormalized by pion-exchange currents.  This
prediction, which comes from soft-pion theorems and current
algebra, is unfortunately hard to test experimentally because
such transitions are difficult to isolate.  There are two possibilities:
one can either examine the small $1/M$ recoil corrections in an
allowed transition, or one can look at transitions in forbidden
decays.  We will consider examples of both.

\subsubsection{Recoil corrections in allowed transitions} \label{Recoil}

The A=12 triad has long been the testing ground of nuclear beta-decay
theories.  It has been used to verify the conserved vector current (CVC)
hypothesis through the identification of weak magnetism terms,
and it has ruled out any possibility of sizeable second-class
currents (see sect.\,\ref{ExWM}).
By yielding the $1/M$ recoil correction in an allowed
Gamow-Teller transition, it has also provided a determination of
the axial-charge matrix element.  Various experiments
\cite{Le78,Br78a,Su78,Br78b,Ma79,Mi86,Mi93}
have measured the correlation between nuclear
alignment and the direction of the emitted electron following
nuclear beta decay from oriented $^{12}$B and $^{12}$N to the
ground state of $^{12}$C (see Table \,\ref{alpexpt}).
The electron's angular distribution
can be represented by a Legendre series, and the coefficient of the
$P_2(\cos \theta)$ Legendre polynomial is the quantity that must be
measured.  If we denote this coefficient as $\alpha_{+}$ for the
positron emission from $^{12}$N and as $\alpha_{-}$ for the
electron emission from $^{12}$B, the
difference, $\alpha_{-} - \alpha_{+}$, isolates the weak
magnetism matrix element (together with the second-class
current term, if present)
while the sum, $\alpha_{-} + \alpha_{+}$, isolates
the axial-charge matrix element.  It is the anti-Hermitian nature
of the axial-charge operator that produces this sign change.
The average of the experimental alignment data in Table \,\ref{alpexpt}
yields
$\alpha_{-} + \alpha_{+} = -(2.823 \pm 0.078)  {\rm GeV}^{-1}$.
Theoretically, the combination
$\alpha_{-} + \alpha_{+}$ is related to nuclear-structure matrix
elements via

\bea
\alpha_{-} + \alpha_{+} & = & - \sfrac{4}{3} \sfrac{1}{2M} \, y
\nonumber  \\
y & = & 1 + 2 \langle {\bf r}(\mbox{\boldmath $\sigma \cdot \nabla$})
\rangle / \langle \mbox{\boldmath $\sigma $} \rangle,
\label{yaxch}
\eea

\noindent where the matrix element
$\langle {\bf r}(\mbox{\boldmath $\sigma \cdot \nabla$}) \rangle $
arises from a projection of the rank-1 multipole from the
impulse-approximation axial-charge current.  The experimental
value for $y$ is $y_{\rm{expt}} = 3.98 \pm 0.11$.

Morita \etal \ \cite{Mo76} calculated these matrix elements in the
impulse approximation with shell-model wavefunctions that
describe the ground states as $0 \hbar \omega$ $(0p)^{-4}$
configurations in the closed-shell at $^{16}$O.  The results
span $y = 3.6 \pm 0.1$ depending on the choice of residual interaction
and are close to the experimental result.  There seemingly
is little evidence for a significant correction arising from
two-body meson-exchange currents.  Later, Koshigiri \etal \ \cite{Ko81}
realized that the shell-model space was quite restrictive and that
enlarging the model space would significantly influence the
$\langle {\bf r}(\mbox{\boldmath $\sigma \cdot \nabla$}) \rangle $
matrix element.  Using perturbation theory, Koshigiri \etal \ \cite{Ko81}
demonstrated that this correction depended principally on the
strength of the residual tensor force.  Their final result is
$y = 3.6 - 1.2 + 1.3 = 3.7$, with $-1.2$ being their core-polarization
estimate with the Hamada-Johnston tensor force and $+1.3$ being
the meson-exchange current enhancement.  Guichon and Samour \cite{GS82}\nl ,
in a similar calculation, obtained $y = 3.0 - 0.4 +1.0 = 3.6$,
where the smaller core-polarization corection of $-0.4$ arises
with the Sussex interaction, which has a weaker tensor
component.  Clearly, details of the analysis
depend on the choice of residual
interactions, model spaces, wavefunctions, and so on, but if the
predicted meson-exchange enhancement is present, it is in all likelihood
masked by the core-polarization correction.

\subsubsection{First-forbidden beta decays} \label{ffbd}

In the case just covered, the axial-charge operator
leads to a small $1/M$ correction in allowed decay, which has to be
isolated from other small $1/M$ recoil corrections.
By contrast, the time part of
the axial-vector current contributes a leading term in first-forbidden
weak transitions.  Unfortunately, the space parts are not negligible:
the familiar Gamow-Teller operator, retarded by one unit of
orbital angular momentum, contributes with comparable magnitude.
In first-forbidden decays, the impulse approximation produces six
one-body operators: one, the axial-charge operator
$\mbox{\boldmath $\sigma \cdot \nabla$}/M$,
from the time part of the axial-vector current; three from
the space part of the axial-vector current,
$[{\bf r},\mbox{\boldmath $\sigma$}]^{(R)}$, with $R = 0,1,$ or $2$
being the tensorial rank; one from the space part of the
weak vector current, $\mbox{\boldmath $\nabla$}/M$; and
one from the time part, ${\bf r}$.  Angular momentum selection
rules limit the number of these operators that can contribute.
For example, in $0^{+}$ to $0^{-}$ transitions, only two rank-zero
operators contribute; in $0^{+}$ to $1^{-}$ the three rank-one
operators contribute; while in $\sfrac{1}{2}^{+}$ to
$\sfrac{1}{2}^{-}$ transitions, the five rank-zero and rank-one
operators are involved.  We limit our discussion to these three cases:
thus, the rank-two operator will play no role.

The notation for the first-forbidden beta-decay matrix elements is
taken from Warburton \etal \ \cite{WBBM88,Wa91} and we define

\bea
M^{T}_{0} & = & \gA \,\langle  J_f \, T_f \tbar \sfrac{1}{M} \mbox{
\boldmath $\sigma \cdot \nabla \tau$} \tbar J_i \, T_i \rangle \, C
\nonumber \\
M^{S}_{0} & = & -\gA \, \langle J_f \, T_f \tbar \mbox{\boldmath
$\sigma \cdot r \tau$} \tbar J_i \, T_i \rangle \, C
\nonumber \\
M^{u}_{1} & = & \gA \sqrt{2} \, \langle J_f \, T_f \tbar [ \mbox{
\boldmath $\sigma , r $}]^{(1)} \mbox{\boldmath $\tau$} \tbar J_i \, T_i
\rangle \, C
\nonumber \\
M^{x}_{1} & = & - \langle J_f \, T_f \tbar \mbox{\boldmath
$ r \tau$} \tbar J_i \, T_i \rangle \, C
\nonumber \\
M^{y}_{1} & = & - \langle J_f \, T_f \tbar \sfrac{1}{M} \mbox{\boldmath
$ \nabla \tau$} \tbar J_i \, T_i \rangle  \, C,
\label{FFme}
 \eea

\noindent with $C = (\hat{J}_f/(\sqrt{2} \hat{J}_i))
\langle T_i \, T_{zi} \, 1 \, \mp 1 \mid T_f \, T_{zf} \rangle$.
The notation is the same as that introduced in Eq.\,\ref{MGT}
for allowed transitions.  The conserved vector current hypothesis
may be used \cite{BB82} to obtain an alternative expression
for $M^{y}_{1}$ in terms of $M^{x}_{1}$:

\be
M^{y}_{1} = E_{\gamma} M^{x}_{1},
\label{yCVCx}
\ee

\noindent where $E_{\gamma}$ is the photon energy of the analogous $E1$
gamma transition.  In the Behrens-B\"{u}hring \cite{BB82} formulation,
which is being followed here, the beta-decay formulae are derived
by expanding the lepton radial wavefunctions in powers of
the lepton's mass and
energy, and of the nuclear charge.  In this treatment,
inclusion in the radial integral of an extra factor
that depends on the nuclear charge distribution introduces
additional matrix elements obtained from $M^S_0$, $M^x_1$ and
$M^u_1$ and denoted $M^{S \prime}_0$, $M^{x \prime}_1$ and
$M^{u \prime}_1$.
The ratios of
the primed to unprimed quantities, denoted $r^{\prime}_S$,
$r^{\prime}_x$ and $r^{\prime}_u$, are insensitive \cite{WBBM88} to
details of the nuclear structure and are roughly 0.7.
In the limit that small lepton-energy-dependent terms are
dropped -- the $\zeta$-approximation of Warburton \cite{Wa91} -- the
formalism for first-forbidden decays exactly matches that for
allowed decays: \viz Eqs.\,\ref{ftGT} and \ref{MGT}. Thus,

\be
ft ( 1 + \delta_R ) = \frac{6146 \pm 6}{B(0)+B(1)} ~~ {\rm s}
\label{ftFF}
\ee

\noindent with $B(0) = \ \mids M(0) \mids^2$,
$B(1) = \ \mids M(1) \mids^2$, and

\bea
M(0) & = & \epsilon_{\rm{mec}} M^T_0 + a_S M^S_0
\nonumber \\
M(1) & = & a_u M^u_1 - a_x M^x_1,
\label{MFF}
\eea

\noindent where

\bea
a_S & = & r^{\prime}_S \xi + \sfrac{1}{3} W_0
\nonumber \\
a_u & = & r^{\prime}_u \xi - \sfrac{1}{3} W_0
\nonumber \\
a_x & = & E_{\gamma} - r^{\prime}_x \xi -\sfrac{1}{3} W_0.
\label{afctrs}
\eea

\noindent  Here $Z$ is the atomic number of the daughter nucleus in
beta decay, $W_0$ the end-point electron energy in electron rest-mass
units,
$\xi = \alpha Z/(2R)$ with
$\alpha$ the fine-structure constant and
$R$ the radius of the nuclear charge distribution in
electron Compton-wavelength units.  The matrix elements given in
Eq.\,\ref{FFme}, and the $a$-factors in Eq.\,\ref{afctrs}
are computed in units that render them dimensionless.

Lastly, we have included in Eq.\,\ref{MFF} a factor, $\epsilon_{\rm
{mec}}$, representing the expected enhancement of the axial-charge
matrix element, $M^T_0$, over the impulse-approximation value
due to meson-exchange currents.  The procedure
is to compute the matrix elements with the best available
shell-model wavefunctions, incorporate core-polariz\-ation corrections,
and, by comparison of theory with experiment for several beta-decay
transitions determine the average value of the enhancement factor.
This has been accomplished by Warburton \etal \ for three mass regions.
In the $A=16$ region, the average \cite{WTB94} from three beta decays and
the $\mu^{-}$ capture on $^{16}$O is
$\epsilon_{\rm{mec}} = 1.61 \pm 0.03$;
in the $A=132$ region, the average \cite{WT92} from three beta transitions is
$\epsilon_{\rm{mec}} = 1.82 \pm 0.07$;
and for eighteen transitions in the $A=208$ region, as re-analysed
by Warburton and Towner \cite{WT94}\nl , the average is
$\epsilon_{\rm{mec}} = 1.79 \pm 0.04$.  In the latter two cases, the
results are dependent on the choice of the residual
interaction used for the core-polarization correction.  The values
given are based on the Bonn nucleon-nucleon $G$-matrix, which has a
weaker tensor component than, for example, the Paris nucleon-nucleon
$G$-matrix.  For the $A=208$ region, the Paris $G$-matrix leads
to a larger enhancement factor of
$\epsilon_{\rm{mec}} = 1.91 \pm 0.04$.  These results show
that a substantial enhancement of the
axial-charge impulse-approximation matrix element is required, and
that there is some evidence for larger enhancement factors
in heavier nuclei than in light nuclei.

Shell-model calculations of $\epsilon_{\rm{mec}}$ have been given
by Towner \cite{To92b} for closed-shell-plus-one configurations,
for which the evaluation of the matrix element of the soft-pion
operator of Eq.\,\ref{A02b} involves the calculation of
two-body matrix elements
between valence and core nucleons and summed over
all core nucleons.  Such a calculation produces some mass dependence
in $\epsilon_{\rm{mec}}$ because the number of core orbitals
is changing from light to heavy nuclei, and the
oscillator frequency parameter used in the radial wavefunction
decreases with increasing nuclear size.  The results
for the soft-pion operator
are $\epsilon_{\rm{mec}} = 1.56$ at $A=16$, and
$\epsilon_{\rm{mec}} = 1.76$ at
both $A=132$ and $A=208$.  This is in
reasonable accord with the values deduced from experiment.

Calculations that go beyond soft pions, invoking heavy mesons
\cite{To92b,KRT92} or chiral perturbation theory \cite{PTK94}\nl ,
and calculations that explain the mass dependence as `Brown-Rho'
scaling \cite{KR91} can be examined in the cited references.

\section{Beyond the Standard Model} \label{BSM}
\subsection{Right-hand currents}  \label{RHC}

Maximal parity violation in weak interactions is built into
the Standard Model as seems to be demanded by experimental
evidence at low energy.  At higher energies, one can envisage
a restoration of parity symmetry through model extensions.
In the charged-current sector, these models introduce additional,
predominantly right-handed charged gauge bosons $W_2^{\pm}$ to
complement the predominantly left-handed gauge bosons
$W_1^{\pm}$.  The right-handed bosons acquire, via spontaneous
symmetry breaking, a mass $m_2$ that is larger than the mass
$m_1 (\sim 81$ GeV) of the observed $W^{\pm}$.  By construction,
the left-handed couplings dominate at low energies with
parity-symmetry restored at a higher-energy scale.  The
weak-interaction eigenstates -- $W_L$, which corresponds to the
Standard Model's $W$ gauge boson, and $W_R$ -- are linear
combinations of the physical bosons $W_1$ and $W_2$ with

\bea
W_1 & = & W_L \cos \zeta - W_R \sin \zeta
\nonumber  \\
W_2 & = & W_L \sin \zeta + W_R \cos \zeta,
\label{Wmix}
\eea

\noindent where $\zeta$ is a real mixing angle.  The simplest of
the model extensions is known as manifest left-right symmetry
\cite{BBMS77}\nl , in which the physical left-handed and right-handed
currents have identical coupling strengths and transformation
properties, and one current is obtained from the other via
$\gamma_5 \rightarrow -\gamma_5$.  In the pure $V-A$ theory,
the relevant part of a semi-leptonic weak interaction can be
written in a schematic form as

\be
T_{fi}  =  \frac{g^2}{8 m_1^2} \left [ (V-A) \right ]
\left [ (V-A) \right ] = \frac{\GF}{\sqrt{2}} V_{ud}
\left [ VV + AA - (VA + AV) \right ],
\label{VAsch}
\ee

\begin{table}[t]
\begin{center}
\caption{Modifications to various observables of $V-A$ theory
through the introduction of right-hand currents evaluated to leading
order$^{a}$ in the small quantities $\zeta$ and $\delta$.
\label{VAmod}}
\vskip 1mm
\begin{tabular}{lcc}
\hline  \\[-3mm]
 & $V-A$ & Modification  \\
\hline  \\[-3mm]
Fermi decay & $\GV^2  \langle \MV \rangle^2$ &
$\GV^2  \langle \MV \rangle^2 (1 + 2 \zeta - 2 \delta )$  \\
GT decay & $\GV^2  \lambda^2 \langle \MA \rangle^2$ &
$\GV^2  \lambda^2 \langle \MA \rangle^2 (1 + 6 \zeta - 2
\delta )$ \\
Muon decay & $\GF^2$ & $\GF^2 (1 + 4 \zeta - 2 \delta )$ \\
Fermi/muon decay &
$ V_{ud}^2 \langle \MV \rangle^2$ &
$ V_{ud}^2 \langle \MV \rangle^2 (1 - 2 \zeta )$  \\
Neutron decay &
$\GV^2 (1 + 3 \lambda^2)$ &
$\GV^2  \left [ (1 + 2 \zeta - 2 \delta ) + 3 \lambda^2
(1 + 6 \zeta - 2 \delta ) \right ] $  \\
Neutron-decay asym., A &
$ -2 \frac{\lambda^2 + \lambda }{1 + 3 \lambda^2}$ &
$ -2 \frac{\lambda^2 (1 + (\zeta + \delta )^2) + \lambda
(1 + \zeta^2 + \delta^2) }{ (1 + (\zeta - \delta )^2)
+ 3 \lambda^2 (1 + (\zeta + \delta )^2) }$  \\
Fermi long. polariz., $P_L^F$ & $1$ & $ 1 - 2(\zeta - \delta)^2$ \\
GT long. polariz., $P_L^{GT}$ & $1$ & $ 1 - 2(\zeta + \delta)^2$ \\
Ratio $P_L^F/P_L^{GT}$ & $1$ & $ 1 + 8 \zeta \delta$ \\
\hline \\
\multicolumn{3}{l}{{\footnotesize $^{a}$Note that radiative corrections are
neglected, and
$\GV = \GF V_{ud}$.}}
\end{tabular}
\end{center}
\end{table}

\noindent where the $(V-A)$ in the first bracket represents
the hadronic current and in the second bracket the leptonic
current.  The modification to incorporate right-hand currents
is

\bea
T_{fi} & = & \frac{g^2}{8 m_1^2} \left [ c(V-A) - s(V+A) \right ]
\left [ c(V-A) - s(V+A) \right ]
\nonumber  \\
&  & ~~ + \frac{g^2}{8 m_2^2} \left [ s(V-A) + c(V+A) \right ]
\left [ s(V-A) + c(V+A) \right ]
\nonumber  \\
& = & \frac{\tilde{\GF}}{\sqrt{2}} V_{ud} \left [ VV + \eta_{AA} AA
+ \eta_{AV} (VA + AV) \right ],
\label{VAman}
\eea

\noindent where $c$ is $\cos \zeta$ and $s$ is $\sin \zeta$, and

\bea
\frac{\tilde{\GF}}{\sqrt{2}} & = & \frac{g^2}{8 m_1^2} (c - s)^2
+ \frac{g^2}{8 m_2^2} (c + s)^2
\rightarrow \frac{\GF}{\sqrt{2}} (1 + \delta - 2 \zeta )
\nonumber  \\
\eta_{AA} & = & \frac{\epsilon^2 + \delta}{\epsilon^2 \delta + 1}
\rightarrow 1 + 4 \zeta
\nonumber  \\
\eta_{AV} & = & - \epsilon \frac{1 - \delta}{\epsilon^2 \delta + 1}
\rightarrow -1 - 2 \zeta + 2 \delta,
\label{modif}
\eea

\noindent with $\epsilon = (1 + \tan \zeta)/(1 - \tan \zeta)$
and $\delta = m_1^2/m_2^2$.  In Table \ref{VAmod} we show the
modifications to various observables in weak interactions
evaluated to leading order in the small quantities
$\zeta$ and $\delta$.  Pure $V-A$ theory is recovered in the
limit that $\zeta \rightarrow 0$ and $ \delta \rightarrow 0$.

{\em Beta asymmetry}.  As discussed in sect.\,\ref{NeutD} on
neutron decay, the asymmetry in the distribution of electrons
relative to a polarization in the parent neutron is governed
by a parameter $A$ that depends on the ratio, $\lambda$,
of axial-vector and vector coupling constants, $\lambda =
\GA^{\prime} / \GV^{\prime}$.  With the inclusion of right-hand currents,
$A$ becomes a function of $\lambda$, $\zeta$ and $\delta$
as given in Table \ref{VAmod}.  Likewise, the ratio of the
neutron's $ft$ value to that of the pure Fermi superallowed
emitters is also given in terms of $\lambda$, $\zeta$ and
$\delta$.  It is then a straightforward matter with these
inputs to eliminate $\lambda$ and, allowing for experimental
uncertainties, to set limits
\cite{HT77,CDH88,GSS90,AFG91}
on $\zeta$ and $\delta$.
A similar analysis \cite{Ca92} can be
performed for the mirror transition in the decay of $^{19}$Ne,
for which the beta asymmetry and lifetime have
been determined.  Carnoy \etal \ \cite{Ca92} have also performed
a combined fit to both data sets and obtain at the 90\%
confidence level: $\delta = 0.12 \pm 0.04$ and $\zeta =
-0.003 \pm 0.009$.  The mass-ratio parameter $\delta$
differs from the pure $V-A$ result by nearly three
standard deviations and this requires the mass of the
predominantly right-hand boson to be in the range 200 to
300 GeV.

{\em Longitudinal polarization}.  Measurements of absolute
electron polarizations in Gamow-Teller decays by van Klinken
\cite{vK66} have a combined statistical precision of 0.9\%
and absolute accuracy of 1.6\%.  The most precise measurements are
for highly hindered decays for which recoil and forbidden
corrections might be important.  However, if the evaluation
of van Klinken \cite{vK66} is taken at face value ($P_L^{GT} =
1.001 \pm 0.012$), a limit of
$\mids (\delta + \zeta) \mids \ \leq 0.10$
is obtained at the 90\% confidence
level.

An entirely different class of experiments is represented by two recent
measurements \cite{Wi87,Ca91} of the ratio of positron
polarizations $P_L^{F}/P_L^{GT}$ for pure Fermi and pure
Gamow-Teller transitions.  These are believed to be
reliable measurements, with precision of less than 0.5\% for
the polarization ratio.  In addition to being free of absolute
analyzing-power uncertainties in the polarimeter, the techniques
employed in these measurements reduce many other systematic
uncertainties.  The results place a limit on the product of
the two parameters: $\mids \zeta \delta \mids \ = 0.00015
\pm 0.00034$.

{\em CKM Unitarity}.  If the $CKM$ matrix is taken to be unitary
in three generations, then a value for $V_{ud}$ can be obtained
from $V_{ud}^2 = 1 - V_{us}^2 - V_{ub}^2$ with
recommended values for $V_{us}$ and
$V_{ub}$ taken from ref.\,\cite{PDG94}\nl .
The result differs from the value determined from
superallowed Fermi decays discussed in sect.\,\ref{Impsm}
(see also Fig.\,\ref{fig2}).
If the difference is ascribed to right-hand currents, then a
value is obtained for the mixing angle of $\zeta =
0.0018 \pm 0.0007            $, which is two standard
deviations from the pure $V-A$ result.

{\em Muon Spectrum}.  Without radiative corrections, the muon
differential
decay rate, $\mu^{+} \rightarrow e^{+} + \nu_e + \overline{\nu}_{\mu}$,
integrated over $e^{+}$ spin directions, is given by

\bea
\frac{ {\rm d}^2 \Gamma}{x^2 {\rm d}x {\rm d}(\cos \theta )} &
= & \left [ (3-2x) + (\sfrac{4}{3} \rho - 1)(4x-3) + 12
\frac{m_e}{m_{\mu}} \frac{x-1}{x} \eta \right ]
\nonumber \\
& & - \left [ (2x-1) + (\sfrac{4}{3} \delta_m -1)(4x-3) \right ]
\xi P_{\mu} \cos \theta.
\label{muspect}
\eea

\noindent Here, $x$ is the reduced electron energy $E_e/E_{\rm max}$,
where $E_{\rm max} = (m_e^2 + m_{\mu}^2 )/(2 m_{\mu} ) =
52.83$ MeV is the maximum energy, and $m_e$ and $m_{\mu}$ are the
particle masses.  The angle between the positron momentum
and the muon polarization vector, $P_{\mu}$, in the $\mu^{+}$
rest frame is $\pi - \theta$.  The muon-decay Michel parameters
are $\rho$, $\eta$, $\xi$ and $\delta_m$, the values of which in a
pure $V-A$ theory are: $\rho = 3/4$, $\eta = 0$, $\xi = 1$ and
$\delta_m = 3/4$.  The experiment of Jodidio \etal \ \cite{Jo86}
was designed to measure the positron spectrum for
$x \geq 0.85$ and $\cos \theta \geq 0.95$, which is sensitive
to the combination of parameters, $\xi P_{\mu} \delta_m / \rho$.
This combination equals 1 for pure $V-A$ and
$1 - 2(2 \delta^2 + 2 \delta \zeta + \zeta^2 )$ with the
inclusion of right-hand currents.  The measured lower limit
\cite{Jo86} of

\be
\xi P_{\mu} \delta_m / \rho \geq 0.99682
\label{mupol}
\ee

\noindent sets limits on the right-hand-current parameters at
the 90\% confidence level of $\delta \leq 0.04$ or
$m_2 \geq 406$ GeV and $-0.056 \leq \zeta \leq 0.040$.

It should be noted that the limits on right-hand currents from
muon decay and $\beta$-asymmetry measurements disagree with
each other at the 90\% confidence level.  This discrepancy,
however, disappears \cite{DQ94} at the $3 \sigma$ level.

\subsection{Scalar Interactions}  \label{ScI}

As has already been noted, the $V-A$ structure of the weak
interaction in the Standard Model has been put in
{\em a priori} to obtain agreement with experiments.  Although
the model agrees with all experimental results at presently
available energies, it remains an interesting quest to
search for deviations from the $V-A$ structure.  In the last
section, we discussed right-hand currents and in this
we consider possible scalar and tensor couplings.  A more
general form of the weak-interaction Hamiltonian was first
written down by Jackson, Treiman and Wyld \cite{JTW57} as

\bea
T_{fi} & = & \frac{\GF }{\sqrt{2}} V_{ud} \left [ \overline{\psi}_p
\gamma_{\mu} \psi_n \  \overline{\psi}_{e^{-}} \gamma_{\mu}
(C_V + C_V^{\prime} \gamma_5 ) \psi_{\overline{\nu}_e} \right.
\nonumber  \\
& & ~~~~~~~~~~ + \overline{\psi}_p \gamma_{\mu} \gamma_5 \psi_n
\  \overline{\psi}_{e^{-}} \gamma_{\mu} \gamma_5
(C_A + C_A^{\prime} \gamma_5 ) \psi_{\overline{\nu}_e}
\nonumber  \\
& & ~~~~~~~~~~ +
\overline{\psi}_p \psi_n \  \overline{\psi}_{e^{-}}
(C_S + C_S^{\prime} \gamma_5 ) \psi_{\overline{\nu}_e}
\nonumber  \\
& & ~~~~~~~~~~ + \left. \sfrac{1}{2} \overline{\psi}_p \sigma_{\mu \nu}
\psi_n \  \overline{\psi}_{e^{-}} \sigma_{\mu \nu}
(C_T + C_T^{\prime} \gamma_5 ) \psi_{\overline{\nu}_e} \right ].
\label{TfiCC}
\eea

\noindent A pseudoscalar contribution has not been included,
since it vanishes in leading order in nuclear beta decay.
Further, the induced terms, such as weak magnetism of Eq.\,\ref{VAc},
are not included here.  Lastly, time-reversal
invariance is assumed so that the coupling constants
$C_i$ and $C_i^{\prime}$ are real.  The Standard Model limit
is recovered by the replacements: $C_V = C_V^{\prime} = \gV$,
$C_A = C_A^{\prime} = \gA$ and $C_S = C_S^{\prime} = C_T =
C_T^{\prime} = 0 $.  Note that this parameterization also includes
right-hand currents when the possibility that $C_i \neq C_i^{\prime}
$ is admitted.  In the nonrelativistic limit, $\overline{\psi}_p
\psi_n \rightarrow \langle \MV \rangle $ and $\overline{\psi}_p
\sigma_{\mu \nu} \psi_n \rightarrow \langle \MA \rangle $
in nuclear beta decay, so the presence of scalar interactions
leads to modifications in Fermi transitions, as tensor
interactions lead to modifications in Gamow-Teller transitions.

In 1984, Boothroyd, Markey and Vogel \cite{BMV84} set limits
on the strength of scalar and tensor interactions in a
least-squares adjustment to 93 pieces of data from nuclear
beta decay that do not depend on nuclear matrix elements.  There
are seven parameters to be adjusted if right-hand currents are
permitted --  namely $C_i/C_V$ and $C_i^{\prime}/C_V$ --  but just
three parameters if $C_i^{\prime} = C_i$.  At the 95\%
confidence level ($2 \sigma $) their results for a
seven-parameter fit are

\bea
\left | \frac{C_S}{C_V} \right | \leq 0.23 &
\displaystyle{\left | \frac{C_S^{\prime}}{C_V} \right | \leq 0.190} &
\left | \frac{C_S + C_S^{\prime}}{C_V} \right | \leq 0.065
\nonumber  \\
\left | \frac{C_T}{C_A} \right | \leq 0.09 &
\displaystyle{\left | \frac{C_T^{\prime}}{C_A} \right | \leq 0.085} &
\left | \frac{C_T + C_T^{\prime}}{C_A} \right | \leq 0.01,
\label{fit7}
\eea

\noindent and for a three-parameter fit

\be
\Bigm{\vert} \frac{C_S}{C_V} \Bigm{\vert} \leq 0.0033 ~~~~
\Bigm{\vert} \frac{C_T}{C_A} \Bigm{\vert} \leq 0.0011.
\label{fit3}
\ee

\noindent
These limits on scalar and tensor interactions come mainly
from measurements on the beta-neutrino angular correlation
coefficient, $a$, and the Fierz interference term, $b$.
If one writes

\be
K_{ij} = C_i C_j + C_i^{\prime} C_j^{\prime},
\label{Kij}
\ee

\noindent then these observables are defined in terms of the
coupling constants as

\bea
a \xi & = & \sfrac{1}{3} \langle \MA \rangle^2 (K_{TT} -
K_{AA}) + \langle \MV \rangle^2 ( K_{VV} - K_{SS})
\nonumber  \\
b \xi & = & \pm 2 \left ( \langle \MA \rangle^2 K_{TA}
+ \langle \MV \rangle^2 K_{SV} \right ),
\label{absi}
\eea

\noindent with

\be
\xi = \langle \MA \rangle^2 (K_{TT} + K_{AA})
+ \langle \MV \rangle^2 ( K_{SS} + K_{VV}).
\label{xisi}
\ee

\noindent The upper sign is for electron emission and the
lower sign for positron emission.
In what follows, we comment on developments since the
Boothroyd \etal \ survey \cite{BMV84}\nl .

{\em $\beta$-$\nu$ angular correlations}.
The classic $e-\nu$ correlation data were
obtained by observation of the lepton-recoil effect on the energy
distribution of stable daughter nuclei from $^{6}$He,
$^{19}$Ne, $^{23}$Ne and $^{35}$Ar decays.  Such experiments
are very difficult because the recoil energies are only a
few hundred eV.  High precision results are only available
for $^{6}$He decay \cite{JPC63} and neutron decay \cite{St78}\nl .
Neither of these are pure Fermi decays and therefore are not
sensitive to scalar couplings.  However,
a recent study of $^{32}$Ar ($0^{+} \rightarrow 0^{+}$) and
$^{33}$Ar ($\sfrac{1}{2}^{+} \rightarrow \sfrac{1}{2}^{+}$)
decays by Schardt and Riisager \cite{SR93} measured with
high precision the shape of the beta-delayed proton peak
following the superallowed decay to its isospin analogue
state in the daughter nucleus.  In this case, the lepton
recoil is delivered to a proton-unbound state, and the
subsequently emitted proton
has a resulting energy spread that exceeds the energy imparted to
the recoil daughter by a factor of $\sim 60$.  Because
its particle decays are isospin-forbidden and the natural
width of the proton-unbound level consequently so small,
the shape of the proton peak predominantly reflects
the $e-\nu -p $ triple-correlation
asymmetry parameter, which for isotropic distributions
equals the $e-\nu $ correlation coefficient.  Adelberger's
analysis \cite{Ad93} of the Schardt-Riisager data
\cite{SR93} improves the limit on the scalar constant from
the seven-parameter fit by a factor of two:

\be
\Bigm{\vert} \frac{C_S}{C_V} \Bigm{\vert} \leq 0.12, ~~~~~~~~
\Bigm{\vert} \frac{C_S^{\prime}}{C_V} \Bigm{\vert} \leq 0.10.
\label{fit7p}
\ee

\noindent  It is evident that isospin-forbidden $\beta$-delayed
proton spectroscopy provides a powerful probe of scalar couplings.

{\em Fierz interference term}.  A measurement of the Fierz interference
term in a pure Fermi transition would also lead to a limit on
scalar couplings.  In sect.\,\ref{SFT}, a discussion is given on the
superallowed Fermi emitters and a statistical rate function, $f$,
is defined in Eq.\,\ref{fint}.  The presence of scalar couplings adds
an additional factor to the integrand in Eq.\,\ref{fint} of
$(1 + \gamma b/W)$, where $\gamma$ is $(1 - (\alpha Z)^2)^{1/2}$
with $\alpha$ the fine-structure constant and $Z$ the atomic
number of the daughter nucleus.  Integration over the electron
spectrum with the introduction of electromagnetic corrections, as in
Eq.\,\ref{scrptft}, leads to the modified expression

\be
(1 - \gamma b \langle W^{-1} \rangle ) =
\frac{K}{2 \GV^{\prime 2} \F t }.
\label{bFT}
\ee

\noindent  The presence of scalar coupling
in pure Fermi superallowed transitions would therefore show up
as a dependence of the $\F t$ values
on $\langle W^{-1} \rangle$ rather than being
constant.  A fit of the data in Table \ref{Exres} to $\gamma \langle
W^{-1} \rangle$ gives a result of $b = -0.005 \pm 0.007$ or
$\mids b \mids \ \leq 0.017$ at the 90\% confidence level.  In the
limit of no right-hand currents, $b$ is given by $-2C_S/C_V$,
with the minus sign for a positron emitter, and hence
$\mids C_S/C_V \mids \ \leq 0.008$.  This result is less restrictive
than earlier fits \cite{Ha90,HT75,SS81,OBH89,CDQ94} of this type, mainly
because the current set of $\F t$ values shows more dispersion
than the earlier data.

{\em Commentary}.  Any evidence for right-hand currents or scalar
interactions mentioned in the last two sections should be taken
with extreme caution.  Generally, such evidence depends on a
number of assumptions, some of which are not always justified.
In the past, when an experimental anomaly has been reported it has
usually receded in the light of subsequent, improved
experiments.  As Deutsch and Quin \cite{DQ94} put it: ``The
Standard Model is discouragingly successful".

\par


\bigskip
\begin{thebibliography} {999}

\bibitem{WSG}
S. Weinberg, \prl {\bf 19}, 1264 (1967);
A. Salam in {\it Elementary Particle Theory}, ed. N. Svartholm (Almquist and
Wiksells, Stockholm, 1969) p.367; S.L. Glashow, J. Iliopoulos and L. Maiani,
\pr {\bf D2}, 1285 (1970)

\bibitem{deWS86}
B. de Wit and J. Smith, {\it Field Theory in Particle Physics} (North
Holland, Amsterdam, 1986)

\bibitem{LLM92}
P. Langacker, M. Luo and A.K. Mann, Rev. Mod. Phys. {\bf 64}, 87 (1992)

\bibitem{KM73}
M. Kobayashi and T. Maskawa, Prog. Theor. Phys. {\bf 49}, 652 (1973)

\bibitem{Cab63}
N. Cabibbo, \prl {\bf 53}, 1802 (1984)

\bibitem{PDG94}
Particle Data Group, \pr {\bf D50}, 1173 (1994)

\bibitem{Ho89}
B.R. Holstein, {\it Weak Interactions in Nuclei}
(Princeton Univ. Press, Princeton, 1989)

\bibitem{Be68}
J. Bernstein, {\it Elementary particles and their currents}
(Freeman, San Francisco, 1968)

\bibitem{Lee81}
T.D. Lee, {\it Particle Physics and Introduction to Field Theory}
(Harwood Academic, New York, 1981)

\bibitem{We58}
S. Weinberg, \pr {\bf 112}, 1375 (1958)

\bibitem{FG58}
R.P. Feynman and M. Gell-Mann, \pr {\bf 109}, 193 (1958)

\bibitem{Ha90}
J.C. Hardy, I.S. Towner, V.T. Koslowsky, E. Hagberg and H. Schmeing,
\np {\bf A509}, 429 (1990)

\bibitem{Ba84}
P.H. Barker and R.E. White, \pr {\bf C29}, 1530 (1984)

\bibitem{Ba88}
P.H. Barker and S.M. Ferguson, \pr {\bf C38}, 1936 (1988)

\bibitem{Ba89}
S.C. Baker, M.J. Brown and P.H. Barker, \pr {\bf C40}, 940 (1989)

\bibitem{Ba90}
P.H. Barker and G.D. Leonard, \pr {\bf C41}, 246 (1990)

\bibitem{Az74}
G. Azuelos, J.E. Crawford and J.E. Kitching, \pr {\bf C9}, 1213 (1974)

\bibitem{Kr91}
M.A. Kroupa, S.J. Freedman, P.H. Barker and S.M. Ferguson,
Nucl. Instr. and Meth., {\bf A310}, 649 (1991)

\bibitem{Na91}
Y. Nagai, K. Kunihiro, T. Toriyama, S. Harada, Y. Torii, A. Yoshida,
T. Nomura, J. Tanaka and T. Shinozuka, \pr {\bf C43}, R9 (1991)

\bibitem{Ro72}
D.C. Robinson, J.M. Freeman and T.T. Thwaites, \np {\bf A181}, 645 (1972)

\bibitem{Sa95}
G. Savard, A. Galindo-Uribarri, E. Hagberg, J.C. Hardy, V.T. Koslowsky,
D.C. Radford
and I.S. Towner, \prl {\bf 74}, 1521 (1995)

\bibitem{Ki91}
S.W. Kikstra, Z. Guo, C. van der Leun, P.M. Endt, S. Raman, T.A. Walkiewicz,
J.W. Starner, E.T. Jueney and I.S. Towner,
\np {\bf A529}, 39 (1991)

\bibitem{Wa92}
T.A. Walkiewicz, S. Raman, E.T. Jurney, J.W. Starner and J.E. Lynn,
\pr {\bf C45}, 1597 (1992)

\bibitem{Ko94a}
V.T. Koslowsky \etal, to be published;  preliminary results appear in
E. Hagberg, V.T. Koslowsky, I.S. Towner, J.G. Hykawy, G. Savard, T. Shinozuka,
P.P. Unger and H. Schmeing, {\it Nuclei far from Stability/Atomic Masses
and Fundamental Constants} (Institute of Physics Conference Series \# 132,
ed. R. Neugart and A. W\"{o}hr, 1994) p.783

\bibitem{Ko94b}
V.T. Koslowsky, E. Hagberg, J.C. Hardy, H.Schmeing and I.S. Towner,
Preprint TASCC-P-94-18, to be published

\bibitem{Br94}
S.A. Brindhaban and P.H. Barker, \pr {\bf C49}, 2401 (1994)

\bibitem{Li94}
S. Lin, S.A. Brindhaban and P.H. Barker, \pr {\bf C49}, 3098 (1994)

\bibitem{Am94}
P.A. Amundsen and P.H. Barker, \pr {\bf C50}, 2466 (1994)

\bibitem{Ha94}
E. Hagberg, V.T. Koslowsky, J.C. Hardy, I.S. Towner, J.G. Hykawy, G. Savard
and T. Shinozuka, \prl {\bf 73}, 396 (1994)

\bibitem{Ko87a}
V.T. Koslowsky, J.C. Hardy, E. Hagberg, R.E. Azuma, G.C. Ball, E.T.H. Clifford,
W.G. Davies, H. Schmeing, U.J. Schrewe and K.S. Sharma,
\np {\bf A472}, 419 (1987)

\bibitem{Ko87b}
V.T. Koslowsky, R.E. Azuma, H.C. Evans, E. Hagberg, J.C. Hardy, J.R. Leslie,
H.Schmeing and C. Virtue, \np {\bf A472}, 408 (1987)

\bibitem{Sc66}
H. Schopper, {\it Weak Interactions and Nuclear Beta Decay}
(North-Holland, Amsterdam, 1966)

\bibitem{To73}
I.S. Towner and J.C. Hardy, \np {\bf A205}, 33 (1973)

\bibitem{Or89}
W.E. Ormand and B.A. Brown, \prl {\bf 62}, 866 (1989); \np {\bf A440},
274 (1985)

\bibitem{To77}
I.S. Towner, J.C. Hardy and M. Harvey, \np {\bf A284}, 269 (1977)

\bibitem{Ma86}
W.J. Marciano and A. Sirlin, \prl {\bf 56}, 22 (1986)

\bibitem{To92}
I.S. Towner, \np {\bf A540}, 478 (1992)

\bibitem{Si67}
A. Sirlin, \prl {\bf 164}, 1767 (1967)

\bibitem{Si87}
A. Sirlin, \pr {\bf D35}, 3423 (1987); W. Jaus and G. Rasche, \pr {\bf D35},
3420 (1987); A. Sirlin and R. Zucchini, \prl {\bf 57}, 1994 (1986)

\bibitem{To94}
I.S. Towner, \pl {\bf B333}, 13 (1994)

\bibitem{GHK92}
A. Garcia, R. Huerta and P. Kielanowski, \pr {\bf D45}, 879 (1992)

\bibitem{Wi90}
D.H. Wilkinson, \np {\bf A511}, 301 (1990); Nucl. Inst. and Method
{\bf 335}, 172,182,201 (1993); Zeit. Phys. {\bf A348}, 129 (1994)

\bibitem{HP77}
W.Y.P. Hwang and H. Primakoff, \pr {\bf C16}, 397 (1977)

\bibitem{BS75}
H. Behrens and L. Szybisz, Zeit. Phys. {\bf A273}, 177 (1975)

\bibitem{De70}
J. Delorme, \np {\bf B19}, 573 (1970)

\bibitem{Ch73}
B.T. Chertok \etal , \pr {\bf C8}, 23 (1973)

\bibitem{Sp72}
E. Spamer \etal , in {\it Proc of Int Conf of Nuclear Electron
Scattering and Photoreaction}, Sendai, Japan, eds. K. Shoda and
H. Ui (Research Report of Lab. of Nuclear Science, Tohoku
Univ., 1972)

\bibitem{De83}
U. Deutschmann, G. Lahm, R. Neuhaussen and J.C. Bergstrom,
\np {\bf A411}, 337 (1983)

\bibitem{De92}
L. De Braeckeleer, \pr {\bf C45}, 1935 (1992)

\bibitem{CH76}
F.P. Calaprice and B.R. Holstein, \np {\bf A273}, 301 (1976)

\bibitem{Aj85}
F. Ajzenberg-Selove, \np {\bf A433}, 1 (1985)

\bibitem{Wi70}
D.H. Wilkinson, \pl {\bf B31}, 447 (1970)

\bibitem{WA71}
D.H. Wilkinson and D.E. Alburger, \prl {\bf 26}, 1127 (1971)

\bibitem{Wi71}
D.H. Wilkinson, \prl {\bf 27}, 1018 (1971)

\bibitem{To73b}
I.S. Towner, \np {\bf A216}, 589 (1973)

\bibitem{LMW63}
Y.K. Lee, L. Mo and C.S. Wu, \prl {\bf 10}, 253 (1963)

\bibitem{WLM77}
C.S. Wu, Y.K. Lee and L. Mo, \prl {\bf 39}, 72 (1977)

\bibitem{Ka77}
W. Kaina, V. Soergel, H. Thies and W. Trost,
\pl {\bf B70}, 411 (1977)

\bibitem{Gr85}
L. Grenacs, Ann. Rev. Nucl. Part. Sci. {\bf 35}, 455 (1985)

\bibitem{MM62}
T. Mayer-Kuckuk and T.C. Michel, \pr {\bf 127}, 545 (1962)

\bibitem{GP63}
N.W. Glass and R.W. Peterson, \pr {\bf 130}, 253 (1963)

\bibitem{Mi89}
T. Minamisono \etal , in {\it Proc. of Yamada Conf. XXIII Nuclear
Weak Process and Nuclear Structure}, Osaka, Japan  eds. M. Morita,
H. Ejiri, H. Ohtsubo and T. Sato (World-Scientific, Singapore,
1989) p.58

\bibitem{Mi73}
T. Minamisono, J. Phys. Soc. Jpn. Suppl. {\bf 34}, 324 (1973)

\bibitem{HM73}
R.C. Haskell and L. Madansky, J. Phys. Soc. Jpn. Suppl. {\bf 34}, 167
(1973)

\bibitem{HCM75}
R.C. Haskell, F.D. Correll and L. Madansky, \pr {\bf B11}, 3268 (1975)

\bibitem{STG75}
K. Sugimoto, I. Tanihata and J. Goring, \prl {\bf 34}, 1533 (1975)

\bibitem{Le78}
P. Lebrun, Ph. Deschepper, L. Grenacs, J. Lehmann, C. Leroy, L. Palffy,
A. Possoz and A. Maio, \prl {\bf 40}, 302 (1978)

\bibitem{Br78a}
H. Br\"{a}ndle, L. Grenacs, J. Lang, L.Ph. Roesch, V.L. Telegdi,
P.Truttmann, A.Weiss and A. Zehnder, \prl {\bf 40}, 306 (1978)

\bibitem{Su78}
K. Sugimoto, T. Minamisono, Y. Nojiri and Y. Masuda,
J. Phys. Soc. Jpn. Suppl. {\bf 44}, 801 (1978)

\bibitem{Br78b}
H. Br\"{a}ndle, G. Miklos, L.Ph. Roesch, V.L. Telegdi,
P.Truttmann, A. Zehnder, L. Grenacs, P.Lebrun and J. Lehmann,
\prl {\bf 41}, 299 (1978)

\bibitem{Ma79}
Y. Masuda, T. Minamisono, Y. Nojiri and K. Sugimoto,
\prl {\bf 43}, 1083 (1979)

\bibitem{Mi86}
T. Minamisono, J. Phys. Soc. Jpn. Suppl. {\bf 55}, 382 (1986)

\bibitem{Mi93}
T. Minamisono, A. Kitagawa, K. Matsuta and Y. Nojiri,
Hyperfine Interactions {\bf 78}, 77 (1993)

\bibitem{De94}
L. De Braeckeleer, \np {\bf A577}, 383 (1994)

\bibitem{BG78}
T.J. Bowles and G.T. Garvey, \pr {\bf C18}, 1447 (1978)

\bibitem{BG82}
T.J. Bowles and G.T. Garvey, \pr {\bf C26}, 2336 (1982)

\bibitem{TG75}
R.E. Tribble and G.T. Garvey, \pr {\bf C12}, 967 (1978)

\bibitem{MGG80}
B.D. McKeown, G.T. Garvey and C.A. Gagliardi, \pr {\bf C22}, 738 (1980)

\bibitem{DR78}
N. Dupuis-Rolin, J. Deutsch, J.P. Favart and R. Prieels,
\pl {\bf 79B}, 359 (1978)

\bibitem{TM78}
R.E. Tribble and D.P. May, \pr {\bf C18}, 2704 (1978)

\bibitem{TMT81}
R.E. Tribble, D.P. May and D.M. Tanner, \pr {\bf C23}, 2245 (1981)

\bibitem{CB83}
E.D. Commins and P.H. Bucksbaum, {\it Weak Interactions of Leptons
and Quarks} (Cambridge University Press, Cambridge, 1983)

\bibitem{Re90}
P. Renton, {\it Electroweak Interactions} (Cambridge University Press,
Cambridge, 1990)

\bibitem{GL60}
M. Gell-Mann and M. Levy, Nuovo Cimento {\bf 16}, 705 (1960)

\bibitem{LN68}
B.W. Lee and H.T. Nieh, \pr {\bf 166}, 1507 (1968)

\bibitem{We67}
S. Weinberg, \prl {\bf 18}, 507 (1967)

\bibitem{IT79}
E. Ivanov and E. Truhlik, \np {\bf A316}, 437 (1979)

\bibitem{BCC73}
D.V. Bugg, A.A. Carter and J.R. Carter, \pl {\bf B44}, 278 (1973)

\bibitem{KH80}
R. Koch and E. Pietarinen, \np {\bf A336}, 331 (1980)

\bibitem{Nij93}
V.G.J. Stoks, R.G.E. Timmermans and J.J. de Swart, \pr {\bf C47}, 512
(1993)

\bibitem{AWP94}
R.A. Arndt,  R.L. Workman and M. Pavan,
\pr {\bf C49}, 2729 (1994)

\bibitem{BM94}
D.V. Bugg and R. Machleidt, preprint {\it $\pi$NN Coupling Constants
from NN Elastic Data between 210 and 800 MeV} (1994)

\bibitem{Ah88}
L.A. Ahrens \etal , \pl {\bf B202}, 284 (1988)

\bibitem{BKM94}
V. Bernard, N. Kaiser and U-G. Meissner, \pr {\bf D50}, 6899 (1994)

\bibitem{Du91}
D. Dubbers, Progress in Particle and Nuclear Physics, {\bf 26}, 173
(1991)

\bibitem{Au93}
G. Audi and A.H. Wapstra, \np {\bf A565}, 1 (1993)

\bibitem{Va93}
R.S. Van Dyck, D.L. Farnham and P.B. Schwinberg, in {\it Nuclei
Far From Stability/ Atomic Masses and Fundamental
Constants} (Institute of Physics Conference Series \# 132,
ed. R. Neugart and A. W\"{o}hr, 1994) p.3

\bibitem{Gr86}
G.L. Greene, E.G. Kessler, R.D. Deslattes and H. B\"{o}rner,
\prl {\bf 56}, 819 (1986)

\bibitem{Sc92}
K. Schreckenbach and W. Mampe, J. Phys. {\bf G18}, 1 (1992)

\bibitem{Ch72}
C.J. Christensen, A. Neilsen, A. Bahnsen, W.K. Brown and B.M. Rustad,
\pr {\bf D5}, 1628 (1972)

\bibitem{Sp88}
P.E. Spivak, Sov. Phys. JETP {\bf 67}, 1735 (1988)

\bibitem{La88}
J. Last, M. Arnold, J. D\"{o}hner, D. Dubbers and S.J. Freedman,
\prl {\bf 60}, 995 (1988)

\bibitem{Ko89}
R. Kossakowski, P.Grivot, P. Liaud, K. Schreckenbach and G. Azuelos,
\np {\bf A503}, 473 (1989)

\bibitem{By90}
J. Byrne, P.G. Dawber, J.A. Spain, A.P. Williams, M.S. Dewey,
D.M. Gilliam, G.L. Greene, G.P. Lamaze, R.D. Scott, J. Pauwels,
R. Eykens and A. Lamberty, \prl {\bf 65}, 289 (1990)

\bibitem{Ko86}
Yu.Yu. Kosvintsev, V.I. Morozov and G.I. Terekhov, JETP Lett.
{\bf 44}, 571 (1986)

\bibitem{Mo89}
V.I. Morozov, Nucl. Inst. and Meth. {\bf A284}, 108 (1989)

\bibitem{Pa89}
W. Paul, F. Anton, L. Paul, S. Paul and W. Mampe, Z. Phys.
{\bf C45}, 25 (1989)

\bibitem{Ma89}
W. Mampe, P.Ageron, J.C. Bates, J.M. Pendlebury and A. Steyerl,
\prl {\bf 63}, 593 (1989) and Nucl. Inst. and Meth. {\bf A284},
111 (1989)

\bibitem{Ne92}
V.V. Nesvizhevskii, A.P. Serebrov, R.R. Tal'daev, A.G. Kharitonov,
V.P. Alfimenkov, A.V. Strelkov and V.N. Shvetsov, Sov. Phys. JETP
{\bf 75}, 405 (1992)

\bibitem{Ma93}
W. Mampe, L.N. Bondarenko, V.I. Morozov, Yu. N. Panin and A.I. Fomin,
JETP Lett. {\bf 57}, 82 (1993)

\bibitem{Kr75}
V.E. Krohn and G. R. Ringo, \pl {\bf 55B}, 175 (1975)

\bibitem{Wi82}
D.H. Wilkinson, \np {\bf A377}, 474 (1982)

\bibitem{Er79}
B.G. Erozolimskii, A.I. Frank, Yu. A. Mostovoi, S.S. Arzumanov
and L.R. Voitzek. Sov. J. Nucl. Phys. {\bf 30}, 356 (1979)

\bibitem{Bo86}
P. Bopp, D. Dubbers, L. Hornig, E. Klemt, J. Last, H. Sch\"{u}tze,
S.J. Freedman and O. Sch\"{a}rpf, \prl {\bf 56}, 919 (1986)

\bibitem{Er91}
B.G. Erozolimskii, I.A. Kuznetsov, I.V. Stepanenko, I.A. Kuida
and Yu. A. Mostovoi, \pl {\bf B263}, 33 (1991)

\bibitem{Sc95}
K. Schreckenbach, P.Liaud, R. Kossakowski, H. Nastoll, A. Bussiere
and J.P. Guillaud, to be published, 1995

\bibitem{St78}
C. Stratowa, R. Dobrozemsky and P. Weinzierl, \pr {\bf D18}, 3970 (1978)

\bibitem{Gr90}
K. Green and D. Thompson, J. Phys. {\bf G16}, L75 (1990)

\bibitem{So87}
X. Song, J. Phys. {\bf G13}, 1023 (1987)

\bibitem{Bl62}
E.J. Bleser, L.M. Lederman, J.L. Rosen, J.E. Rothberg and
E. Zavattini, \prl {\bf 8}, 288 (1962)

\bibitem{Ro63}
J.E. Rothberg, E.W. Anderson, E.J. Bleser, L.M. Lederman, L.S.Meyer,
J.L. Rosen and I.T. Wang, \pr {\bf 132}, 2664 (1963)

\bibitem{Al69}
A. Alberigi Quaranta, A. Bertin, G. Matone, F.Palmonari, G. Torelli,
P.Dalpiaz, A. Placci and E. Zavattini, \pr {\bf 117}, 2118 (1969)

\bibitem{By74}
V.M. Bystritskii \etal, Sov. Phys. JETP {\bf 39}, 19 (1974)

\bibitem{Ba81a}
G. Bardin \etal, \np {\bf A352}, 365 (1981)

\bibitem{Ba81b}
G. Bardin \etal, \pl {\bf B104}, 320 (1981)

\bibitem{BF87}
D.S. Beder and H. Fearing, \pr {\bf D35}, 2130 (1987)

\bibitem{Ha93}
M.D. Hasinoff \etal, {\it Proc. of Third Int. Symp. on Weak and
Electromagnetic
Interactions in Nuclei}, ed. Ts.D. Vylov (World Scientific, Singapore,
1993) p.312

\bibitem{LL87}
H.J. Lipkin and T.S.H. Lee, \pl {\bf B183}, 22 (1987)

\bibitem{EFT73}
M. Ericson, A. Figureau and C. Thevevet, \pl {\bf B45}, 19 (1973)

\bibitem{Rho74}
M. Rho, \np {\bf A231}, 493 (1974)

\bibitem{OW74}
K. Ohta and M. Wakamatsu, \np {\bf A234}, 445 (1974)

\bibitem{SIA78}
K. Shimizu, M. Ichimura and A. Arima, \np {\bf A226}, 282 (1978)

\bibitem{BH82}
G.F. Bertsch and I. Hamamoto, \pr {\bf C26}, 607 (1982)

\bibitem{BS68}
D.M. Brink and G.R. Satchler, {\it Angular Momentum}
(Clarendon, Oxford, 1968)

\bibitem{Br92}
B.A. Brown, \prl {\bf 69}, 1034 (1992)

\bibitem{Wi74}
D.H. Wilkinson, \np {\bf A225}, 365 (1974)

\bibitem{BW85}
B.A. Brown and B.H. Wildenthal, At. Data Nucl. Data Tables {\bf 33}, 347 (1985)

\bibitem{ASBH87}
A. Arima, K. Shimizu, W. Bentz and H. Hyuga, Advances in Nuclear Physics
{\bf 18}, 1 (1987)

\bibitem{TK79}
I.S. Towner and F.C. Khanna, \prl {\bf 42}, 51 (1979)

\bibitem{To87}
I.S. Towner, Phys. Reports  {\bf 155}, 263 (1987)

\bibitem{CR71}
M. Chemtob and M. Rho, \np {\bf A163}, 1 (1971)

\bibitem{Os92}
F. Osterfeld, Rev. Mod. Phys. {\bf 64}, 491 (1992)

\bibitem{Ta87}
T.T. Taddeucci \etal, \np {\bf A469}, 125 (1987)

\bibitem{Os82}
F. Osterfeld, \pr {\bf C26}, 762 (1982)

\bibitem{OCS85}
F. Osterfeld, D. Cha and J. Speth, \pr {\bf C31}, 375 (1985)

\bibitem{Ba80}
D.E. Bainum \etal, \prl {\bf 44}, 1751 (1980)

\bibitem{Go80}
C.D. Goodman \etal, \prl {\bf 44}, 1755 (1980)

\bibitem{Ho80}
D.J. Horen \etal, \pl {\bf B95}, 27 (1980)

\bibitem{Ga81}
C. Gaarde \etal, \np {\bf A369}, 258 (1981)

\bibitem{An80}
B.D. Anderson \etal, \prl {\bf 45}, 699 (1980)

\bibitem{IFF63}
K.I. Ikeda, S. Fujii and J.I. Fujita, \pl {\bf 3}, 271 (1963)

\bibitem{Ga83}
C. Gaarde, \np {\bf A396}, 127c (1983)

\bibitem{BM81}
A. Bohr and B. Mottelson, \pl {\bf B100}, 10 (1981)

\bibitem{SW87}
R.D. Smith and J. Wambach, \pr {\bf C36}, 2704 (1987)

\bibitem{Ve89}
M.C. Vetterli \etal, \pr {\bf C40}, 559 (1989)

\bibitem{Ra83}
J. Rapaport \etal, \np {\bf A410}, 371 (1983)

\bibitem{Ro93}
T. R\"{o}nnqvist \etal, \np {\bf A563}, 225 (1993)

\bibitem{Go90}
C.D. Goodman, J. Rapaport and S.D. Bloom, \pr {\bf C42}, 1150 (1990)

\bibitem{Ga91}
A. Garc\'{\i}a \etal, \prl {\bf 67}, 3654 (1991);
E.G. Adelberger, A. Garc\'{\i}a, P.V. Magnus and D.P. Wells,
 \prl {\bf 67}, 3658 (1991)

\bibitem{Ra81}
J. Rapaport \etal, \prl {\bf 47}, 1518 (1981)

\bibitem{Au92}
M.B. Aufderheide, S.D. Bloom, D.A. Resler and C.D. Goodman,
 \pr {\bf C46}, 2251 (1992);
C.D. Goodman, M.B. Aufderheide, S.D. Bloom and D.A. Resler,
 \prl {\bf 69}, 2445 (1992)

\bibitem{Ra92}
J. Rapaport and E.R. Sugarbaker, \prl {\bf 69}, 2444 (1992)

\bibitem{Ga95}
A. Garc\'{\i}a \etal, \pr {\bf C51}, 439 (1995)

\bibitem{Wi84}
B.H. Wildenthal, Prog. Part. Nucl. Phys. {\bf 11}, 5 (1984)

\bibitem{RT65}
H.P.C. Rood and H.A. Tolhoek, \np {\bf 70}, 658 (1965)

\bibitem{Fe66}
H.W. Fearing, \pr {\bf 146}, 723 (1966)

\bibitem{BD70}
E. Borchi and S. De Gennaro, \pr {\bf C2}, 1012 (1970)

\bibitem{Ch81}
P. Christillin, \np {\bf A362}, 391 (1981)

\bibitem{GT87}
M. Gmitro and P. Tru\"{o}l, Adv. Nucl. Phys. {\bf 18}, 241 (1987)

\bibitem{CRS80}
P. Christillin, M. Rosa-Clot and S. Servadio, \np {\bf A345}, 331 (1980)

\bibitem{FW92}
H.W. Fearing and M.S. Welsh, \pr {\bf C46}, 2077 (1992)

\bibitem{CG85}
P. Christillin and M. Gmitro, \pl {\bf B150}, 50 (1985)

\bibitem{RN90}
F. Roig and J. Navarro, \pl {\bf B236}, 393 (1990)

\bibitem{GOT86}
M. Gmitro, A.A. Ovchinnikova and T.V. Tetereva, \np {\bf 453}, 685 (1986)

\bibitem{Do88}
M. D\"{o}beli \etal, \pr {\bf C37}, 1633 (1988)

\bibitem{Ar91}
D.S. Armstrong \etal, \pr {\bf C43}, 1425 (1991)

\bibitem{Ar92}
D.S. Armstrong \etal, \pr {\bf C46}, 1094 (1992)

\bibitem{Ro82}
L.Ph. Roesch, V.L. Telegdi, P. Truttmann, A. Zehnder, L. Grenacs and
L. Palffy, \prl {\bf 46}, 1507 (1981);
Helv. Phys. Acta {\bf 55}, 74 (1982)

\bibitem{Pa75}
L. Palffy \etal, \prl {\bf 34}, 212 (1975)

\bibitem{Ga83}
C.A. Gagliardi, G.T. Garvey, J.R. Wrobel and S.J. Freedman, \pr
{\bf C28}, 2423 (1983);
H. Heath and G.T. Garvey, \pr {\bf C31}, 2190 (1985);
L.A. Hamil, L. Lessard, H. Jeremie and J. Chauvin, Zeit. Phys.
{\bf A321}, 439 (1985);
T. Minamisono, K. Takeyama, T. Ishigai, H. Takeshima, Y. Nojiri and
K. Asahi, \pl {\bf B130}, 1 (1983)

\bibitem{Wa94}
E.K. Warburton, I.S. Towner and B.A. Brown, \pr {\bf C49}, 824 (1994)

\bibitem{HJ90}
W.C. Haxton and C. Johnson, \prl {\bf 65}, 1325 (1990)

\bibitem{Go94}
T.P. Gorringe \etal, \prl {\bf 72}, 3472 (1994)

\bibitem{KLV94}
E. Kolbe, K. Langanke and P. Vogel, \pr {\bf C50}, 2576 (1994)

\bibitem{Er71}
M. Ericson, Ann. Phys. {\bf 63}, 562 (1971)

\bibitem{DEFT76}
J. Delorme, M. Ericson, A. Figureau and C. Thevenet, Ann. Phys.
{\bf 102}, 273 (1976)

\bibitem{Er78}
M. Ericson, Prog. Part. Nucl. Phys. {\bf 1}, 67 (1978)

\bibitem{DE94}
J. Delorme and M. Ericson, \pr {\bf C49},1763 (1994)

\bibitem{KR94}
M. Kirchbach and D.O. Riska, \np {\bf A578}, 511 (1994)

\bibitem{AD66}
S.L. Adler and Y. Dothan, \pr {\bf 151}, 1267 (1966)

\bibitem{AD68}
S.L. Adler and R.F. Dashen, {\it Current Algebras}
(Benjamin, New York, 1968)

\bibitem{AFFR73}
V. de Alfaro, S. Fubini, G. Furlan and C. Rosetti, {\it Currents in
Hadron Physics} (North-Holland, Amsterdam, 1973)

\bibitem{KDR78}
K. Kubodera, J. Delorme and M. Rho, \prl {\bf 40}, 755 (1978)

\bibitem{To92b}
I.S. Towner, \np {\bf A542}, 631 (1992); \np {\bf A569}, 834 (1994)

\bibitem{To86}
I.S. Towner, Ann. Rev. Nucl. Part. Sci. {\bf 36}, 115 (1986);
I.S. Towner, Comments Nucl. Part. Phys. {\bf 15}, 145 (1986)

\bibitem{De82}
J. Delorme, \np {\bf A374}, 541 (1982)

\bibitem{Mo76}
M. Morita, M. Nishimura, A. Shimizu, H. Ohtsubo and K. Kubodera,
Prog. Theor. Phys. Suppl. {\bf 60}, 1 (1978);
M. Morita, M. Nishimura and H. Ohtsubo, \pl {\bf 73B}, 17 (1978)

\bibitem{Ko81}
K. Koshigiri, H. Ohtsubo and M. Morita, Prog. Theor. Phys.
{\bf 66}, 358 (1981);
K. Koshigiri, K. Kubodera, H. Ohtsubo and M. Morita,
{\it Proc. Int. Conf. on Nuclear Weak Process and Nuclear Structure},
eds. M. Morita, H. Ejiri, H. Ohtsubo and T. Sato (World-Scientific,
Singapore, 1989) p.52

\bibitem{GS82}
P.A.M. Guichon and C. Samour, \np {\bf A382}, 461 (1982)

\bibitem{WBBM88}
E.K. Warburton, J.A. Becker, B.A. Brown and D.J. Millener,
Ann. Phys. {\bf 187}, 471 (1988)

\bibitem{Wa91}
E.K. Warburton, \pr {\bf C44}, 233 (1991); \prl {\bf 66}, 1823 (1991)

\bibitem{BB82}
H. Behrens and W. B\"{u}hring, {\it Electron Radial Wave Functions
and Nuclear Beta Decay} (Clarendon, Oxford, 1982)

\bibitem{WTB94}
E.K. Warburton, I.S. Towner and B.A. Brown, \pr {\bf C49}, 824 (1994)

\bibitem{WT92}
E.K. Warburton and I.S. Towner, \pl {\bf B294}, 1 (1992)

\bibitem{WT94}
E.K. Warburton and I.S. Towner, Phys. Reports  {\bf 242}, 103 (1994)

\bibitem{KRT92}
M. Kirchbach, D.O. Riska and K. Tsushima, \np {\bf A542}, 616 (1992)

\bibitem{PTK94}
T.-S. Park, I.S. Towner and K. Kubodera, \np  {\bf A579}, 381 (1994)

\bibitem{KR91}
K. Kubodera and M. Rho, \prl {\bf 67}, 3479 (1991)

\bibitem{BBMS77}
M.A.B. B\'{e}g, R.V. Budny, R. Mohapatra and A. Sirlin,
\prl {\bf 38}, 1252 (1977)

\bibitem{HT77}
B.R. Holstein and S.B. Treiman, \pr {\bf D16}, 2369 (1977)

\bibitem{CDH88}
A.-S. Carnoy, J. Deutsch and B.R. Holstein, \pr {\bf D38}, 1636 (1988)

\bibitem{GSS90}
Yu. V. Gaponov, P.E. Spivak and N.B. Shul'gina, Sov. J. Nucl. Phys.
{\bf 52}, 1042 (1990)

\bibitem{AFG91}
M. Aquino, A. Fernandez and A. Garcia, \pl {\bf B261}, 280 (1991)

\bibitem{Ca92}
A.-S. Carnoy, J. Deutsch, R. Prieels, N. Severijns and P.A. Quin,
J. Phys. G: Nucl. Part. Phys. {\bf 18}, 823 (1992)

\bibitem{vK66}
J. van Klinken, \np {\bf 75}, 145 (1966)

\bibitem{Wi87}
V.A. Wichers \etal , \prl {\bf 58}, 1821 (1987)

\bibitem{Ca91}
A.-S. Carnoy, J. Deutsch, T.A. Girard and R. Prieels, \prl {\bf 65},
3249 (1990); \pr {\bf C43}, 2825 (1991)

\bibitem{Jo86}
A. Jodidio \etal , \pr {\bf D34}, 1967 (1986); \pr {\bf D37}, 237(E)
(1988)

\bibitem{DQ94}
J. Deutsch and P. Quin, in {\em Precision Tests of the Standard
Electroweak Model} ed. P. Langacker (World-Scientific, Singapore, 1994)
to be published

\bibitem{JTW57}
J.D. Jackson, S.B. Treiman and H.W. Wyld, \pr {\bf 106}, 517 (1957)

\bibitem{BMV84}
A.I. Boothroyd, J. Markey and P. Vogel, \pr {\bf C29}, 603 (1984)

\bibitem{JPC63}
C.H. Johnson, F. Pleasanton and T.A. Carlson, \pr {\bf 132}, 1149 (1963)

\bibitem{Ad93}
E.G. Adelberger, \prl {\bf 70}, 2856 (1993)

\bibitem{SR93}
D. Schardt and K. Riisager, Z. Phys. {\bf A345}, 265 (1993)

\bibitem{HT75}
J.C. Hardy and I.S. Towner, \np {\bf A254}, 221 (1975)

\bibitem{SS81}
L. Szybisz and V. Silbergleit, J. Phys. {\bf G7}, L201 (1981)

\bibitem{OBH89}
W.E. Ormand, B.A. Brown and B.R. Holstein, \pr {\bf C40}, 2914 (1989)

\bibitem{CDQ94}
A.S. Carnoy, J. Deutsch and P. Quin, \np {\bf A568}, 265 (1994)

\end{thebibliography}
\end{document}